\documentclass[11pt, a4paper]{article}
\usepackage{amsmath, amssymb, amsthm}
\usepackage{graphicx}
\usepackage{booktabs}
\usepackage{algorithm}
\usepackage{algpseudocode}
\usepackage[margin=1in]{geometry}
\usepackage{hyperref}
\usepackage{url}
\usepackage{multirow}
\usepackage{subcaption}
\usepackage[table]{xcolor}

\graphicspath{ {.} }

\newtheorem{remark}{Remark}

\title{\textbf{Order Out of Noise and Disorder: \\
 Fate of the Frustrated Manifold}}

\author{Igor Halperin\thanks{Simulation codes and videos are available at https://github.com/ighalp/frustrated-brownian-particles-manifolds. Simulations and analyzes presented in this work were done using Claude Code with Opus 4.5. Email for correspondence: ighalp@gmail.com.}\thanks{I thank Andrey Itkin and Michael Varenberg for interesting discussions. All remaining errors are my own.}}
\date{\today}

\begin{document}

\maketitle

\begin{abstract}
We study Langevin dynamics of $N$ Brownian particles on two-dimensional Riemannian manifolds, interacting through pairwise potentials linear in geodesic distance with quenched random couplings. These \emph{frustrated Brownian particles} experience competing demands of random attractive and repulsive interactions while confined to curved surfaces. We consider three geometries: the sphere $S^2$, torus $T^2$, and bounded cylinder. Our central finding is disorder-induced dimension reduction with spontaneous rotational symmetry breaking: order emerges from two sources of randomness (thermal noise and quenched disorder), with manifold topology determining the character of emerging structures. Glassy relaxation drives particles from 2D distributions to quasi-1D structures: bands on $S^2$, rings on $T^2$, and localized clusters on the cylinder. Unlike conventional symmetry breaking, the symmetry-breaking direction is not frozen but evolves slowly via thermal noise. On the sphere, the structure normal precesses diffusively on the Goldstone manifold with correlation time $\tau_c \approx 18$, a classical realization of type-A dissipative Nambu-Goldstone dynamics. The model requires no thermodynamic gradients, no fine-tuning, and no slow external input. We discuss connections to spin glass theory, quantum field theory, astrophysical structure formation, and self-organizing systems. The model admits a large-$N$ limit yielding statistical field theory on Riemannian surfaces, while remaining experimentally realizable in colloidal and soft matter systems.
\end{abstract}

\tableofcontents

\section{Introduction}

The relaxation dynamics of disordered systems toward low-energy configurations is a central theme in statistical physics, with spin glasses serving as the paradigmatic example \cite{mezard1987}. A universal feature of such glassy systems is the emergence of complex energy landscapes with numerous metastable states, leading to slow, history-dependent relaxation and aging phenomena. Less explored is the spatial manifestation of this relaxation: how does the geometry of particle configurations evolve as frustrated systems seek equilibrium on curved spaces?

In this work, we investigate a model of glassy non-equilibrium dynamics that combines the essential physics of spin glasses (random competing interactions) with geometric constraints provided by compact Riemannian manifolds. We study $N$ Brownian particles confined to move on compact two-dimensional surfaces, interacting through pairwise potentials linear in geodesic distance with quenched random coupling constants $\phi_{ij}$. These \emph{frustrated Brownian particles} experience standard thermal noise from their environment while subject to the competing demands of random couplings, making the model physically realizable in soft matter and colloidal systems. We consider three geometries: the sphere $S^2$, torus $T^2$ (closed manifolds without boundary), and bounded cylinder $S^1 \times [0, H]$ (manifold with boundary).

Our central finding is \textit{disorder-induced dimension reduction accompanied by spontaneous breaking of rotational symmetry}. The glassy relaxation dynamics drives particles from configurations spanning the full two-dimensional manifold to non-equilibrium steady states (NESS) \cite{qian2006, seifert2012, sekimoto2010} concentrated on lower-dimensional submanifolds:

\begin{itemize}
    \item \textbf{Sphere $S^2 \to S^1$}: Particles collapse onto a band near a great circle, breaking SO(3) to SO(2).
    \item \textbf{Bounded cylinder $S^1 \times [0, H] \to \mathbb{Z}_2$}: Particles form localized clusters near boundaries, breaking SO(2) to $\mathbb{Z}_2$.
    \item \textbf{Torus $T^2 \to \coprod_2 S^1$}: Particles coalesce into two diametrically opposite rings, breaking SO(2)$\times$SO(2) to SO(2)$\times\mathbb{Z}_2$.
\end{itemize}

The topological distinction between closed manifolds and manifolds with boundary proves fundamental: on closed manifolds, particles form extended structures (bands, rings) wrapping closed geodesics with continuous residual symmetries, while boundaries induce localized clusters with discrete residual symmetries.

Unlike conventional spontaneous symmetry breaking, the symmetry-breaking direction is not frozen but slowly drifts via thermal noise, while the reduced-dimensional structure persists. On the sphere, the structure normal $\hat{\mathbf{n}}(t)$ (the direction perpendicular to the band plane) slowly precesses on the Goldstone manifold $S^2 \cong \text{SO}(3)/\text{SO}(2)$, analogous to the magnetization direction in a ferromagnet. This slow precession, with correlation time $\tau_c \approx 18$ far exceeding the microscopic thermal time, constitutes a classical realization of dissipative Nambu-Goldstone modes, specifically the type-A diffusion modes in the Minami-Hidaka classification \cite{minami2018, hidaka2020}. The torus and cylinder show more complex behavior due to their different symmetry breaking patterns and geometric constraints.

The model connects to several areas of fundamental physics. The rapid structure formation from localized initial conditions resembles instanton-like transitions between near-zero energy states, linking to instanton physics in quantum field theory \cite{polyakov1975} and spin glasses \cite{lopatin1999}. The dimensional confinement via a potential linear in geodesic distance provides an analogy to quark confinement in QCD \cite{thooft1978, mandelstam1976, polyakov1977}. The spontaneous emergence of ordered patterns from disordered interactions relates our model to self-organizing systems \cite{prigogine1977, cross1993}, though with quenched disorder rather than deterministic instabilities as the organizing principle. Geometric (though not mechanistic) parallels exist to astrophysical structure formation, where angular momentum conservation drives dimensional reduction in disk and ring systems \cite{mo1998, cuzzi2010}; our disorder-driven mechanism provides an alternative route to similar geometric outcomes.

The paper is organized as follows. Section 2 develops the theoretical framework for Langevin dynamics on Riemannian manifolds. Section 3 describes the numerical implementation. Section 3.4 presents simulation results documenting the dynamic dimension reduction. Section 4 discusses connections to spin glass physics, instanton physics, QCD confinement, astrophysical structure formation, and self-organizing systems. Section 5 summarizes our findings. Appendices provide technical details on geometry-specific derivations, rotational diffusion, covariant integration schemes, and analytical methods for non-equilibrium glassy dynamics.

\section{Theoretical Framework}
\label{sec:theory}

\subsection{Langevin Dynamics on Riemannian Manifolds}
\label{sec:langevin_manifolds}

We develop the Langevin dynamics for particles confined to a Riemannian manifold following the approach of Zinn-Justin \cite{zinnjustin2002}: we start with the Langevin equation in the embedding Euclidean space and project onto the manifold to obtain a covariant formulation. This approach is equivalent to the intrinsic formulation pioneered by It\^{o} \cite{ito1962}, who established that the infinitesimal generator of Brownian motion on a Riemannian manifold $\mathcal{M}$ is one-half the Laplace-Beltrami operator, $\frac{1}{2}\Delta_{\mathcal{M}}$. The embedding-space method we employ, following Stroock \cite{stroock1971} and others, provides a constructive approach to implementing Brownian motion on manifolds.

\paragraph{Notation and conventions.}
We adopt standard tensor notation throughout. Greek indices ($\alpha, \beta, \ldots = 1, \ldots, n$) label coordinates in the embedding Euclidean space $\mathbb{R}^n$, while Latin indices ($i, j, \ldots = 1, \ldots, d$) label intrinsic coordinates on the $d$-dimensional manifold $\mathcal{M}$. Latin indices from the beginning of the alphabet ($a, b, \ldots = 1, \ldots, d$) label components in an orthonormal frame. The Einstein summation convention is employed: repeated indices (one upper, one lower) are implicitly summed. Indices are raised and lowered using the metric tensor $g_{ij}$ and its inverse $g^{ij}$, satisfying $g^{ik}g_{kj} = \delta^i_j$. For example, $v^i = g^{ij}v_j$ and $v_i = g_{ij}v^j$. The covariant derivative is denoted $\nabla_i$, and partial derivatives are written $\partial_i \equiv \partial/\partial\varphi^i$.

\paragraph{Derivation of the Covariant Langevin Equation.}

\subparagraph{Manifolds Embedded in Euclidean Space.}
Consider a smooth $d$-dimensional manifold $\mathcal{M}$ embedded in $\mathbb{R}^n$ (with $n > d$). The manifold is defined by a set of constraint equations:
\begin{equation}
    E^s(\chi_\alpha) = 0, \quad s = 1, \ldots, n-d
    \label{eq:constraint}
\end{equation}
where $\chi_\alpha$ ($\alpha = 1, \ldots, n$) are Euclidean coordinates. Locally, we can solve these constraints and express some components $\sigma^s$ as functions of $d$ independent coordinates $\varphi^i$ ($i = 1, \ldots, d$):
\begin{equation}
    \chi_\alpha \equiv \{\sigma^s(\varphi), \varphi^i\}
    \label{eq:coordinates}
\end{equation}

The induced metric tensor on $\mathcal{M}$ is obtained from the embedding:
\begin{equation}
    g_{ij}(\varphi) = \delta_{ij} + \partial_i \sigma^s \partial_j \sigma^s
    \label{eq:induced_metric}
\end{equation}
where $\partial_i \equiv \partial/\partial \varphi^i$. This metric encodes how distances in the embedding space translate to distances on the manifold.

\paragraph{Langevin Equation in the Embedding Space.}
In the embedding space $\mathbb{R}^n$, the overdamped Langevin equation for a particle subject to a potential $A(\chi)$ takes the standard form:
\begin{equation}
    \dot{\chi}_\alpha(t) = -\frac{1}{\gamma} \frac{\partial A}{\partial \chi_\alpha} + \nu_\alpha(t)
    \label{eq:langevin_embedding}
\end{equation}
where $\gamma$ is the friction coefficient, and $\nu_\alpha(t)$ is Gaussian white noise with correlations given by the fluctuation-dissipation relation:
\begin{equation}
    \langle \nu_\alpha(t) \nu_\beta(t') \rangle = 2D \, \delta(t - t') \delta_{\alpha\beta}
    \label{eq:noise_correlation}
\end{equation}
where $D = T/\gamma$ is the diffusion constant and $T$ is the temperature (in units where $k_B = 1$).

\paragraph{Projection onto the Manifold.}
To constrain the dynamics to the manifold, we project the velocity $\dot{\chi}_\alpha$ onto the tangent space $T_\chi \mathcal{M}$. The variations $\delta\chi_\alpha$ tangent to the manifold satisfy:
\begin{equation}
    \frac{\partial E^s}{\partial \chi_\alpha} \delta\chi_\alpha = 0
    \label{eq:tangent_constraint}
\end{equation}

We introduce an orthonormal basis $\{e_a^\alpha\}$ ($a = 1, \ldots, d$) for the tangent space, satisfying:
\begin{equation}
    e_a^\alpha e_b^\alpha = \delta_{ab}, \quad \frac{\partial E^s}{\partial \chi_\alpha} e_a^\alpha = 0
    \label{eq:orthonormal_basis}
\end{equation}

The projection operator onto the tangent space is:
\begin{equation}
    P_{\alpha\beta} = e_a^\alpha e_a^\beta = \delta_{\alpha\beta} - n_\alpha^s n_\beta^s
    \label{eq:projection_operator}
\end{equation}
where $n_\alpha^s$ are unit normals to the manifold.

Projecting the Langevin equation (\ref{eq:langevin_embedding}) yields:
\begin{equation}
    \dot{\chi}_\alpha = e_a^\alpha e_a^\beta \left( -\frac{1}{\gamma}\frac{\partial A}{\partial \chi_\beta} + \nu_\beta \right)
    \label{eq:langevin_projected}
\end{equation}

\paragraph{Covariant Form.}
In terms of the intrinsic coordinates $\varphi^i$, the projected Langevin equation becomes:
\begin{equation}
    \dot{\varphi}^i = -\frac{1}{\gamma} g^{ij} \partial_j A + g^{ij} t_j^\alpha \nu_\alpha
    \label{eq:langevin_covariant}
\end{equation}
where $g^{ij}$ is the inverse metric tensor, and $t_i^\alpha = \partial \chi_\alpha / \partial \varphi^i$ relates the intrinsic and embedding coordinates. Explicitly, the components of $t_i^\alpha$ are:
\begin{equation}
    t_i^\alpha \equiv \begin{cases} t_i^j = \delta_{ij} & \text{(intrinsic directions)} \\ t_i^s = \partial_i \sigma^s & \text{(constrained directions)} \end{cases}
    \label{eq:t_definition}
\end{equation}
The quantities $t_i^\alpha$ satisfy:
\begin{equation}
    t_i^\alpha t_j^\alpha = g_{ij}
    \label{eq:t_metric}
\end{equation}
which follows directly from (\ref{eq:induced_metric}) and (\ref{eq:t_definition}).

The effective noise in intrinsic coordinates, $\xi^i = g^{ij} t_j^\alpha \nu_\alpha$, has correlations:
\begin{equation}
    \langle \xi^i(t) \xi^j(t') \rangle = 2D \, g^{ij} \, \delta(t - t')
    \label{eq:intrinsic_noise}
\end{equation}
where $D = k_B T / \gamma$ is the diffusion constant. This shows that the noise is ``metric-weighted,'' i.e., its covariance is determined by the inverse metric tensor.

The covariant Langevin equation (\ref{eq:langevin_covariant}) can be written more compactly as:
\begin{equation}
    d\varphi^i = -\frac{1}{\gamma} g^{ij} \partial_j A \, dt + \sqrt{2D} \, dW^i_{\mathcal{M}}
    \label{eq:langevin_intrinsic}
\end{equation}
where $dW^i_{\mathcal{M}}$ denotes Brownian motion on $\mathcal{M}$ with covariance $\langle dW^i_{\mathcal{M}} dW^j_{\mathcal{M}} \rangle = g^{ij} dt$. Since $D = T/\gamma$, the relative strength of drift and diffusion is controlled by the temperature $T$.

\begin{remark}[Geometric interpretation]
The factor $g^{ij}$ in front of the force term ensures that the dynamics respects the manifold geometry: forces are measured using the metric, and the particle moves along geodesics in the absence of potential gradients and noise. The metric-weighted noise ensures detailed balance with respect to the Boltzmann distribution $\propto \sqrt{g} \exp(-A/k_B T)$, where $g = \det(g_{ij})$ is the metric determinant.
\end{remark}

\paragraph{The It\^{o}-Stratonovich Choice on Manifolds.}
\label{sec:ito_stratonovich}

When the noise amplitude depends on the system's state, as occurs for position-dependent metrics, the stochastic integral requires a choice of discretization. Consider the general SDE with multiplicative noise:
\begin{equation}
    d\varphi^i = a^i(\varphi) \, dt + \sum_k b^{ik}(\varphi) \, dW^k
    \label{eq:general_sde}
\end{equation}
where $a^i$ is the drift and $b^{ik}$ is the noise amplitude. The stochastic integral $\int b^{ik} dW^k$ is defined as the limit of a discrete sum $\sum_n b^{ik}(\varphi(t_n^*)) \Delta W^k_n$, where $t_n^* \in [t_n, t_{n+1}]$ is the evaluation point. Unlike ordinary integrals, this limit depends on the choice of $t_n^*$. The It\^{o} prescription evaluates at the \emph{start} of each interval ($t_n^* = t_n$), while the Stratonovich prescription uses the \emph{midpoint} ($t_n^* = (t_n + t_{n+1})/2$, denoted $\circ\, dW^k$).

As established by Hunt and Ross \cite{huntross1981}, both formulations yield identical physical predictions when applied consistently; they differ only in how the noise term is discretized and correspondingly how the drift must be defined. The two are related by:
\begin{equation}
    a^i_{\text{It\^{o}}} = a^i_{\text{Strat}} + \frac{1}{2} \sum_{j,k} b^{jk} \frac{\partial b^{ik}}{\partial \varphi^j}
    \label{eq:strat_to_ito}
\end{equation}
The additional term is called the \emph{noise-induced drift}. This is not a correction for an error but rather the drift that must accompany the It\^{o} discretization to produce the same dynamics as Stratonovich.

The Stratonovich formulation is natural for physical systems because it arises as the white-noise limit of systems driven by colored (finite-correlation-time) noise, and it preserves the standard chain rule of calculus. In contrast, It\^{o}'s non-anticipatory prescription (where the integrand is independent of the noise increment by construction) yields martingales, making it preferred in mathematical finance and probability theory. For our purposes, since the Langevin equation describes the physical limit of thermal fluctuations, we adopt the Stratonovich interpretation.

The Euler-Maruyama numerical scheme is the natural discretization corresponding to the It\^{o} integral definition. It evaluates both drift and noise coefficients at the left endpoint of each time step:
\begin{equation}
    \varphi^{i,(n+1)} = \varphi^{i,(n)} + a^i(\varphi^{(n)}) \Delta t + \sum_k b^{ik}(\varphi^{(n)}) \Delta W^k_n
    \label{eq:euler_maruyama}
\end{equation}
This left-point evaluation mirrors the definition of the It\^{o} stochastic integral as a limit of sums $\sum_n b(X_{t_n})(W_{t_{n+1}} - W_{t_n})$. Since Euler-Maruyama is designed to discretize It\^{o} SDEs, simulating Stratonovich dynamics requires first converting to the equivalent It\^{o} form by adding the noise-induced drift (\ref{eq:strat_to_ito}) to the drift coefficient $a^i$.

\paragraph{Noise-Induced Drift for Our Geometries.}
The noise-induced drift vanishes when the noise amplitude is state-independent. For our three geometries:

\begin{itemize}
    \item \textbf{Sphere $S^2$}: The embed-and-project method generates noise in $\mathbb{R}^3$ and projects onto the tangent plane. While the projection operator $P_{\alpha\beta} = \delta_{\alpha\beta} - x_\alpha x_\beta$ is position-dependent, the noise amplitude in the embedding space is constant. The projection and renormalization steps have been shown to produce the correct equilibrium distribution $\rho_{\text{eq}} \propto \sqrt{g} \, e^{-U/k_BT}$ without requiring an explicit drift correction \cite{ciccotti2008, lelievre2010}.
    
    \item \textbf{Bounded cylinder $S^1 \times [0,H]$}: The metric $g_{ij} = \text{diag}(R^2, 1)$ is \emph{constant}. The noise amplitude $b^{ik} = \text{diag}(1/R, 1)$ is also constant, so $\partial_j b^{ik} = 0$ and the It\^{o}-Stratonovich correction vanishes identically. No drift correction is needed.
    
    \item \textbf{Torus $T^2$}: The metric coefficient $g_{\theta\theta} = (R + r\cos\varphi)^2$ depends on the poloidal angle $\varphi$. The noise amplitude in the $\theta$-direction is $b^{\theta} = 1/(R + r\cos\varphi)$, which varies with position. This generates a non-trivial noise-induced drift that must be included for correct equilibrium statistics.
\end{itemize}

\paragraph{Explicit Correction for the Torus.}
For the torus with Cholesky factor $L = \text{diag}(1/R(\varphi), 1/r)$ where $R(\varphi) = R + r\cos\varphi$, the noise-induced drift in the $\theta$-direction is:
\begin{equation}
    \Delta a^\theta = D \sum_{j,k} L^{jk} \frac{\partial L^{\theta k}}{\partial \varphi^j} = D \cdot \frac{1}{r} \cdot \frac{\partial}{\partial\varphi}\left(\frac{1}{R(\varphi)}\right) = D \cdot \frac{1}{r} \cdot \frac{r\sin\varphi}{R(\varphi)^2} = \frac{D \sin\varphi}{R(\varphi)^2}
    \label{eq:torus_drift_correction}
\end{equation}
This correction pushes particles toward the outer equator ($\varphi = 0$, larger $R$), counteracting the tendency of It\^{o} noise to accumulate particles where the noise amplitude is largest (the inner equator, smaller $R$). The net effect ensures uniform equilibrium density with respect to the Riemannian area element $dA = R(\varphi) \cdot r \, d\theta \, d\varphi$. Geometrically, at the inner equator where $R(\varphi)$ is smallest, angular steps in $\theta$ cover less arc length, so pure diffusion would over-sample this region; the Stratonovich correction adds systematic drift away from the inner equator, exactly compensating for this geometric bias.

\subsection{Application to Particle Systems with Random Interactions}
\label{sec:particle_system}

We now apply the general framework to a system of $N$ particles on a compact two-dimensional manifold $\mathcal{M}$ with positions $\{q_i\}_{i=1}^N \subset \mathcal{M}$. The particles interact through a potential energy:
\begin{equation}
    U(q_1, \ldots, q_N) = \sum_{i < j} \phi_{ij} \, d_g(q_i, q_j)
    \label{eq:potential_general}
\end{equation}
where $d_g(q_i, q_j)$ denotes the geodesic distance on $\mathcal{M}$ and $\phi_{ij}$ are coupling constants. When $\phi_{ij} > 0$, the interaction is repulsive (particles prefer to be far apart); when $\phi_{ij} < 0$, it is attractive.

\paragraph{The Force from Linear-Distance Potentials.}
For the potential (\ref{eq:potential_general}), the force on particle $i$ from particle $j$ requires computing the gradient of the geodesic distance. A key result from Riemannian geometry states that for $q_i \neq q_j$ not in the cut locus of each other:
\begin{equation}
    \nabla_{q_i} d_g(q_i, q_j) = -\hat{\mathbf{t}}_{ij}
\end{equation}
where $\hat{\mathbf{t}}_{ij} \in T_{q_i}\mathcal{M}$ is the unit tangent vector at $q_i$ pointing toward $q_j$ along the minimizing geodesic.

Therefore, the force on particle $i$ from particle $j$ is:
\begin{equation}
    \mathbf{F}_{ij} = -\phi_{ij} \nabla_{q_i} d_g(q_i, q_j) = \phi_{ij} \, \hat{\mathbf{t}}_{ij}
    \label{eq:force_general}
\end{equation}

This remarkably simple result (the force magnitude is simply $|\phi_{ij}|$, independent of distance) holds for any Riemannian manifold. The total force on particle $i$ is:
\begin{equation}
    \mathbf{F}_i = \sum_{j \neq i} \phi_{ij} \, \hat{\mathbf{t}}_{ij}
    \label{eq:total_force_general}
\end{equation}

\paragraph{Random Coupling Constants.}
The coupling constants $\phi_{ij}$ are drawn independently for each pair $(i,j)$ with $i < j$, and symmetrized: $\phi_{ji} = \phi_{ij}$. Two distributions are considered: (i) \emph{Gaussian couplings}: $\phi_{ij} \sim \mathcal{N}(0, \sigma^2)$, where $\sigma$ controls the interaction strength scale; and (ii) \emph{Discrete binary couplings}: $\phi_{ij} = +1$ with probability $p$ and $\phi_{ij} = -1$ with probability $1-p$. When $p = 0.5$, this gives an unbiased mixture of attractive and repulsive interactions, maximizing frustration.

\subsection{Manifold-Specific Details of Langevin Equations}
\label{sec:theory_summary}

The general framework developed above (covariant Langevin dynamics (\ref{eq:langevin_intrinsic}) with constant-magnitude forces (\ref{eq:force_general}) along geodesic directions) applies to any Riemannian manifold. In this work, we specialize to three canonical two-dimensional geometries: the sphere $S^2$, the bounded cylinder $S^1 \times [0,H]$, and the torus $T^2$. Each manifold requires explicit expressions for the induced metric tensor $g_{ij}$ and its inverse $g^{ij}$, the geodesic distance $d_g(q_i, q_j)$ between particle pairs, the unit tangent vector $\hat{\mathbf{t}}_{ij}$ at $q_i$ pointing toward $q_j$ along the minimizing geodesic, and the projection of 3D Brownian noise onto the tangent space.

Appendix~\ref{app:geometry} provides the complete derivations for each geometry, including:
\begin{itemize}
    \item \textbf{Sphere $S^2$}: The constraint $|\mathbf{x}|^2 = 1$ yields the projection operator $P_{\alpha\beta} = \delta_{\alpha\beta} - x_\alpha x_\beta$ and geodesic distance $d = \arccos(\mathbf{x}_i \cdot \mathbf{x}_j)$. Forces lie in the tangent plane and point along great circles.
    
    \item \textbf{Bounded cylinder $S^1 \times [0,H]$}: The intrinsically flat metric $ds^2 = R^2 d\theta^2 + dz^2$ with boundary conditions in the axial direction. Geodesics are straight lines in the $(\theta, z)$ plane (with appropriate wrapping).
    
    \item \textbf{Torus $T^2$}: The position-dependent metric $ds^2 = (R + r\cos\varphi)^2 d\theta^2 + r^2 d\varphi^2$ requires careful treatment of geodesic distances. We employ a midpoint-metric approximation for computational efficiency.
\end{itemize}

A key result common to all geometries is that symmetric pairwise forces contribute zero net torque. Specifically, on the sphere $S^2$, where each particle position $\mathbf{x}_i$ is a unit vector and the force on particle $i$ from particle $j$ is $\mathbf{F}_{ij} = \phi_{ij} \hat{\mathbf{t}}_{ij}$ with $\hat{\mathbf{t}}_{ij}$ the unit tangent vector pointing along the geodesic toward $\mathbf{x}_j$, the total torque from all pairwise interactions vanishes:
\begin{equation}
    \sum_{i=1}^{N} \mathbf{x}_i \times \mathbf{F}_i = \sum_{i=1}^{N} \mathbf{x}_i \times \sum_{j \neq i} \phi_{ij} \hat{\mathbf{t}}_{ij} = \mathbf{0}
    \label{eq:zero_torque}
\end{equation}
regardless of the coupling constants $\phi_{ij}$, provided only that they are symmetric ($\phi_{ij} = \phi_{ji}$). The proof follows from the antisymmetry of the cross product: the torque contribution from the pair $(i,j)$ is $\phi_{ij} \mathbf{x}_i \times \hat{\mathbf{t}}_{ij} = \phi_{ij} (\mathbf{x}_i \times \mathbf{x}_j)/\sin\theta_{ij}$, where $\theta_{ij}$ is the geodesic angle between particles. By symmetry, the contribution from $(j,i)$ is $\phi_{ji} (\mathbf{x}_j \times \mathbf{x}_i)/\sin\theta_{ij} = -\phi_{ij} (\mathbf{x}_i \times \mathbf{x}_j)/\sin\theta_{ij}$, so each pair's contribution cancels exactly. The same result holds for the cylinder since $\rho = R$ is constant. For the torus, the cancellation is exact only when interacting particles have equal distance from the $z$-axis; for ring configurations where particles within each ring share the same poloidal angle, the zero-torque property holds approximately (see Appendix~\ref{app:rotational_diffusion} for details). This zero-torque property leaves thermal noise as the dominant driver of orientation evolution.

\subsection{Rotational Dynamics and Orientation Diffusion}
\label{sec:rotational_dynamics}

Having established that pairwise forces contribute zero net torque, we now characterize the rotational dynamics of the particle system. The theoretical framework follows the standard treatment of rotational Brownian motion \cite{vankampen2007, gardiner2009, hofling2025, vanLengerich2020}, adapted to our multi-particle setting.

\paragraph{From Underdamped to Overdamped Rotational Dynamics.}
For a rigid body or particle system with moment of inertia $I$ and friction coefficient $\gamma$, the angular momentum $\mathbf{L}$ satisfies the Langevin equation
\begin{equation}
    \frac{d\mathbf{L}}{dt} = -\frac{\gamma}{I} \mathbf{L} + \boldsymbol{\tau}_\xi(t)
    \label{eq:L_langevin}
\end{equation}
where $\boldsymbol{\tau}_\xi(t)$ is the stochastic torque from thermal fluctuations. This is an Ornstein-Uhlenbeck process with relaxation time $\tau_L = I/\gamma$. As shown by van Kampen \cite{vankampen2007}, in the overdamped limit ($\gamma \to \infty$ or equivalently $\tau_L \to 0$), the angular momentum becomes a ``fast variable'' that thermalizes instantaneously. The physically meaningful slow variable is the \emph{orientation} of the rigid body in 3D space, which undergoes slow rotational diffusion. Appendix~\ref{app:rotational_diffusion} provides a detailed derivation.

\paragraph{Two Observables: Fast Cumulative Rotation and Slow Orientation.}
For our multi-particle system, we analyze two distinct observables that capture different aspects of collective rotational dynamics:

\begin{enumerate}
    \item \textbf{Cumulative rotation} $\Theta(t)$: The total amount of rotation accumulated around a fixed axis (the $z$-axis in our geometries). Since $\Theta(t)$ is driven by the instantaneous angular momentum, which is a fast variable in the overdamped limit, we expect $\Theta(t)$ to exhibit fast diffusive behavior similar to angular momentum itself. It grows without bound and does not represent a slow collective mode.
    
    \item \textbf{Structure normal} $\hat{\mathbf{n}}(t)$: A unit vector specifying the instantaneous orientation of the emergent structure (band, rings, or clusters) in 3D space. This is the direct implementation of van Kampen's ``orientation angle'' for our collective structures. Unlike $\Theta(t)$, the structure normal is a bounded quantity (it lives on $S^2$) and represents the true slow variable that undergoes rotational diffusion on a much longer timescale than the thermal velocity correlations.
\end{enumerate}

While $\Theta(t)$ answers ``how much has the structure rotated?'', $\hat{\mathbf{n}}(t)$ answers ``where is the structure pointing now?'' We analyze both observables in Section~\ref{sec:results} to obtain a complete characterization of the rotational dynamics.

\paragraph{Cumulative Rotation: Definition and Expected Behavior.}
We define the cumulative rotation as
\begin{equation}
    \Theta(t) := \int_0^t \omega(s) \, ds
    \label{eq:cumulative_rotation}
\end{equation}
where $\omega(t)$ is the instantaneous angular velocity of the collective structure. The zero-torque result (\ref{eq:zero_torque}) implies that pairwise interaction forces produce no deterministic drift in $\Theta(t)$. In the absence of drift, the cumulative rotation evolves according to
\begin{equation}
    d\Theta = \sqrt{2D_{\text{rot}}} \, dW
    \label{eq:orientation_diffusion}
\end{equation}
where $D_{\text{rot}}$ is the effective rotational diffusion coefficient and $dW$ is a standard Wiener process. The mean-squared displacement therefore grows linearly:
\begin{equation}
    \langle \Delta\Theta^2(t) \rangle = 2 D_{\text{rot}} \, t
    \label{eq:msd_rotation}
\end{equation}
This diffusive scaling ($\alpha = 1$ in $\langle \Delta\Theta^2 \rangle \propto t^\alpha$) is the expected ``fast'' behavior when thermal torques are temporally uncorrelated. Deviations from $\alpha = 1$ signal persistent correlations induced by the evolving particle configuration or global constraints from boundaries and metric structure.

\paragraph{Geometry-Specific Formulations.}
To connect with our simulations, we express the cumulative rotation in terms of particle coordinates. For particles on a manifold embedded in $\mathbb{R}^3$, the instantaneous rate of change of cumulative rotation about the $z$-axis is $\dot{\Theta}_z = \sum_i \rho_i^2 \dot{\theta}_i$, where $\rho_i$ is the distance from the $z$-axis and $\theta_i$ is the azimuthal angle. We define the cumulative rotation as
\begin{equation}
    \Theta_z(t) := \int_0^t \sum_{i=1}^{N} \rho_i^2(s) \, d\theta_i(s)
    \label{eq:cumulative_rotation_z}
\end{equation}
This stochastic integral has dimensions of [length]$^2$ and represents the time-integrated angular momentum about the $z$-axis.

On curved manifolds, the Langevin equation for the azimuthal coordinate takes the form $d\theta_i = (F_i^\theta / g_{\theta\theta}) \, dt + \sqrt{2D/g_{\theta\theta}} \, dW_i^\theta$, where the metric factor $g_{\theta\theta} = \rho_i^2$. Substituting into (\ref{eq:cumulative_rotation_z}) and using the zero-torque property $\sum_i F_i^\theta = 0$:
\begin{equation}
    d\Theta_z = \sum_i \rho_i^2 \, d\theta_i = \sum_i \rho_i^2 \cdot \frac{\sqrt{2D}}{\rho_i} \, dW_i^\theta = \sqrt{2D} \sum_{i=1}^{N} \rho_i \, dW_i^{\theta}
    \label{eq:dTheta_general}
\end{equation}
The factor of $\rho_i^2$ from the angular momentum definition partially cancels with the $1/\rho_i$ noise amplitude from the metric, leaving $\rho_i$ to the first power. Table~\ref{tab:rotation_diffusion} summarizes the rotational diffusion properties for each geometry.

\begin{table}[h!]
\centering
\caption{Rotational diffusion characteristics. For each manifold, we list the relevant rotation variable(s), their stochastic evolution, and the effective rotational diffusion coefficient $D_{\text{rot}}$ defined by $\langle (d\Theta)^2 \rangle = 2D_{\text{rot}} \, dt$.}
\label{tab:rotation_diffusion}
\begin{tabular}{@{}llll@{}}
\toprule
\textbf{Manifold} & \textbf{Variable} & \textbf{Evolution} & \textbf{Effective $D_{\text{rot}}$} \\ \midrule
Sphere $S^2$ & $\Theta_z$ (azimuthal) & $d\Theta_z = \sqrt{2D} \sum_i \sin\theta_i \, dW_i^{\phi}$ & $\frac{2ND}{3}$ (uniform dist.) \\[0.5em]
Cylinder & $\Theta_z$ (azimuthal) & $d\Theta_z = \sqrt{2D} \, R \sum_i dW_i^{\theta}$ & $D R^2 N$ \\[0.5em]
Torus $T^2$ & $\Theta_\theta$ (toroidal) & $d\Theta_\theta = \sqrt{2D} \sum_i (R + r\cos\varphi_i) \, dW_i^{\theta}$ & $DN(R^2 + 3r^2/2)$ \\[0.5em]
 & $\Theta_\varphi$ (poloidal) & $d\Theta_\varphi = \sqrt{2D} \, r \sum_i dW_i^{\varphi}$ & $D r^2 N$ \\ \bottomrule
\end{tabular}
\end{table}

\paragraph{Physical Interpretation.}
On all three manifolds, the cumulative rotation $\Theta(t)$ evolves with \emph{no deterministic drift} from pairwise interactions. For uniform particle distributions with uncorrelated thermal noise, one expects diffusive MSD scaling ($\alpha = 1$) for this fast variable. On the sphere, this is confirmed: the cumulative rotation MSD yields ensemble-averaged $\alpha \approx 0.9$. However, the cylinder and torus exhibit anomalous scaling (ballistic with $\alpha \approx 2$ and superdiffusive with $\alpha \approx 1.7$, respectively), indicating that the non-uniform particle distribution after structure formation induces persistent correlations in the effective thermal torques. The structure normal $\hat{\mathbf{n}}(t)$ (the slow Goldstone mode variable) exhibits different dynamics characterized by the correlation time $\tau_c$, as described in Section~\ref{sec:orientation_theory}.

\subsection{Structure Orientation from Position Covariance}
\label{sec:orientation_theory}

We now provide the technical definition of the structure normal $\hat{\mathbf{n}}(t)$ introduced above. This observable implements van Kampen's slow orientation variable for our collective structures.

\paragraph{Definition via Eigendecomposition.}
For each geometry, we define the structure normal $\hat{\mathbf{n}}(t)$ through the position covariance matrix in the 3D embedding space,
\begin{equation}
    M_{ij} = \frac{1}{N} \sum_{k=1}^{N} r_k^{(i)} r_k^{(j)}, \quad i,j \in \{x, y, z\},
    \label{eq:position_covariance}
\end{equation}
where $\mathbf{r}_k = (r_k^{(x)}, r_k^{(y)}, r_k^{(z)})$ are the 3D Cartesian coordinates of particle $k$. For the sphere, these are simply the unit sphere coordinates; for the cylinder and torus, we use the standard embeddings $\mathbf{r}(\theta, z) = (R\cos\theta, R\sin\theta, z)$ and $\mathbf{r}(\theta, \varphi) = ((R + r\cos\varphi)\cos\theta, (R + r\cos\varphi)\sin\theta, r\sin\varphi)$, respectively.

The eigendecomposition of $M$ yields three eigenvalues $\lambda_1 \geq \lambda_2 \geq \lambda_3$ with corresponding eigenvectors. The \emph{structure normal} $\hat{\mathbf{n}}$ is defined as the eigenvector associated with the \emph{smallest} eigenvalue $\lambda_3$: this is the direction along which the particle distribution has minimal extent, i.e., the direction perpendicular to the plane of the band, ring, or cluster configuration. We define the \emph{structure thickness} as
\begin{equation}
    \mathcal{T} = \sqrt{\frac{\lambda_3}{\lambda_1}},
    \label{eq:thickness}
\end{equation}
which measures the aspect ratio of the particle distribution. For a perfect planar configuration, $\mathcal{T} = 0$; for an isotropic distribution, $\mathcal{T} = 1$.

\paragraph{Expected Dynamics.}
For a well-formed structure ($\mathcal{T} \ll 1$), the normal $\hat{\mathbf{n}}(t)$ lives on the unit sphere $S^2$. Its dynamics reflect the slow reorientation of the collective structure under thermal fluctuations. The autocorrelation function $C(\tau) = \langle \hat{\mathbf{n}}(t) \cdot \hat{\mathbf{n}}(t+\tau) \rangle$ defines a correlation time $\tau_c$ that measures how quickly the structure loses memory of its orientation. We expect $\tau_c$ to be much longer than the microscopic thermal correlation time, reflecting the adiabatic nature of orientation dynamics in the overdamped regime.

\subsection{Summary: Two Complementary Observables}
\label{sec:observables_summary}

The theoretical framework developed in this section provides two complementary observables for characterizing collective rotational dynamics:

\begin{enumerate}
    \item \textbf{Cumulative rotation} $\Theta(t)$: A fast, unbounded quantity tracking total angular displacement around a fixed axis. Driven by the instantaneous angular momentum, its mean-squared displacement $\langle \Delta\Theta^2 \rangle \propto t^\alpha$ is expected to show diffusive scaling ($\alpha = 1$) when thermal torques are uncorrelated.
    
    \item \textbf{Structure normal} $\hat{\mathbf{n}}(t)$: The slow, bounded unit vector specifying instantaneous orientation in 3D. Its autocorrelation defines a correlation time $\tau_c \gg \tau_{\text{thermal}}$, and its power spectrum $S_{n_z}(\omega)$ probes the frequency content of orientation fluctuations.
\end{enumerate}

In Section~\ref{sec:results}, we present numerical results for both observables across all three geometries. For each manifold, we first describe the structure formation process and then analyze the cumulative rotation and orientation dynamics in the non-equilibrium steady state.

\section{Numerical Implementation}
\label{sec:simulation}

The numerical schemes presented in this section are based on discretizing the Langevin dynamics in the embedding space $\mathbb{R}^3$, as formulated in equations (\ref{eq:langevin_embedding})--(\ref{eq:langevin_projected}) of Section~\ref{sec:langevin_manifolds}, rather than directly discretizing the covariant intrinsic formulation (\ref{eq:langevin_intrinsic}). This embed-and-project approach generates noise in the embedding Euclidean space, projects it onto the manifold tangent space, and then enforces the manifold constraint through normalization or coordinate wrapping. As we demonstrate in Section~\ref{sec:covariant_integration}, the two approaches yield equivalent numerical schemes for the geometries considered here, validating the consistency between the embedding-space and intrinsic formulations.

\subsection{Euler-Maruyama Integration with Manifold Projection}

The stochastic differential equations are discretized using the Euler-Maruyama method. For embedded manifolds like the sphere, we employ the \emph{project-and-normalize} approach: update in the embedding space, then project back to the manifold \cite{ciccotti2008, lelievre2010}.

\paragraph{Sphere.}
\begin{equation}
    \mathbf{x}_i^{(n+1)} = \mathcal{N}\left( \mathbf{x}_i^{(n)} + \mathbf{F}_i^{(n)} \Delta t + \sqrt{2D \Delta t} \, \boldsymbol{\xi}_i^{\perp} \right)
    \label{eq:euler_sphere}
\end{equation}
where $\boldsymbol{\xi}_i \sim \mathcal{N}(0, \mathbf{I}_3)$ is a 3D standard Gaussian random vector, $\boldsymbol{\xi}_i^{\perp} = \boldsymbol{\xi}_i - (\boldsymbol{\xi}_i \cdot \mathbf{x}_i^{(n)}) \mathbf{x}_i^{(n)}$ is the projection onto the tangent space, and $\mathcal{N}(\mathbf{v}) = \mathbf{v} / |\mathbf{v}|$ denotes normalization to the unit sphere.

\paragraph{Bounded Cylinder.}
\begin{equation}
    \begin{aligned}
        \theta_i^{(n+1)} &= \left[ \theta_i^{(n)} + F_i^{\theta,(n)} \Delta t + \sqrt{\frac{2D \Delta t}{R^2}} \, \xi_i^{\theta} \right] \mod 2\pi \\
        z_i^{(n+1)} &= \mathcal{B}_z\left( z_i^{(n)} + F_i^{z,(n)} \Delta t + \sqrt{2D \Delta t} \, \xi_i^{z} \right)
    \end{aligned}
    \label{eq:euler_cylinder}
\end{equation}
where $\xi_i^{\theta}, \xi_i^{z} \sim \mathcal{N}(0,1)$ independently, and $\mathcal{B}_z$ applies the boundary condition (reflective or periodic) in the $z$-direction.

\paragraph{Torus.}
The position-dependent metric requires special care in the discretization. As discussed in Section~\ref{sec:ito_stratonovich}, the noise amplitude $b^\theta = 1/(R + r\cos\varphi)$ depends on the poloidal angle $\varphi$, creating an It\^{o}-Stratonovich ambiguity. For physically correct equilibrium statistics, we must include the noise-induced drift correction (\ref{eq:torus_drift_correction}).

The complete Euler-Maruyama scheme with Stratonovich correction is:
\begin{equation}
    \begin{aligned}
        \theta_i^{(n+1)} &= \left[ \theta_i^{(n)} + \left(F_i^{\theta,(n)} + \frac{D \sin\varphi_i^{(n)}}{(R + r\cos\varphi_i^{(n)})^2}\right) \Delta t + \sqrt{\frac{2D \Delta t}{(R + r\cos\varphi_i^{(n)})^2}} \, \xi_i^{\theta} \right] \mod 2\pi \\
        \varphi_i^{(n+1)} &= \left[ \varphi_i^{(n)} + F_i^{\varphi,(n)} \Delta t + \sqrt{\frac{2D \Delta t}{r^2}} \, \xi_i^{\varphi} \right] \mod 2\pi
    \end{aligned}
    \label{eq:euler_torus}
\end{equation}
where $\xi_i^{\theta}, \xi_i^{\varphi} \sim \mathcal{N}(0,1)$ independently. The second term in the $\theta$-update, $D\sin\varphi/(R + r\cos\varphi)^2$, is the noise-induced drift that ensures the correct equilibrium distribution. Without this correction, particles would spuriously accumulate near the inner equator ($\varphi = \pi$) where the noise amplitude is largest. For comparison, the sphere and cylinder do not require such corrections: for the sphere, the embed-and-project method with constant-amplitude noise in the embedding space produces correct equilibrium statistics \cite{ciccotti2008}, while for the cylinder the constant metric means ItÃ´ and Stratonovich formulations coincide.

The force computation follows (\ref{eq:total_force_general}) specialized to each geometry. The computational cost is $O(N^2)$ per time step due to the pairwise nature of the interactions.

\subsection{Alternative: Intrinsic Covariant Integration}
\label{sec:covariant_integration}

The discretization schemes above use different approaches depending on the geometry. The \textbf{sphere} employs true embed-and-project: noise is generated in $\mathbb{R}^3$ with constant amplitude, projected onto the tangent space, and the position is normalized back to the unit sphere. This approach has been proven to produce the correct equilibrium distribution without explicit drift correction \cite{ciccotti2008, lelievre2010}. The \textbf{cylinder} and \textbf{torus} use intrinsic coordinates with the covariant formulation (\ref{eq:langevin_intrinsic}), generating metric-weighted noise. For the cylinder, the constant metric means ItÃ´ and Stratonovich formulations coincide, so no drift correction is needed. For the torus, the position-dependent metric requires the explicit noise-induced drift (\ref{eq:torus_drift_correction}) to ensure correct equilibrium statistics.

For the three geometries considered in this work, both approaches yield \emph{equivalent} dynamics because the inverse metric is diagonal in natural coordinates. For the \textbf{sphere}, the embed-and-project method is actually \emph{preferable} because it avoids the coordinate singularity at the poles (where the Cholesky factor $\csc\theta \to \infty$). For the \textbf{cylinder} and \textbf{torus}, working in intrinsic coordinates is natural since no singularities arise.

The covariant formulation's value lies in: (1) theoretical clarity, making manifest that dynamics depends only on intrinsic Riemannian geometry; (2) generalizability to manifolds with \emph{non-diagonal} metrics, where genuine correlations between noise components arise; and (3) extensibility to higher-order geometric integrators. Complete details of the covariant integration scheme, including explicit Cholesky factors for each geometry and the treatment of non-diagonal metrics, are provided in Appendix~\ref{app:covariant}.

\subsection{Initial Conditions}

While uniform random distributions on each manifold provide one natural choice of initial conditions (which we implemented but do not report here), all simulations presented in this work employ a ``Big Bang''-like initial condition: all $N$ particles are initialized as samples from a Gaussian distribution with small variance $\sigma_0^2 \ll 1$ centered at a randomly chosen point on the manifold. We focus on the Big Bang scenario for two reasons: first, this concentrated initial condition connects to a broad class of problems in physics involving expansion from localized states; second, the phase of random manifold coverage that would characterize uniform initial conditions arises naturally as an intermediate stage during the rapid initial expansion in the Big Bang scenario across all three geometries we study.

Specifically, the Big Bang initialization proceeds as follows:
\begin{itemize}
    \item \textbf{Sphere}: Choose a random center $\mathbf{c} \in S^2$, apply Gaussian perturbations in the tangent plane $T_{\mathbf{c}}S^2$, then project back onto $S^2$ by normalization.
    \item \textbf{Bounded Cylinder}: Choose a random center $(\theta_0, z_0)$, add Gaussian noise with variance scaled by the metric components.
    \item \textbf{Torus}: Choose a random center $(\theta_0, \varphi_0)$, add Gaussian noise with variance scaled by the metric components.
\end{itemize}

This concentrated initial condition has an important energetic consequence: the system starts with nearly zero total energy. In the overdamped Langevin dynamics, the kinetic energy contribution is negligible. The potential energy $U = \sum_{i<j} \phi_{ij} \, d_g(q_i, q_j)$ is also nearly zero initially because: (i) all geodesic distances $d_g(q_i, q_j)$ are of order $\sigma_0 \ll 1$ when particles are tightly clustered, and (ii) the quenched coupling constants $\phi_{ij}$ have a nearly balanced mixture of positive and negative values (for the discrete distribution with $p = 0.5$, or zero mean for the Gaussian distribution). The sum over $O(N^2)$ small, randomly signed terms thus produces a total potential energy close to zero.

Since the thermal bath at temperature $T$ is the only source of energy fluctuations in our setting, the subsequent dynamical evolution proceeds with total energy fluctuating around zero. This near-zero energy constraint, combined with the competing attractive and repulsive interactions, drives the system toward the frustrated, low-dimensional structures we observe.

\subsection{Simulation Parameters}

All simulations across the three geometries (sphere, cylinder, torus) used identical parameters as listed in Table~\ref{tab:parameters}. This ensures that observed differences in structure formation and dynamics are attributable solely to differences in manifold geometry and topology, not to variations in physical conditions.

\begin{table}[h!]
\centering
\caption{Default simulation parameters. All quantities are expressed in dimensionless units with characteristic length scale set by the manifold radius ($R_S = R_C = 1$), characteristic energy scale set by the coupling strength ($J = 1$), and characteristic time scale set by the friction coefficient ($\gamma = 1$). The cumulative rotation $\Theta(t)$ defined in Eq.~(\ref{eq:cumulative_rotation}) has units of $[\text{length}]^2$.}
\label{tab:parameters}
\begin{tabular}{@{}llll@{}}
\toprule
\textbf{Parameter} & \textbf{Symbol} & \textbf{Value} & \textbf{Notes} \\ \midrule
Number of particles & $N$ & 400 & \\
Temperature & $T$ & 0.40 & \\
Friction coefficient & $\gamma$ & 1.0 & \\
Diffusion constant & $D = T/\gamma$ & 0.40 & \\
Time step & $\Delta t$ & 0.0025 & \\
Sphere radius & $R_S$ & 1.0 & \\
Bounded cylinder radius & $R_C$ & 1.0 & \\
Bounded cylinder height & $H$ & 3.0 & \\
Torus major radius & $R$ & 1.5 & \\
Torus minor radius & $r$ & 0.6 & \\
Coupling std (Gaussian) & $\sigma$ & 1.0 & \\
Probability of $+1$ (discrete) & $p$ & 0.5 & \\ \bottomrule
\end{tabular}
\end{table}

\paragraph{Natural Units and Physical Interpretation.}
All simulations use a natural unit system where the manifold radius, coupling strength, and friction coefficient are set to unity. This choice defines: (1) the \emph{length scale} through the manifold radius $R = 1$, so all distances are measured in units of $R$; (2) the \emph{energy scale} through the coupling strength $J = 1$, noting that since the potential $U = \phi_{ij} d_g$ is linear in geodesic distance, energy has units of $[J \cdot R] = [\text{length}]$ in this system; (3) the \emph{time scale} through the friction coefficient $\gamma = 1$, so the natural time unit is $\gamma R^2 / J$, which is the time for a particle to traverse a distance $R$ under a force of magnitude $J$. Angular momentum, being $L = \mathbf{x} \times \dot{\mathbf{x}}$ with $|\mathbf{x}| = R$ and $|\dot{\mathbf{x}}| \sim R/t$, has units of $[\text{length}]^2/[\text{time}] = R^2 \gamma / J$ in natural units. Throughout this work, all times are reported in these natural units; for reference, the integration time step $\Delta t = 0.0025$ corresponds to 400 steps per unit time.
\label{sec:results}

\subsection{Overview: From 2D Manifolds to Lower-Dimensional Structures}

The simulations reveal a universal phenomenon: \emph{dynamic dimension reduction}. Starting from arbitrary initial conditions on two-dimensional manifolds, the system relaxes into configurations concentrated on lower-dimensional submanifolds. This process is analogous to relaxation in glassy systems, where the system navigates a complex energy landscape toward low-energy configurations, but here the relaxation has a clear geometric signature.\footnote{Simulation codes and videos are available at https://github.com/ighalp/frustrated-brownian-particles-manifolds.}

The dimension reduction can be summarized as:
\begin{align}
    S^2 &\to S^1 \quad \text{(sphere to great-circle band)} \label{eq:dim_red_sphere} \\
    S^1 \times [0,H] &\to \mathbb{Z}_2 \quad \text{(bounded cylinder to discrete clusters)} \label{eq:dim_red_cylinder} \\
    S^1 \times S^1 &\to \coprod_2 S^1 \quad \text{(torus to two minor-circle rings)} \label{eq:dim_red_torus}
\end{align}
where $\coprod$ denotes disjoint union. On the bounded cylinder, the $\mathbb{Z}_2$ notation reflects that clusters form at two height levels, with each level containing two clusters at diametrically opposite azimuthal positions; because the clusters are approximately vertically aligned, a single $180^\circ$ rotation exchanges both pairs simultaneously, yielding a single $\mathbb{Z}_2$ residual symmetry.

This dimensional collapse is \emph{approximate}: residual fluctuations persist due to two sources:
\begin{enumerate}
    \item \textbf{Thermal fluctuations}: The stochastic noise in the Langevin dynamics causes particles to fluctuate around the low-dimensional attractor.
    \item \textbf{Interaction-induced fluctuations}: The random couplings create local force imbalances that prevent perfect alignment along the submanifold.
\end{enumerate}

The relaxation proceeds through phases characteristic of glassy dynamics:
\begin{itemize}
    \item \textbf{Initial rapid relaxation} ($t \lesssim \tau_{\text{micro}}$): Fast energy decrease as particles escape high-energy configurations; analogous to the $\beta$-relaxation in glasses.
    \item \textbf{Intermediate plateau}: System becomes trapped in metastable configurations; exploration of the energy landscape.
    \item \textbf{Slow structural relaxation} ($t \sim \tau_{\text{relax}}$): Gradual organization into low-dimensional structures; analogous to $\alpha$-relaxation.
    \item \textbf{Quasi-equilibrium drift} ($t \gg \tau_{\text{relax}}$): Persistent slow motion (precession, rotation) of the final structures; aging-like behavior.
\end{itemize}

\subsection{Results on the Sphere: $S^2 \to S^1$}
\label{sec:results_sphere}

\begin{figure}[!ht]
    \centering
    \includegraphics[width=0.6\textwidth]{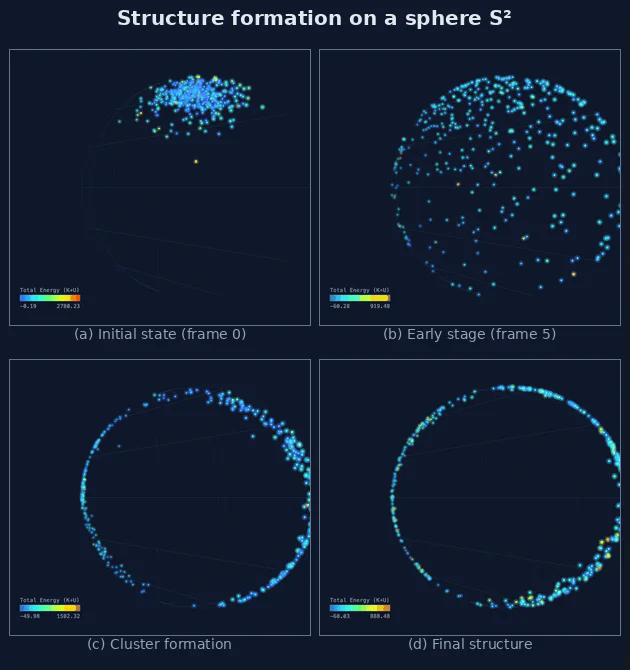}
    \caption{Dynamic dimension reduction on the sphere $S^2 \to S^1$. (a) Initial Gaussian cluster. (b) Early dispersion phase. (c) Intermediate band formation. (d) Final quasi-one-dimensional band along a great circle. The reduction from the 2D sphere to a 1D ring structure is clearly visible. Color indicates total energy per particle.}
    \label{fig:sphere_evolution}
\end{figure}

Figure~\ref{fig:sphere_evolution} illustrates the dynamic dimension reduction $S^2 \to S^1$ on the sphere.

\paragraph{Band Formation: Collapse to a Great Circle.}
The dimensional reduction $S^2 \to S^1$ manifests as band formation along a dynamically selected great circle. Table~\ref{tab:sphere_phases} summarizes the structure evolution phases.

\begin{table}[h!]
\centering
\caption{Structure formation phases on the sphere from simulation with $N=400$ particles, $T=0.4$, total time $t=0$ to $50$. Band formation is identified by the onset of strong correlation between total and band cumulative rotation.}
\label{tab:sphere_phases}
\begin{tabular}{@{}llll@{}}
\toprule
\textbf{Phase} & \textbf{Time} & \textbf{Duration} & \textbf{Description} \\ \midrule
Initial cluster & $t = 0$ & ;  & Localized Gaussian blob \\
Dispersion & $t = 0$--$2$ & $\sim 2$ & Particles spread across sphere \\
Band assembly & $t = 2$--$5$ & $\sim 3$ & Band structure emerges \\
\textbf{Steady state} & $t > 5$ & Indefinite & Persistent band with thermal fluctuations \\ \bottomrule
\end{tabular}
\end{table}

The key timescale for band formation is approximately $t \approx 5$ time units. Band quality can be quantified through the asphericity $A = (\lambda_1 - \lambda_2)/(\lambda_1 + \lambda_2)$, where $\lambda_{1,2}$ are eigenvalues of the 2D projected configuration tensor. Peak asphericity of 0.956 (very close to the ideal value of 1 for a perfect line) was achieved at minimum band width.

\paragraph{Rotational Dynamics and Band Precession.}
\label{sec:sphere_precession}

The band geometry emerges from energy minimization in the frustrated interaction landscape, not from angular momentum dynamics. Particles concentrate along a great circle because this configuration minimizes the total potential energy for typical realizations of the random couplings. The zero-torque property (\ref{eq:zero_torque}) ensures that the band orientation, once established, evolves only through thermal noise with no deterministic drift. Two observables characterize this evolution: the cumulative rotation $\Theta_z(t)$ measures total angular displacement around the $z$-axis (a fast variable), while the structure normal $\hat{\mathbf{n}}(t)$ specifies the instantaneous band orientation in 3D (the slow Goldstone mode).

\paragraph{Precession Rate.}
Frame-by-frame tracking of the band orientation during the steady-state phase reveals:
\begin{itemize}
    \item Band orientation angle: nearly constant at $\approx -70^\circ$ (in projection)
    \item Precession rate: $\sim 34^\circ$ per unit time
    \item Full precession period (180$^\circ$): $\sim 5$ time units
\end{itemize}

The slow precession rate reflects the absence of deterministic torque: orientation changes arise solely from accumulated thermal fluctuations, consistent with a random walk in orientation space (the Goldstone manifold $S^2$).

\paragraph{Cumulative Rotation Dynamics.}
\label{sec:sphere_L_analysis}

The cumulative rotation $\Theta_z(t) = \int_0^t \sum_i \rho_i^2 \, d\theta_i$ (defined in Eq.~(\ref{eq:cumulative_rotation_z})) exhibits large-amplitude fluctuations characteristic of a random walk, as predicted by the theoretical result (\ref{eq:dTheta_sphere}). Table~\ref{tab:sphere_L_stats} summarizes the statistics.

\begin{table}[h!]
\centering
\caption{Cumulative rotation statistics on the sphere separated by phase (dimensionless units). The formation phase shows larger fluctuations; the steady state exhibits reduced variance once the band has formed.}
\label{tab:sphere_L_stats}
\begin{tabular}{@{}lccc@{}}
\toprule
\textbf{Phase} & \textbf{Time Range} & $\langle \Theta_z \rangle$ & $\sigma_{\Theta_z}$ \\ \midrule
Formation & $t = 0$--$5$ & $-3.2$ & $205.1$ \\
Steady state & $t = 5$--$50$ & $-9.6$ & $148.7$ \\ \bottomrule
\end{tabular}
\end{table}

The mean cumulative rotation is small compared to its fluctuations ($|\langle \Theta_z \rangle| \ll \sigma_{\Theta_z}$), consistent with rotational symmetry: there is no preferred direction for net rotation. The theoretical prediction $d\Theta_z = \sqrt{2D} \sum_i \sin\theta_i \, dW_i^{\phi}$ implies pure diffusion with no deterministic drift. The standard deviation \emph{decreases} from the formation phase ($\sigma = 205$) to the steady state ($\sigma = 149$), indicating that band formation partially constrains the rotational fluctuations.

\paragraph{MSD Scaling and Ensemble Averaging.}
The mean-square displacement of the cumulative rotation exhibits power-law scaling $\langle \Delta \theta^2 \rangle \sim t^{\alpha}$. Ensemble averaging over five independent simulations (each with different quenched disorder $\phi_{ij}$) yields $\alpha = 0.9 \pm 0.3$, confirming \emph{diffusive} dynamics on the sphere. Individual runs show $\alpha$ ranging from $0.55$ to $1.13$, reflecting the different free energy landscapes explored by different realizations. The autocorrelation of $\Theta_z$ decays rapidly ($\tau < 0.2$), consistent with the white-noise character of Langevin dynamics.

The relaxation dynamics exhibits glassy features: two-stage relaxation (fast dispersion followed by band assembly), metastability, and fluctuation reduction upon structure formation.

\begin{figure}[!ht]
    \centering
    \includegraphics[width=\textwidth]{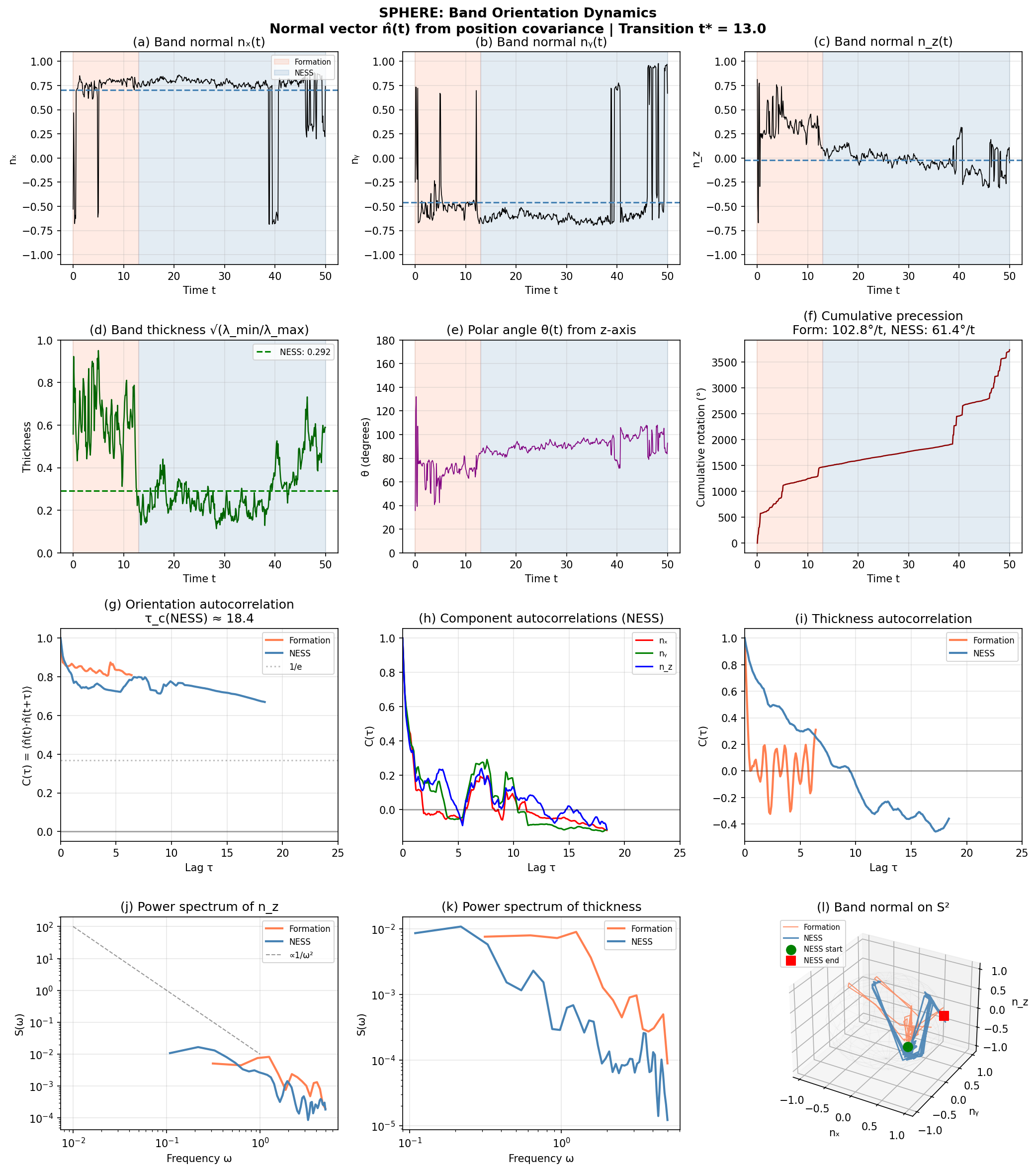}
    \caption{Sphere $S^2$: Band orientation dynamics for $N = 400$ particles, $T = 0.4$, full time series $t = 0$ to $50$. Top row: Band normal components $n_x(t)$, $n_y(t)$, $n_z(t)$ showing slow precession; coral shading indicates formation phase ($t < 13$), blue shading indicates NESS. Second row: Band thickness $\mathcal{T}(t)$ decreasing from $\sim 0.6$ to $\sim 0.3$ as the band forms; polar angle $\theta(t)$ of the band normal; cumulative angular displacement showing steady precession at $\sim 61^\circ$/time in NESS. Third row: Orientation autocorrelation $C(\tau) = \langle \hat{\mathbf{n}}(t) \cdot \hat{\mathbf{n}}(t+\tau) \rangle$ with correlation time $\tau_c \approx 18$; component autocorrelations in NESS; thickness autocorrelation. Bottom row: Power spectra showing $S_{n_z}(\omega) \propto \omega^{-1.6}$; 3D trajectory of the band normal on $S^2$ demonstrating isotropic exploration of the Goldstone manifold.}
    \label{fig:sphere_orientation}
\end{figure}

\paragraph{Band Orientation Dynamics.}
We now analyze the structure normal $\hat{\mathbf{n}}(t)$ as defined in Section~\ref{sec:orientation_theory}. On the sphere, the band normal undergoes slow diffusive precession on the unit sphere $S^2 \cong \mathrm{SO}(3)/\mathrm{SO}(2)$, which is precisely the Goldstone manifold for the broken rotational symmetry. Figure~\ref{fig:sphere_orientation} presents the complete orientation dynamics.

\begin{itemize}
    \item \textbf{Structure formation}: The thickness decreases from $\mathcal{T} \approx 0.62$ (diffuse, nearly isotropic) during formation to $\mathcal{T} \approx 0.29$ (well-defined band) in NESS, confirming dimensional collapse.
    
    \item \textbf{Isotropic orientation}: The mean $\langle n_z \rangle \approx 0$ with comparable standard deviations in all components ($\sigma(n_x) \approx \sigma(n_y) \approx \sigma(n_z) \approx 0.1$--$0.3$), indicating that the band normal explores all directions on $S^2$ without preferred orientation.
    
    \item \textbf{Long correlation time}: The orientation autocorrelation $C(\tau) = \langle \hat{\mathbf{n}}(t) \cdot \hat{\mathbf{n}}(t+\tau) \rangle$ decays to $1/e$ at $\tau_c \approx 18$ time units, much longer than the thermal velocity correlation time ($\tau_v < 0.2$). This separation of timescales confirms the adiabatic nature of orientation dynamics.
    
    \item \textbf{Cumulative precession}: The band normal accumulates approximately $2300^\circ$ of angular displacement during the NESS phase ($t = 13$ to $50$), corresponding to a mean precession rate of $\sim 61^\circ$ per unit time. This continuous exploration of the Goldstone manifold distinguishes our ``disorder-induced symmetry breaking'' from true spontaneous symmetry breaking where the order parameter would be frozen.
    
    \item \textbf{Power spectrum}: The structure normal power spectrum $S_{n_z}(\omega) \propto \omega^{-1.6}$ is shallower than the $1/\omega^2$ prediction for type-A Goldstone diffusion modes. This may reflect the bounded nature of $\hat{\mathbf{n}}$ on $S^2$.
\end{itemize}

\paragraph{Summary: Sphere Dynamics.}
On the sphere, dimensional reduction produces a single equatorial band that forms by $t \approx 5$ and persists indefinitely. The cumulative rotation $\Theta_z(t)$ exhibits diffusive scaling with MSD exponent $\alpha = 0.9 \pm 0.3$, consistent with fast, thermally-driven dynamics. The structure normal $\hat{\mathbf{n}}(t)$ (the Goldstone mode variable) precesses slowly and isotropically on $S^2$ with correlation time $\tau_c \approx 18$, far exceeding the thermal correlation time. Both observables reflect the zero-torque property: no deterministic drift from pairwise forces. The sphere thus provides the cleanest example of disorder-induced symmetry breaking, where the structure forms rapidly but its orientation evolves stochastically on the Goldstone manifold.

\subsection{Results on the Bounded Cylinder: $S^1 \times [0,H] \to \mathbb{Z}_2$}
\label{sec:results_cylinder}

\begin{figure}[!ht]
    \centering
    \includegraphics[width=0.6\textwidth]{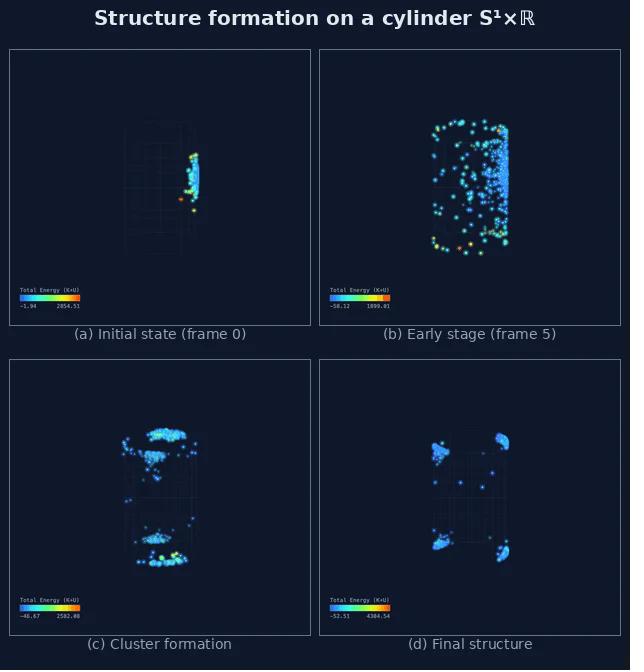}
    \caption{Dynamic dimension reduction on the bounded cylinder $S^1 \times [0,H] \to \mathbb{Z}_2$. (a) Initial Gaussian cluster. (b) Early dispersion. (c) Emergence of vertically separated clusters. (d) Final configuration showing four localized clusters arranged in two vertical pairs at approximately opposite azimuthal positions. The continuous axial direction has collapsed to a discrete set of levels, with further azimuthal fragmentation driven by interaction frustration. The vertically-aligned cluster pairs at diametrically opposite positions reflect the residual $\mathbb{Z}_2$ symmetry under $180^\circ$ rotation.}
    \label{fig:cylinder_evolution}
\end{figure}

Figure~\ref{fig:cylinder_evolution} illustrates the dynamic dimension reduction on the bounded cylinder.

\paragraph{Cluster Formation: From Rings to Localized Structures.}

On the bounded cylinder, particles organize into vertically separated clusters at different heights $z$ along the cylinder axis. However, detailed frame-by-frame analysis of 414 simulation frames reveals that the final structures are not continuous rings but rather \emph{localized clusters} concentrated at specific azimuthal positions. 

As shown in Figure~\ref{fig:cylinder_evolution}(d), the system typically forms \textbf{four distinct clusters} arranged in two vertical pairs:
\begin{itemize}
    \item Two clusters near the top of the cylinder (at different azimuthal angles $\theta$)
    \item Two clusters near the bottom of the cylinder (at corresponding azimuthal angles)
\end{itemize}

Table~\ref{tab:cylinder_phases} summarizes the frequency of different cluster configurations observed across the simulation.

\begin{table}[h!]
\centering
\caption{Distribution of cluster configurations on the bounded cylinder across 414 simulation frames.}
\label{tab:cylinder_phases}
\begin{tabular}{@{}lcc@{}}
\toprule
\textbf{Configuration} & \textbf{Frames} & \textbf{Percentage} \\ \midrule
1 top + 1 bottom cluster & 201 & 48.6\% \\
2 top + 2 bottom clusters & 114 & 27.5\% \\
1 top + 2 bottom clusters & 37 & 8.9\% \\
Other configurations & 62 & 15.0\% \\ \bottomrule
\end{tabular}
\end{table}

The vertical separation arises from the interplay between attractive and repulsive interactions: particles with net attractive coupling tend to cluster together, while those with net repulsive coupling are pushed to different vertical positions. The azimuthal fragmentation (the splitting of what might have been continuous rings into localized clusters) reflects the frustration induced by the random couplings, which creates local neighborhoods of predominantly attractive or repulsive interactions.

\paragraph{Correlated Dynamics of Vertically Aligned Clusters.}
\label{sec:cylinder_correlations}

A key question regarding the bounded cylinder dynamics concerns the relative motion of clusters at different positions. Frame-by-frame analysis reveals that the clusters do not move independently.

\paragraph{Four-Cluster Regime Analysis.}
During the 114 frames exhibiting the four-cluster configuration (two top, two bottom), we tracked the azimuthal positions of each cluster, labeled as Left-Top (LT), Right-Top (RT), Left-Bottom (LB), and Right-Bottom (RB) based on their angular positions. The position statistics are summarized in Table~\ref{tab:cluster_positions}.

\begin{table}[h!]
\centering
\caption{Cluster position statistics in the four-cluster regime (114 frames). Positions measured in image pixels; the visible cylinder spans approximately 240 pixels in the azimuthal direction, corresponding to $\sim 180^\circ$.}
\label{tab:cluster_positions}
\begin{tabular}{@{}lccc@{}}
\toprule
\textbf{Cluster} & \textbf{Mean Position} & \textbf{Std. Dev.} & \textbf{Angular Position} \\ \midrule
Left-Top (LT) & 240.2 px & 6.3 px & $\sim 45^\circ$ from center \\
Right-Top (RT) & 359.1 px & 6.5 px & $\sim 45^\circ$ opposite \\
Left-Bottom (LB) & 236.2 px & 5.7 px & $\sim 48^\circ$ from center \\
Right-Bottom (RB) & 360.1 px & 6.8 px & $\sim 45^\circ$ opposite \\ \bottomrule
\end{tabular}
\end{table}

The angular separation between left and right clusters at the same vertical level is approximately $90^\circ$--$100^\circ$, indicating that the cluster pairs are located on roughly opposite sides of the cylinder.

\paragraph{Correlation Structure.}
Computing the Pearson correlation coefficients between the time series of cluster positions reveals a clear pattern (Table~\ref{tab:correlation_matrix}).

\begin{table}[h!]
\centering
\caption{Correlation matrix for cluster azimuthal positions in the four-cluster regime. Strong positive correlations ($r > 0.8$) between vertically aligned clusters indicate coupled motion; negative correlations between left and right clusters reflect the geometric constraint of the cylindrical projection.}
\label{tab:correlation_matrix}
\begin{tabular}{@{}lcccc@{}}
\toprule
 & \textbf{L-Top} & \textbf{R-Top} & \textbf{L-Bottom} & \textbf{R-Bottom} \\ \midrule
\textbf{L-Top} & 1.000 & $-0.502$ & \cellcolor{green!20}$\mathbf{0.819}$ & $-0.702$ \\
\textbf{R-Top} & $-0.502$ & 1.000 & $-0.466$ & \cellcolor{green!20}$\mathbf{0.866}$ \\
\textbf{L-Bottom} & \cellcolor{green!20}$\mathbf{0.819}$ & $-0.466$ & 1.000 & $-0.578$ \\
\textbf{R-Bottom} & $-0.702$ & \cellcolor{green!20}$\mathbf{0.866}$ & $-0.578$ & 1.000 \\ \bottomrule
\end{tabular}
\end{table}

The correlation structure reveals two key findings:

\begin{enumerate}
    \item \textbf{Vertically aligned clusters move together}: The correlations between Left-Top and Left-Bottom ($r = 0.819$) and between Right-Top and Right-Bottom ($r = 0.866$) are strongly positive. This indicates that clusters at the same azimuthal position but different heights exhibit \emph{coupled}, not independent, motion.
    
    \item \textbf{Opposite-side clusters are anti-correlated}: The negative correlations between left and right clusters (ranging from $-0.47$ to $-0.70$) reflect a combination of the geometric projection (as one side rotates into view, the other rotates out) and possible collective oscillation modes.
\end{enumerate}

\paragraph{Vertical Alignment Persistence.}
The mean offset between vertically aligned clusters is remarkably small:
\begin{equation}
    \langle \theta_{\text{LT}} - \theta_{\text{LB}} \rangle = 4.0 \pm 6.0 \text{ pixels}, \quad
    \langle \theta_{\text{RT}} - \theta_{\text{RB}} \rangle = -1.0 \pm 6.5 \text{ pixels}
\end{equation}
corresponding to angular differences of less than $5^\circ$. This near-perfect vertical alignment persists throughout the simulation, further supporting the conclusion that top and bottom clusters at similar azimuths form coherent vertical structures.

\paragraph{Time Evolution and Dynamical Phases.}
Analysis of the full time series from $t = 0$ to $t = 50$ reveals rapid cluster formation followed by a long steady-state phase. During the \emph{formation phase} ($t = 0$--$1$), after the initial Gaussian cluster disperses, the system undergoes rapid reorganization into four distinct clusters within approximately $t \approx 1$ time units. This remarkably fast formation reflects the strong boundary pinning effect. In the \emph{steady state} ($t > 1$), the four-cluster configuration \textbf{persists indefinitely} throughout the remainder of the simulation (tested to $t = 50$), with clusters maintaining positions at the boundaries while undergoing collective rotational diffusion.

\paragraph{Consistency with Zero-Torque Property.}
The observed correlation structure is fully consistent with the theoretical framework of Section~\ref{sec:rotational_dynamics}. Since pairwise forces produce zero net torque about the cylinder axis (by the same argument as (\ref{eq:zero_torque}) for the sphere), the total cumulative rotation $\Theta_z = R^2 \int_0^t \sum_i \dot{\theta}_i \, ds$ evolves according to (\ref{eq:dTheta_cylinder}):
\begin{equation}
    d\Theta_z = \sqrt{2D} \cdot R \sum_{i=1}^{N} dW_i^{\theta}
\end{equation}
with no deterministic drift. For uncorrelated thermal noise and uniform particle distributions, this Langevin equation would predict diffusive MSD scaling ($\alpha = 1$). However, the actual MSD exhibits \emph{ballistic} scaling ($\alpha \approx 2$), as shown below. This deviation from the naive prediction indicates that the boundary-induced cluster configuration creates persistent correlations in the effective thermal torques that are not captured by the simple local Langevin theory. The strong correlations between vertically aligned clusters ($r > 0.8$) suggest that the system behaves approximately like a rigid body with long-lived collective rotational modes.

\paragraph{Effects of the Manifold Boundary.}
The bounded cylinder is a manifold with boundary (circular edges at $z = 0$ and $z = H$). With reflective dynamics, clusters near the edges experience effective ``image'' forces that pin them to specific heights. The boundaries play a crucial role in the four-cluster structure: the edges effectively confine the vertical extent of clusters, preventing them from spreading into continuous axial distributions. Combined with azimuthal fragmentation from coupling frustration, this produces the characteristic ``four corners'' arrangement. If periodic $z$-boundaries were imposed (making the manifold topologically a torus $T^2$), we would expect continuous ring structures rather than discrete clusters.

\paragraph{Cumulative Rotation Dynamics.}
\label{sec:cylinder_L_analysis}

The total cumulative rotation about the cylinder axis, $\Theta_\theta = R^2 \sum_i \dot{\theta}_i \, dt$, exhibits large-amplitude fluctuations characteristic of pure noise. Table~\ref{tab:cylinder_L_stats} summarizes the statistics.

\begin{table}[h!]
\centering
\caption{Cumulative rotation statistics on the cylinder separated by phase (dimensionless units). Unlike the sphere, the standard deviation remains essentially constant across phases.}
\label{tab:cylinder_L_stats}
\begin{tabular}{@{}lccc@{}}
\toprule
\textbf{Phase} & \textbf{Time Range} & $\langle \Theta_\theta \rangle$ & $\sigma_{\Theta_\theta}$ \\ \midrule
Formation & $t = 0$--$1$ & $-11.7$ & $240.4$ \\
Steady state & $t = 1$--$50$ & $45.8$ & $243.8$ \\ \bottomrule
\end{tabular}
\end{table}

The standard deviation $\sigma \approx 244$ is substantially larger than for the sphere ($\sigma \approx 149$), reflecting both the different metric structure and the additive contributions from four independent clusters. The autocorrelation decays rapidly ($\tau < 0.2$), and inter-cluster angular velocity correlations are weakly negative ($\rho \approx -0.06$ to $-0.21$), consistent with the zero-torque property.

\paragraph{Ballistic Rotation.}
Unlike the sphere, the cylinder exhibits \emph{ballistic} MSD scaling:
\begin{equation}
    \langle \Delta \theta^2 \rangle \sim t^{\alpha}, \quad \alpha \approx 2.0
\end{equation}
Ensemble averaging over five runs confirms $\alpha = 2.00 \pm 0.05$. This ballistic behavior indicates persistent collective motions that do not decorrelate over the simulation timescale, possibly arising from clusters pinned at the reflective $z$-boundaries or the discrete $\mathbb{Z}_2$ residual symmetry preventing continuous exploration of a Goldstone manifold.

\paragraph{Cluster Spatial Distribution.}
The four clusters localize at the reflective boundaries ($z \approx \pm 1.5$), with azimuthal separation of approximately $90^\circ$--$180^\circ$ reflecting coupling frustration.

\begin{figure}[!ht]
    \centering
    \includegraphics[width=\textwidth]{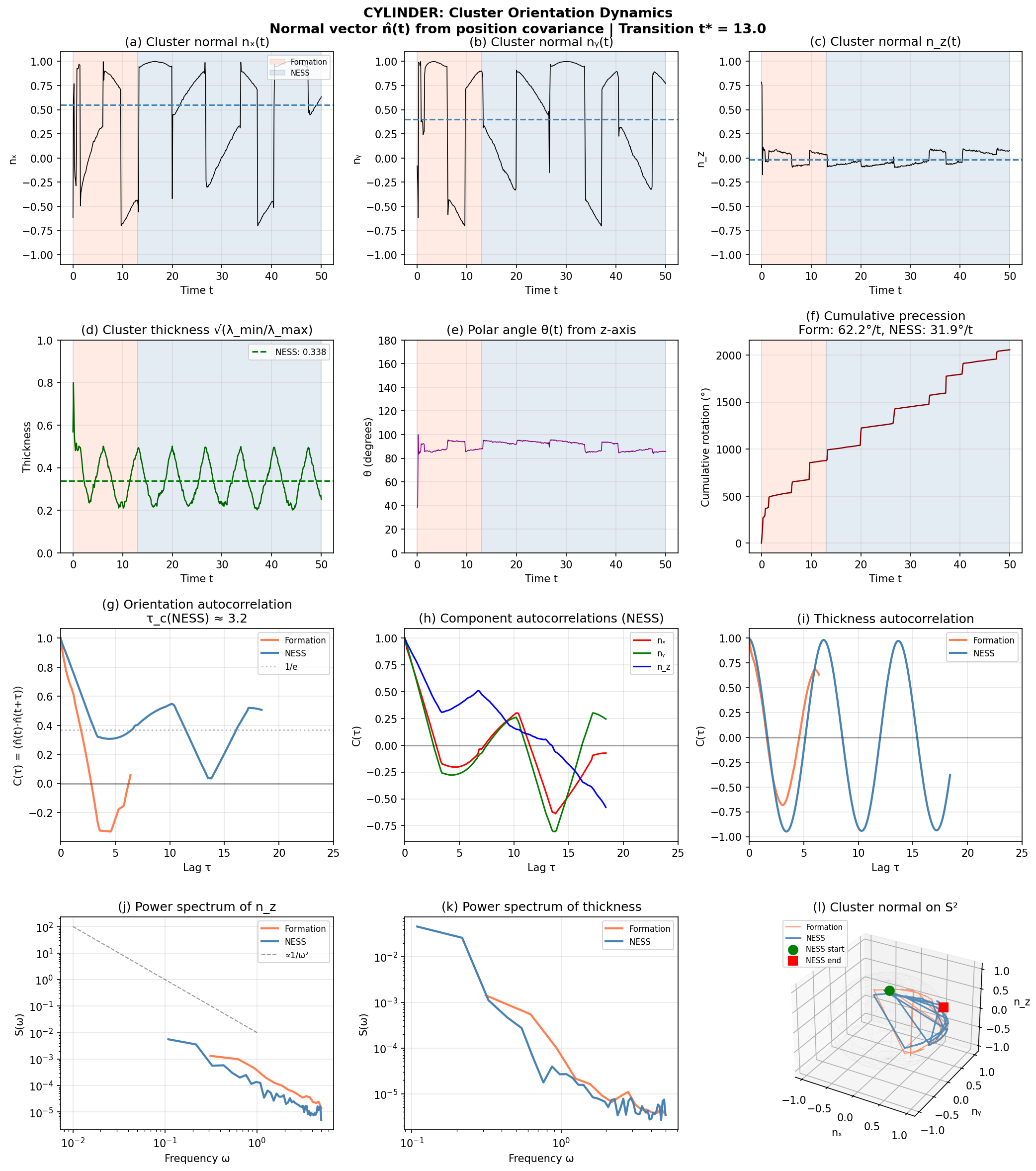}
    \caption{Cylinder $S^1 \times [0,H]$: Cluster orientation dynamics for $N = 400$ particles, $T = 0.4$, full time series $t = 0$ to $50$. Top row: Cluster normal components showing confinement to the $xy$-plane ($n_z \approx 0$). Second row: Thickness remains approximately constant ($\mathcal{T} \approx 0.34$); polar angle $\theta \approx 90^\circ$ confirming horizontal orientation; cumulative precession at $\sim 32^\circ$/time. Third row: Orientation autocorrelation with short correlation time $\tau_c \approx 3.2$; component autocorrelations; thickness autocorrelation. Bottom row: Power spectra with $S_{n_z}(\omega) \propto \omega^{-2.0}$, the only geometry approaching the theoretical $1/\omega^2$ prediction; 3D trajectory confined to the equatorial band of $S^2$.}
    \label{fig:cylinder_orientation}
\end{figure}

\paragraph{Cluster Orientation Dynamics.}
On the cylinder, the cluster normal $\hat{\mathbf{n}}(t)$ is confined to lie approximately in the $xy$-plane (perpendicular to the cylinder axis), with $\langle n_z \rangle \approx 0$ and $\sigma(n_z) \approx 0.07$. This reflects the geometric constraint: clusters arranged along the $z$-axis produce a distribution whose minimal-extent direction lies in the horizontal plane. Figure~\ref{fig:cylinder_orientation} shows the orientation dynamics.

\begin{itemize}
    \item \textbf{Rapid reorientation}: The correlation time $\tau_c \approx 3.2$ is significantly shorter than on the sphere or torus, indicating faster reorientation of the cluster arrangement. This may reflect the discrete nature of the cluster configuration (four clusters) compared to the continuous band on the sphere.
    
    \item \textbf{Planar precession}: The cumulative precession of $\sim 1200^\circ$ over the NESS phase corresponds to a rate of $\sim 32^\circ$ per unit time, approximately half the sphere's rate despite the shorter correlation time. The precession is confined to the $xy$-plane rather than exploring the full $S^2$.
    
    \item \textbf{Constant thickness}: Unlike the sphere, the cylinder shows minimal change in thickness between formation ($\mathcal{T} \approx 0.36$) and NESS ($\mathcal{T} \approx 0.34$), suggesting that the cluster configuration establishes quickly and maintains a consistent geometry.
    
    \item \textbf{Power spectrum}: The structure normal power spectrum $S_{n_z}(\omega) \propto \omega^{-2.0}$ is the only geometry approaching the theoretical $1/\omega^2$ prediction for type-A Goldstone diffusion modes.
\end{itemize}

\paragraph{Summary: Cylinder Dynamics.}
On the bounded cylinder, dimensional reduction produces four localized clusters (two vertical pairs) that form rapidly by $t \approx 1$. The dynamics differ qualitatively from the sphere: the cumulative rotation exhibits \emph{ballistic} scaling ($\alpha = 2.0$) rather than diffusive, and the orientation correlation time is short ($\tau_c \approx 3.2$) compared to the sphere ($\tau_c \approx 18$). The ballistic MSD suggests persistent collective modes that do not decorrelate, while the rapid orientation reorientation reflects the discrete nature of the cluster configuration. The boundaries play a key role, pinning clusters at $z \approx \pm 1.5$ and creating the discrete $\mathbb{Z}_2$ residual symmetry.

\subsection{Results on the Torus: $S^1 \times S^1 \to \coprod_2 S^1$}
\label{sec:torus}
\label{sec:results_torus}

\begin{figure}[!ht]
    \centering
    \includegraphics[width=0.6\textwidth]{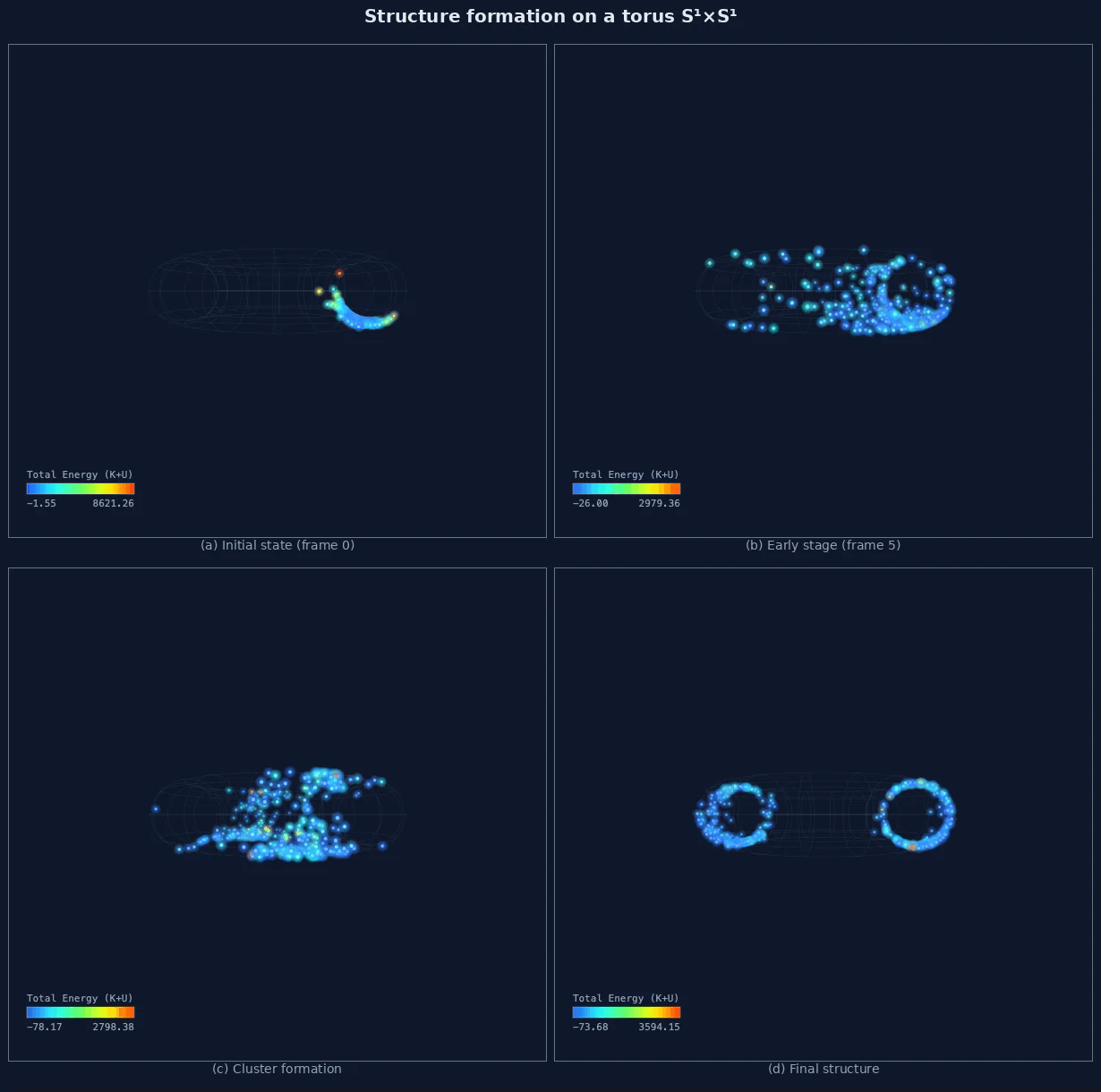}
    \caption{Dynamic dimension reduction on the torus $S^1 \times S^1 \to \coprod_2 S^1$. (a) Initial arc-shaped cluster. (b) Early dispersion across the torus surface. (c) Emergence of a metastable three-ring structure. (d) Final configuration with two well-separated rings along minor circles (constant $\theta$) after coalescence from the three-ring state. The two-torus has collapsed to a disjoint union of two circles.}
    \label{fig:torus_evolution}
\end{figure}

Figure~\ref{fig:torus_evolution} illustrates the dynamic dimension reduction on the torus.

\paragraph{Ring Formation: Torus to Disjoint Circles.}

The dimensional reduction $S^1 \times S^1 \to \coprod_2 S^1$ proceeds through a characteristic formation process. Analysis of the full time series from $t = 0$ to $t = 50$ reveals that ring formation occurs over approximately $t \approx 13$ time units.

Table~\ref{tab:torus_phases} summarizes the formation timeline based on the full history analysis.

\begin{table}[h!]
\centering
\caption{Ring formation phases on the torus from simulation with $N=400$ particles, $T=0.4$, $R=1.5$, $r=0.6$, total time $t=0$ to $50$.}
\label{tab:torus_phases}
\begin{tabular}{@{}llll@{}}
\toprule
\textbf{Phase} & \textbf{Time} & \textbf{Duration} & \textbf{Description} \\ \midrule
Initial cluster & $t = 0$ & ;  & Localized Gaussian blob \\
Dispersion & $t = 0$--$5$ & $\sim 5$ & Particles spread across torus \\
Ring assembly & $t = 5$--$13$ & $\sim 8$ & Ring structures emerge \\
\textbf{Steady state} & $t > 13$ & Indefinite & Persistent two-ring configuration \\ \bottomrule
\end{tabular}
\end{table}

The stable two-ring structure emerges by approximately $t \approx 13$ and \textbf{persists indefinitely} throughout the remainder of the simulation (tested to $t = 50$).

This two-stage relaxation has a geometric interpretation: the minor circles at the outer equator ($\varphi = 0$) and inner equator ($\varphi = \pi$) are geodesics on the torus, making them natural attractors. The initial three-ring state represents a local energy minimum with particles distributed across three toroidal positions. However, for balanced attractive/repulsive interactions, the global minimum corresponds to two diametrically opposite rings, which maximizes the separation between particles with net repulsive interactions between the groups while minimizing frustration.

\paragraph{Cumulative Rotation Dynamics.}
\label{sec:torus_correlations}

Table~\ref{tab:torus_L_stats} summarizes the cumulative rotation statistics for the torus.

\begin{table}[h!]
\centering
\caption{Cumulative rotation statistics on the torus separated by phase (dimensionless units). The standard deviation \emph{increases} from formation to steady state, indicating that the two-ring structure allows larger collective fluctuations.}
\label{tab:torus_L_stats}
\begin{tabular}{@{}lccc@{}}
\toprule
\textbf{Phase} & \textbf{Time Range} & $\langle \Theta_\theta \rangle$ & $\sigma_{\Theta_\theta}$ \\ \midrule
Formation & $t = 0$--$13$ & $57.1$ & $359$ \\
Steady state & $t = 13$--$50$ & $80.5$ & $396.0$ \\ \bottomrule
\end{tabular}
\end{table}

The torus exhibits the largest rotational fluctuations among the three geometries ($\sigma \approx 396$ vs.\ $\sigma \approx 149$ for sphere and $\sigma \approx 244$ for cylinder), consistent with the theoretical predictions of Table~\ref{tab:rotation_diffusion}. The inter-ring angular velocity correlation is $\rho = -0.27$, indicating that when one ring accelerates, the other tends to decelerate. This negative correlation is stronger than on the cylinder ($\rho \approx -0.1$ to $-0.2$) because the torus has only two structures versus four clusters.

\paragraph{Superdiffusive Ring Rotation with Systematic Drift.}
The MSD of the cumulative azimuthal rotation exhibits power-law scaling:
\begin{equation}
    \langle \Delta \theta^2 \rangle \sim t^{\alpha}, \quad \alpha \approx 1.7
\end{equation}
The torus exhibits \emph{superdiffusive} behavior, intermediate between the diffusive sphere ($\alpha \approx 1$) and the ballistic cylinder ($\alpha \approx 2$). Ensemble averaging over five independent realizations yields $\alpha = 1.73 \pm 0.10$.

The torus also shows a \emph{systematic positive drift}: all five realizations exhibit $\bar{\omega} > 0$, with ensemble mean $\langle \bar{\omega} \rangle = +0.075$ (significantly different from zero, $p < 0.001$). This is in stark contrast to the sphere and cylinder, where $\bar{\omega}$ fluctuates around zero. The drift arises from a \emph{stochastic ratchet mechanism}: particles undergo thermal fluctuations in $\varphi$ around their mean ring positions, and because the effective noise amplitude $\sigma_\theta(\varphi) = \sqrt{2D}/(R + r\cos\varphi)$ varies nonlinearly with $\varphi$, these fluctuations are rectified into directed toroidal motion. The asymmetric ring positions (one ring on the outer torus surface, one on the inner surface) break detailed balance, creating a geometry-induced Brownian motor \cite{reimann2002}.

The contrast between torus and cylinder dynamics is summarized in Table~\ref{tab:torus_vs_cylinder}.

\begin{table}[h!]
\centering
\caption{Comparison of collective dynamics on the bounded cylinder versus the torus (dimensionless units). Both geometries exhibit weak anti-correlations between structures, but show qualitatively different MSD scaling.}
\label{tab:torus_vs_cylinder}
\begin{tabular}{@{}lcc@{}}
\toprule
\textbf{Property} & \textbf{Bounded Cylinder} & \textbf{Torus} \\ \midrule
Final structure & 4 localized clusters & 2 quasi-continuous rings \\
Formation time & $t \approx 1$ & $t \approx 13$ \\
$\omega$ correlation & $\rho \approx -0.1$ to $-0.2$ & $\rho = -0.27$ \\
MSD exponent & $\alpha \approx 2.0$ (ballistic) & $\alpha \approx 1.7$ (superdiffusive) \\
Systematic drift & No ($\langle\bar{\omega}\rangle \approx 0$) & Yes ($\langle\bar{\omega}\rangle = +0.075$) \\ \bottomrule
\end{tabular}
\end{table}

\paragraph{Ring Width and Curvature Effects.}
The quality of dimensional reduction can be assessed through the ring widths, which show similar spread in both toroidal and poloidal directions ($\sigma \approx 20$--$26$ pixels), consistent with the quasi-1D reduction $T^2 \to \coprod_2 S^1$.

\begin{figure}[!ht]
    \centering
    \includegraphics[width=\textwidth]{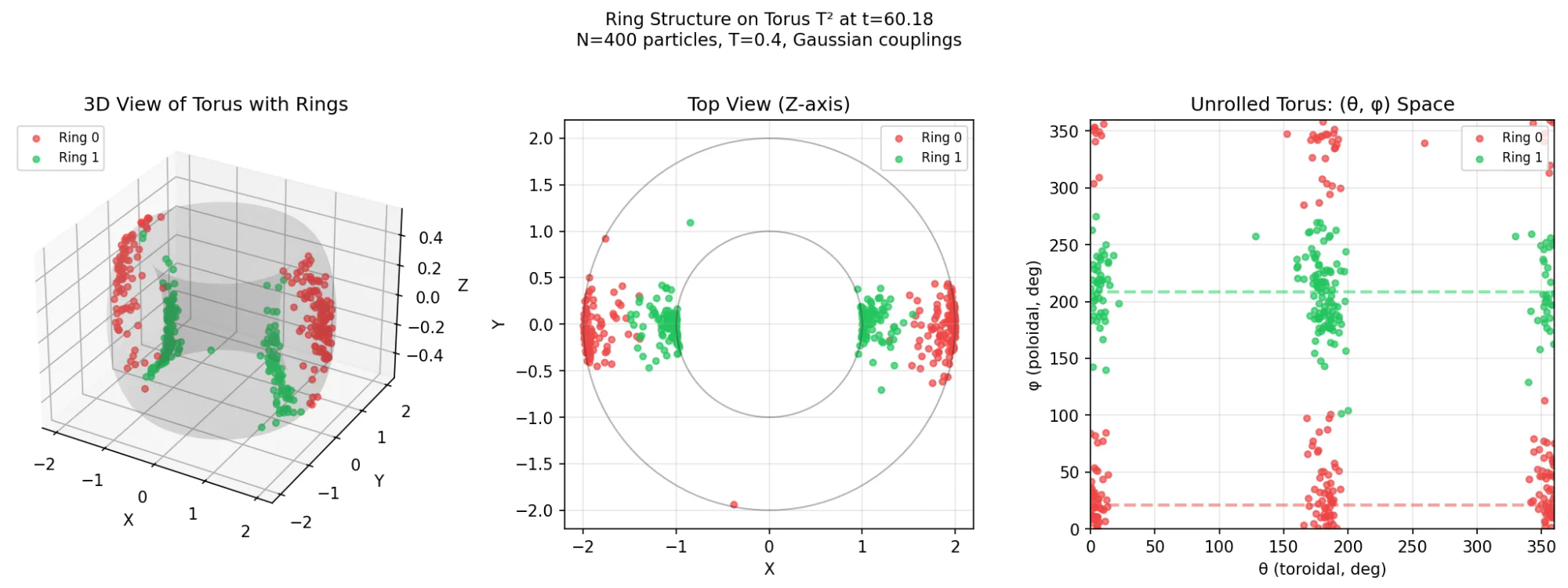}
    \caption{Visualization of two-ring structure on torus $T^2$ at $t = 50$. $N = 400$ particles, $T = 0.4$, Gaussian couplings. Left: 3D view showing two rings (red and green) wrapped around the torus. Center: Top view (Z-axis) showing ring positions. Right: Unrolled torus in $(\theta, \varphi)$ space, with horizontal dashed lines marking ring poloidal positions.}
    \label{fig:torus_rings_viz}
\end{figure}

\begin{figure}[!ht]
    \centering
    \includegraphics[width=\textwidth]{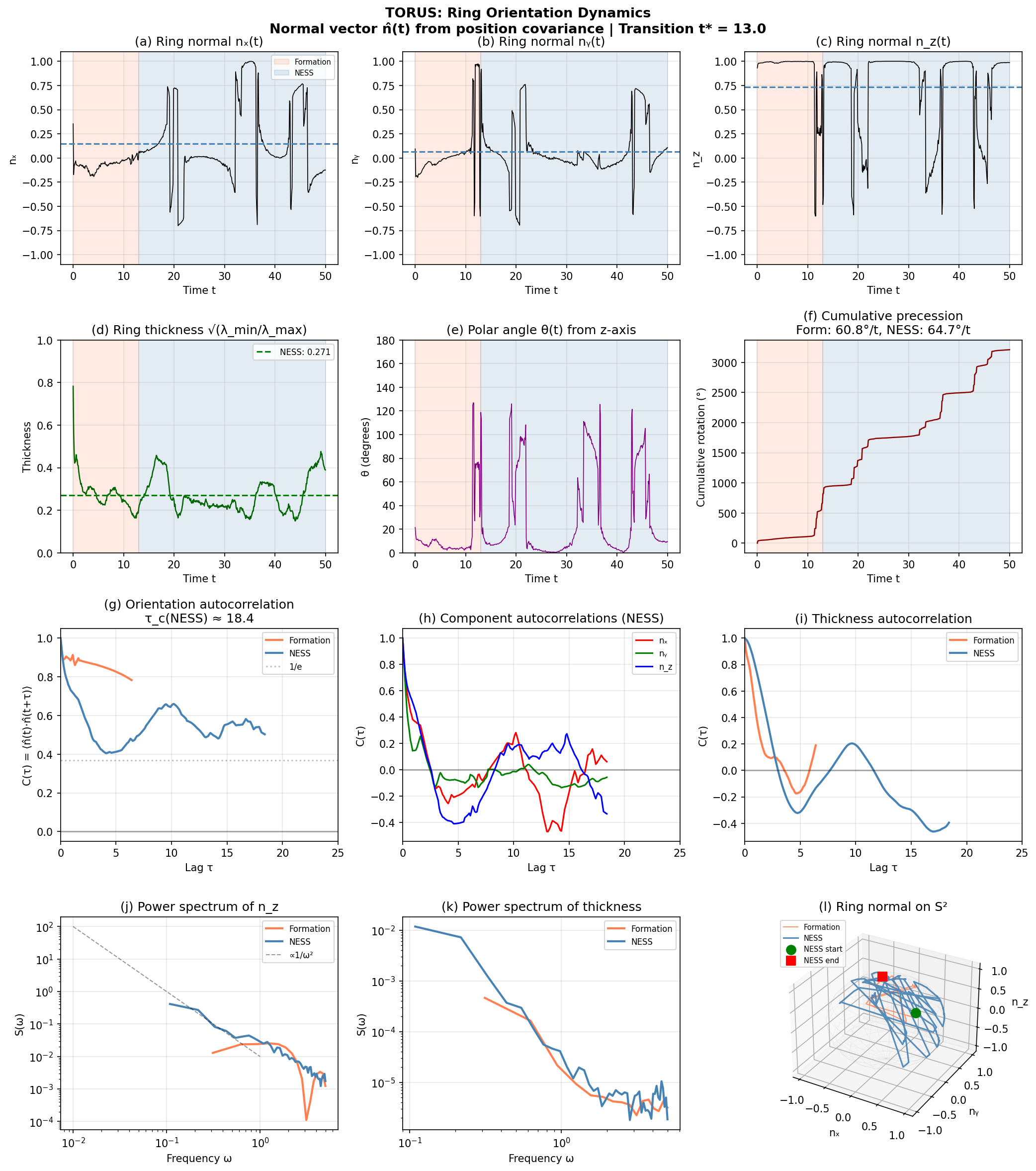}
    \caption{Torus $T^2$: Ring orientation dynamics for $N = 400$ particles, $T = 0.4$, full time series $t = 0$ to $50$. Top row: Ring normal components showing strong $z$-bias ($\langle n_z \rangle \approx 0.73$) with large fluctuations. Second row: Thickness remains constant ($\mathcal{T} \approx 0.27$); polar angle $\theta \approx 43^\circ$ reflecting the tilted ring plane; cumulative precession at $\sim 65^\circ$/time. Third row: Orientation autocorrelation with correlation time $\tau_c \approx 18$, comparable to the sphere; component autocorrelations showing different decay rates; thickness autocorrelation. Bottom row: Power spectra with $S_{n_z}(\omega) \propto \omega^{-1.4}$ scaling; 3D trajectory on $S^2$ showing exploration biased toward the northern hemisphere.}
    \label{fig:torus_orientation}
\end{figure}

\paragraph{Ring Orientation Dynamics.}
On the torus, the ring normal exhibits qualitatively different behavior: a strong bias toward the $z$-axis with $\langle n_z \rangle \approx 0.73$ and large fluctuations $\sigma(n_z) \approx 0.44$. This reflects the torus geometry: when particles form rings at constant poloidal angle $\varphi$, the plane containing these rings is tilted relative to the $xy$-plane, producing a ring normal with significant $z$-component. Figure~\ref{fig:torus_orientation} illustrates this behavior.

\begin{itemize}
    \item \textbf{Geometric constraint}: The mean $\langle n_z \rangle \approx 0.73$ corresponds to a ring-plane tilt of $\arccos(0.73) \approx 43^\circ$ from horizontal. This is consistent with rings forming at poloidal positions $\varphi \approx \pm 45^\circ$ from the outer equator of the torus.
    
    \item \textbf{Large orientation fluctuations}: The standard deviation $\sigma(n_z) \approx 0.44$ is substantially larger than for the sphere ($\sigma(n_z) \approx 0.11$), indicating that the ring plane can tilt significantly as particles redistribute between the two rings.
    
    \item \textbf{Long correlation time}: Despite the large fluctuations, the correlation time $\tau_c \approx 18$ matches the sphere, suggesting similar underlying dynamics for the slow orientation evolution.
    
    \item \textbf{Rapid precession}: The cumulative precession of $\sim 2400^\circ$ (rate $\sim 65^\circ$/time) slightly exceeds the sphere, consistent with the systematic drift observed in the cumulative rotation.
    
    \item \textbf{Power spectrum}: The structure normal power spectrum $S_{n_z}(\omega) \propto \omega^{-1.4}$ is the shallowest among the three geometries, deviating most from the theoretical $1/\omega^2$ prediction for type-A Goldstone diffusion modes.
\end{itemize}

\paragraph{Summary: Torus Dynamics.}
On the torus, dimensional reduction produces two rings at diametrically opposite poloidal positions, forming by $t \approx 13$ (the slowest among the three geometries). The cumulative rotation exhibits \emph{superdiffusive} scaling ($\alpha = 1.7$) and systematic positive drift ($\langle\bar{\omega}\rangle = +0.075$), arising from a stochastic ratchet mechanism created by the non-uniform metric. The ring normal shows a strong $z$-bias ($\langle n_z \rangle \approx 0.73$) reflecting the tilted ring-plane geometry, with long correlation time ($\tau_c \approx 18$) comparable to the sphere. The torus provides the clearest example of how non-uniform local geometry can induce non-equilibrium steady states with directed motion from thermal fluctuations.

\subsection{Comparison: Manifold-Dependent Dimension Reduction}

Table~\ref{tab:comparison} summarizes the dynamic dimension reduction across the three geometries, revealing both shared features and manifold-dependent differences.

\begin{table}[h!]
\centering
\small
\caption{Dynamic dimension reduction across manifolds (dimensionless units). Statistics are from full time series analysis ($t = 0$ to $50$) for $N = 400$ particles at $T = 0.4$ with Gaussian couplings.}
\label{tab:comparison}
\begin{tabular}{@{}p{2.8cm}p{2.8cm}p{3.2cm}p{3.2cm}@{}}
\toprule
\textbf{Feature} & \textbf{Sphere $S^2$} & \textbf{Cylinder $S^1 \times [0,H]$} & \textbf{Torus $T^2$} \\ \midrule
Manifold topology & Closed & With boundary & Closed \\
Dimension reduction & $S^2 \to S^1$ & $S^1 \times [0,H] \to \mathbb{Z}_2$ & $T^2 \to \coprod_2 S^1$ \\
Target space & Continuous & Discrete & Continuous \\
Final structure & 1 equatorial band & 4 clusters (2 vertical pairs) & 2 opposite rings \\
Formation time & $t \approx 5$ & $t \approx 1$ & $t \approx 13$ \\
Inter-structure corr. & N/A (single band) & $\rho \approx -0.1$ to $-0.2$ & $\rho = -0.27$ \\
MSD exponent $\alpha$ & $0.9$ (diffusive) & $2.0$ (ballistic) & $1.7$ (superdiffusive) \\ \bottomrule
\end{tabular}
\end{table}

While dimension reduction from 2D to quasi-1D structures occurs on all three manifolds, the character of the reduction depends strongly on the specific manifold. The shared features are: (i) 2D $\to$ 1D collapse to quasi-1D structures; (ii) approximate reduction with residual width and fluctuations from thermal noise; and (iii) glassy relaxation through multiple timescales.

The manifold-specific differences reflect the interplay of topology and local geometry:
\begin{itemize}
    \item On the \textbf{sphere}, uniform local geometry and simple topology produce the cleanest behavior: diffusive rotation ($\alpha \approx 0.9$) with no systematic drift, directly confirming the pure-diffusion prediction.
    
    \item On the \textbf{bounded cylinder}, boundaries dramatically alter the dynamics. The four-cluster configuration shows \emph{ballistic} behavior ($\alpha \approx 2.0$), suggesting boundary conditions create persistent collective modes that do not decorrelate.
    
    \item On the \textbf{torus}, non-uniform local geometry produces \emph{superdiffusive} behavior ($\alpha \approx 1.7$) with systematic positive drift arising from a stochastic ratchet mechanism.
\end{itemize}

\paragraph{Role of Topology: Closed Manifolds vs.\ Manifolds with Boundary.}
A clear distinction emerges between \emph{closed manifolds} (compact without boundary) and \emph{manifolds with boundary}. The sphere and torus are topologically closed: every direction ``closes on itself,'' with no edges or endpoints. On these manifolds, dimension reduction produces \emph{continuous} target spaces: particles collapse onto extended structures (bands, rings) that wrap around closed geodesics, inheriting continuous residual symmetries (SO(2) on the sphere; SO(2)$\times\mathbb{Z}_2$ on the torus).

The bounded cylinder is topologically distinct: it has circular boundaries at $z = 0$ and $z = H$. These boundaries act as impenetrable barriers, causing axial localization (particles accumulate near edges) and azimuthal fragmentation (clusters at specific angles rather than continuous rings). The resulting discrete residual symmetry ($\mathbb{Z}_2$, corresponding to $180^\circ$ rotation exchanging both cluster pairs) reflects this boundary-induced localization.

This topological dependence suggests a general principle: \emph{dimension reduction on closed manifolds preserves continuous symmetries, while boundaries break continuous symmetry down to discrete subgroups}. The nature of emergent structures (extended loops versus localized clusters) is dictated by whether particles can form topologically non-trivial configurations that wrap around the manifold.

\paragraph{Role of Local Geometry: Uniform vs.\ Non-Uniform Metrics.}
Beyond topology, the local geometry, encoded in the metric tensor $g_{ab}$, significantly affects the dynamics. The sphere has uniform local geometry: constant Gaussian curvature $K = 1/R^2$ everywhere. This uniformity ensures that the noise amplitude in angular coordinates is position-independent, leading to pure diffusive behavior ($\alpha \approx 1$) with no systematic drift.

The torus, despite being topologically closed like the sphere, has non-uniform local geometry: the Gaussian curvature varies from positive (outer equator) to negative (inner equator), and the metric factor $(R + r\cos\varphi)^2$ creates position-dependent noise amplitudes. This geometric heterogeneity is responsible for the superdiffusive scaling ($\alpha \approx 1.7$) and the systematic drift, effects absent on the geometrically uniform sphere.

The cylinder has uniform local geometry (zero Gaussian curvature, flat metric), yet its dynamics differs qualitatively from both sphere and torus due to the boundary effects. This comparison illustrates that topology and local geometry contribute independently to the collective dynamics: the cylinder's ballistic behavior ($\alpha \approx 2$) arises from topological boundary effects despite its uniform local geometry.

\subsection{Summary: Orientation Dynamics Across Geometries}
\label{sec:orientation_summary}

Table~\ref{tab:orientation_stats} summarizes the structure orientation statistics for all three geometries.

\begin{table}[h!]
\centering
\caption{Structure orientation statistics across geometries. The formation phase shows larger thickness (diffuse structure) and larger orientation fluctuations. In NESS, structures are well-formed (small $\mathcal{T}$) with characteristic orientation dynamics. Simulations with $N = 400$ particles, $T = 0.4$, transition time $t^* = 13$.}
\label{tab:orientation_stats}
\begin{tabular}{@{}llccc@{}}
\toprule
\textbf{Geometry} & \textbf{Phase} & \textbf{Thickness $\mathcal{T}$} & $\boldsymbol{\sigma(n_z)}$ & \textbf{Correlation time $\tau_c$} \\ \midrule
\multirow{2}{*}{Sphere $S^2$} & Formation & $0.62 \pm 0.15$ & $0.19$ & --- \\
 & NESS & $0.29 \pm 0.12$ & $0.11$ & $18.4$ \\
\midrule
\multirow{2}{*}{Cylinder} & Formation & $0.36 \pm 0.11$ & $0.09$ & --- \\
 & NESS & $0.34 \pm 0.09$ & $0.07$ & $3.2$ \\
\midrule
\multirow{2}{*}{Torus $T^2$} & Formation & $0.27 \pm 0.07$ & $0.32$ & --- \\
 & NESS & $0.27 \pm 0.08$ & $0.44$ & $18.4$ \\
\bottomrule
\end{tabular}
\end{table}

The structure normal analysis reveals a clear hierarchy of orientation dynamics:

\begin{enumerate}
    \item \textbf{Sphere}: Full 3D precession on $S^2$ with isotropic exploration, long correlation time ($\tau_c \approx 18$), and diffusive dynamics. The band normal is a true order parameter living on the Goldstone manifold $\mathrm{SO}(3)/\mathrm{SO}(2) \cong S^2$.
    
    \item \textbf{Torus}: Constrained precession with $z$-bias, reflecting the intrinsic geometry of ring configurations on $T^2$. Long correlation time comparable to the sphere, but with larger amplitude fluctuations due to the discrete (two-ring) structure.
    
    \item \textbf{Cylinder}: Planar precession confined to the $xy$-plane, with short correlation time ($\tau_c \approx 3.2$) reflecting rapid reorganization of the discrete cluster configuration. The boundary-induced localization constrains orientation dynamics to a 1D subspace of $S^2$.
\end{enumerate}

\paragraph{Connection to Dissipative Nambu-Goldstone Modes.}
The structure normal $\hat{\mathbf{n}}(t)$ is the Goldstone mode variable: it specifies where the emergent structure is oriented on the Goldstone manifold $S^2$ (for the sphere) or related spaces (for other geometries). The autocorrelation function $C(\tau) = \langle \hat{\mathbf{n}}(t) \cdot \hat{\mathbf{n}}(t+\tau) \rangle$ and its Fourier transform, the structure normal power spectrum $S_{n_z}(\omega)$, characterize this Goldstone mode dynamics. For type-A diffusion modes in dissipative systems, theory predicts $S(\omega) \propto 1/\omega^2$ at low frequencies \cite{minami2018, hidaka2020}. Our measurements reveal geometry-dependent power law exponents: $S_{n_z}(\omega) \propto \omega^{-\beta}$ with $\beta \approx 1.6$ on the sphere, $\beta \approx 2.0$ on the cylinder, and $\beta \approx 1.4$ on the torus. Only the cylinder approaches the theoretical $1/\omega^2$ prediction, but the cylinder has no continuous Goldstone manifold (SO(2) $\to \mathbb{Z}_2$). The deviations on the sphere and torus may reflect finite-size effects or the bounded nature of $\hat{\mathbf{n}}$ on $S^2$.

The long correlation time $\tau_c \approx 18$ on the sphere and torus is the characteristic timescale for the Goldstone mode, far exceeding the microscopic thermal correlation time ($\tau_v < 0.2$). This separation of timescales confirms the adiabatic nature of orientation dynamics: the structure can be treated as quasi-static on timescales shorter than $\tau_c$. The short correlation time on the cylinder ($\tau_c \approx 3.2$) reflects the absence of a continuous Goldstone manifold; the discrete $\mathbb{Z}_2$ symmetry allows rapid reorientation between equivalent configurations.

\subsection{Symmetry Analysis: Steady States and Disorder-Induced Rotational Symmetry Breaking}
\label{sec:symmetry_analysis}

The emergent structures observed in our simulations (bands on the sphere, clusters on the cylinder, and rings on the torus) can be understood through the lens of symmetry breaking. Each manifold possesses continuous rotational symmetries that are \emph{spontaneously broken} by the instantaneous particle configurations. However, as we shall argue, this phenomenon differs fundamentally from spontaneous symmetry breaking (SSB) as understood in quantum field theory and equilibrium statistical mechanics.

\paragraph{Symmetry Groups and Their Violation.}
Each manifold possesses a characteristic continuous symmetry group: the sphere $S^2$ is invariant under the full rotation group SO(3); the bounded cylinder $S^1 \times [0,H]$ under SO(2) rotations about its axis; and the torus $T^2$ under the product group SO(2) $\times$ SO(2) (toroidal and poloidal rotations). The structures that emerge from our simulations \emph{instantaneously} break these symmetries. On the \textbf{sphere}, band formation along a specific great circle breaks SO(3) to SO(2), the subgroup of rotations about the axis perpendicular to the band plane. On the \textbf{cylinder}, clusters at specific azimuthal positions break SO(2) to $\mathbb{Z}_2$ (a $180^\circ$ rotation exchanges both cluster pairs). On the \textbf{torus}, the two rings at diametrically opposite toroidal positions break SO(2) $\times$ SO(2) to SO(2)$\times\mathbb{Z}_2$: the continuous SO(2) of poloidal rotations is preserved (both rings rotate synchronously around the minor circle), while the toroidal SO(2) breaks to $\mathbb{Z}_2$ (only $180^\circ$ rotations exchanging the two rings remain).

\begin{table}[h!]
\centering
\caption{Symmetry groups and their breaking by emergent structures. The instantaneous configurations break the full symmetry of each manifold to a smaller residual symmetry.}
\label{tab:symmetry_breaking}
\begin{tabular}{@{}lccc@{}}
\toprule
\textbf{Manifold} & \textbf{Full Symmetry} & \textbf{Residual Symmetry} & \textbf{Broken Generators} \\ \midrule
Sphere $S^2$ & $\mathrm{SO}(3)$ & $\mathrm{SO}(2)$ & 2 (tilts of band plane) \\
Cylinder $S^1 \times [0,H]$ & $\mathrm{SO}(2)$ & $\mathbb{Z}_2$ & 1 (azimuthal rotation) \\
Torus $T^2$ & $\mathrm{SO}(2) \times \mathrm{SO}(2)$ & $\mathrm{SO}(2) \times \mathbb{Z}_2$ & 1 (toroidal $\to \mathbb{Z}_2$) \\ \bottomrule
\end{tabular}
\end{table}

\paragraph{Distinction from True Spontaneous Symmetry Breaking.} Despite the apparent similarity, our phenomenon differs fundamentally from spontaneous symmetry breaking (SSB) as understood in quantum field theory \cite{zinnjustin2002}. In true SSB, the ground state does not share the Hamiltonian's symmetry, the symmetry-breaking direction is fixed once selected, and Goldstone modes emerge corresponding to motion along the degenerate ground-state manifold. In our system, the symmetry-breaking direction (band orientation, cluster positions, ring locations) is not fixed but \textit{slowly evolves} in time, driven by thermal noise. Pairwise forces produce zero net torque (equation (\ref{eq:zero_torque}) and its analogues for other geometries), so there is no deterministic restoring force that would freeze the orientation. True SSB also requires the thermodynamic limit $N \to \infty$; our finite-$N$ structures are precursors to any putative SSB.

\paragraph{Identifying the Goldstone Mode: Structure Normal vs.\ Cumulative Rotation.}
A key conceptual point requires clarification. In our system, two distinct observables characterize rotational dynamics:

\begin{enumerate}
    \item \textbf{Structure normal $\hat{\mathbf{n}}(t)$}: The slow variable that specifies where the emergent structure is oriented in 3D space. This is the \emph{Goldstone mode} variable that explores the manifold of degenerate configurations (the Goldstone manifold).
    
    \item \textbf{Cumulative rotation $\Theta(t)$}: The fast variable that measures total angular displacement around a fixed axis. This is \emph{not} a Goldstone mode; it is driven by the instantaneous angular momentum, which thermalizes rapidly in the overdamped limit.
\end{enumerate}

The analogy with ferromagnetism makes this distinction clear. In a ferromagnet, SO(3) $\to$ SO(2) breaking produces one Goldstone mode: the slow reorientation of the magnetization direction $\hat{\mathbf{M}}(t)$ on $S^2 \cong \text{SO}(3)/\text{SO}(2)$. The total spin $|\mathbf{M}|$ is fixed; only its direction fluctuates. Similarly, on our sphere, the band normal $\hat{\mathbf{n}}(t)$ plays the role of $\hat{\mathbf{M}}$, slowly precessing on the Goldstone manifold $S^2$. The cumulative rotation $\Theta_z(t)$ measures something different: how much the system has rotated around the external $z$-axis, regardless of band orientation. This is analogous to measuring total angular momentum transfer, not magnetization direction.

\paragraph{Goldstone Mode Counting.}
The number of Goldstone modes depends on the symmetry breaking pattern:

\begin{itemize}
    \item \textbf{Sphere}: SO(3) $\to$ SO(2) breaks two generators. The Goldstone manifold is $S^2$, a 2-dimensional space. In dissipative systems, this yields one diffusive mode: the slow precession of $\hat{\mathbf{n}}(t)$ on $S^2$, with correlation time $\tau_c \approx 18$.
    
    \item \textbf{Torus}: SO(2) $\times$ SO(2) $\to$ SO(2) $\times \mathbb{Z}_2$ breaks one continuous generator (toroidal SO(2) $\to \mathbb{Z}_2$). The Goldstone manifold is $S^1/\mathbb{Z}_2$, corresponding to the collective toroidal position of the ring pair. The structure normal $\hat{\mathbf{n}}(t)$ captures a related but distinct aspect: the orientation of the ring plane, which has a preferred direction due to torus geometry ($\langle n_z \rangle \approx 0.73$).
    
    \item \textbf{Cylinder}: SO(2) $\to \mathbb{Z}_2$ breaks continuous symmetry to discrete. There is no continuous Goldstone manifold; the discrete $\mathbb{Z}_2$ symmetry allows only finite rotations ($180^\circ$) between equivalent configurations. This explains the short correlation time ($\tau_c \approx 3.2$) and ballistic MSD: the system does not explore a continuous manifold but instead exhibits persistent rotational modes.
\end{itemize}

\paragraph{Dissipative Goldstone Mode Classification.} When spacetime symmetries are broken in dissipative systems, the Minami-Hidaka classification \cite{minami2018, hidaka2020} applies:

\begin{itemize}
    \item \textbf{Type-A diffusion modes}: Relaxation rate $\Gamma \propto Dk^2$. At $k = 0$, this gives a gapless diffusive mode. The power spectrum prediction is $S(\omega) \propto 1/\omega^2$ at low frequencies.
    
    \item \textbf{Type-B diffusion modes}: Relaxation rate $\Gamma \propto D'k^4$ or gapped.
\end{itemize}

For our sphere, the slow precession of $\hat{\mathbf{n}}(t)$ should exhibit type-A behavior. Our measurements of the structure normal power spectrum $S_{n_z}(\omega) \propto \omega^{-\beta}$ yield:
\begin{itemize}
    \item Sphere: $\beta \approx 1.6$, somewhat shallower than the $1/\omega^2$ prediction
    \item Cylinder: $\beta \approx 2.0$, consistent with $1/\omega^2$ (but this geometry has no continuous Goldstone manifold)
    \item Torus: $\beta \approx 1.4$, the shallowest spectrum
\end{itemize}

The deviations from $\beta = 2$ on the sphere and torus may reflect: (i) finite-size effects; (ii) the bounded nature of $\hat{\mathbf{n}}$ on $S^2$, which constrains long-time fluctuations; (iii) coupling between different components of $\hat{\mathbf{n}}$; or (iv) the non-uniform metric on the torus creating additional correlations.

\paragraph{The Fast Cumulative Rotation.}
The cumulative rotation $\Theta(t)$ exhibits different behavior that should not be interpreted as Goldstone physics:
\begin{itemize}
    \item On the sphere, $\Theta_z(t)$ shows diffusive MSD ($\alpha \approx 0.9$), consistent with the fast, thermally-driven rotational dynamics predicted by the overdamped Langevin equation.
    \item On the cylinder, $\Theta_z(t)$ shows ballistic MSD ($\alpha \approx 2.0$), indicating persistent collective modes that do not decorrelate.
    \item On the torus, $\Theta_\theta(t)$ shows superdiffusive MSD ($\alpha \approx 1.7$) with systematic drift from the stochastic ratchet mechanism.
\end{itemize}

The diffusive scaling $\alpha \approx 1$ on the sphere confirms that $\Theta(t)$ is driven by white thermal noise, as expected for the fast angular momentum variable in the overdamped limit. This is the behavior of a driven system, not the slow exploration of a Goldstone manifold.

\begin{table}[h!]
\centering
\caption{Summary of Goldstone mode characteristics across geometries. The structure normal $\hat{\mathbf{n}}(t)$ represents the Goldstone mode (slow orientation dynamics), while the cumulative rotation $\Theta(t)$ represents fast angular momentum dynamics. The sphere provides the cleanest Goldstone behavior; the cylinder has no continuous Goldstone manifold; the torus has geometric complications.}
\label{tab:goldstone_summary}
\begin{tabular}{lcccc}
\toprule
\textbf{Geometry} & \textbf{Goldstone manifold} & \textbf{$\tau_c$ of $\hat{\mathbf{n}}$} & \textbf{$S_{n_z} \propto \omega^{-\beta}$} & \textbf{MSD $\alpha$ of $\Theta$} \\ \midrule
Sphere $S^2$ & $S^2$ (continuous) & $18.4$ & $\beta \approx 1.6$ & $0.9$ (diffusive) \\
Torus $T^2$ & $S^1/\mathbb{Z}_2$ (w/ drift) & $18.4$ & $\beta \approx 1.4$ & $1.7$ (superdiffusive) \\
Cylinder & None (discrete) & $3.2$ & $\beta \approx 2.0$ & $2.0$ (ballistic) \\ \bottomrule
\end{tabular}
\end{table}

\paragraph{How Anomalous Diffusion Can Arise from Ordinary Langevin Dynamics}
The superdiffusive ($\alpha \approx 1.7$) and ballistic ($\alpha \approx 2$) MSD scaling observed in our model may appear paradoxical, since ordinary Brownian motion yields $\alpha = 1$. However, established literature confirms that anomalous diffusion emerges naturally from Langevin equations with state-dependent noise, without requiring fractional derivatives, L\'{e}vy flights, or memory kernels. Lacasta \textit{et al.} \cite{lacasta2004} demonstrated that standard underdamped Langevin dynamics on periodic potentials produces the full range from subdiffusion to superdiffusion. Cherstvy and Metzler \cite{cherstvy2013} showed that \emph{heterogeneous diffusion processes} with position-dependent diffusivity $D(x)$ yield anomalous MSD exponents; Vitali \textit{et al.} \cite{vitali2018} obtained similar results for distributed friction parameters; and the review by Metzler \textit{et al.} \cite{metzler2014} catalogs such mechanisms within standard Langevin frameworks. On curved manifolds, the metric tensor $g_{ab}(q)$ naturally creates position-dependent noise amplitude, placing our model squarely within this class. The torus, with its $\varphi$-dependent metric factor $(R + r\cos\varphi)^2$, exemplifies a heterogeneous diffusion process; the cylinder's boundary-pinned clusters create persistent collective modes analogous to the underdamped regime of \cite{lacasta2004}. These precedents establish that our observed anomalous scaling is a genuine physical effect of geometry-dependent Langevin dynamics, not a numerical artifact.

\paragraph{Connections to Adiabatic Invariants and Spin Glass Phenomenology.}
The rotational dynamics resembles \emph{adiabatic invariants} in classical mechanics \cite{arnold1989, landau1976}. The separation of timescales is dramatic: band formation occurs at $t \approx 5$ while diffusive exploration of the orientation manifold continues indefinitely. This enables a quasi-static treatment of slow collective variables. The slowly evolving symmetry-breaking structures also connect to spin glass physics: band orientations and ring positions label different members of a continuous family of degenerate steady states, analogous to the ultrametric organization in mean-field spin glasses \cite{mezard1987}. We characterize this ``disorder-induced rotational symmetry breaking'' by: instantaneous symmetry breaking; slow collective drift driven by thermal noise; ergodic exploration of all symmetry-equivalent configurations over long times; and adiabatic separation of timescales. This phenomenon bridges true SSB (frozen order parameter) and trivially symmetric states, closely related to weak ergodicity breaking in spin glasses \cite{bouchaud1998}.

\section{Discussion: Connections to Glassy Systems, Quantum Field Theory, and Other Models in Physics}
\label{sec:discussion}

The dynamic dimension reduction observed in our simulations connects to several major areas of physics: (1) disordered systems, particularly spin glasses, where random interactions lead to frustration and complex relaxation dynamics; (2) quantum field theory, where instanton transitions and confinement mechanisms share structural similarities with our structure formation process; and (3) astrophysical structure formation, where angular momentum plays a central role in determining final configurations. These connections are not merely analogies; they suggest that our model may serve as a tractable laboratory for studying phenomena that remain analytically challenging in their original contexts. We explore these connections in depth, identifying both the parallels and the crucial differences that distinguish our geometric model from each reference system.

\subsection{Connections to Spin Glass Physics}
\label{sec:spin_glass}

Our model is fundamentally a \emph{geometric spin glass}: particles with continuous degrees of freedom (positions on a manifold) interact through random couplings, creating the frustration and metastability characteristic of glassy systems. The potential energy $U = \sum_{i<j} \phi_{ij} \, d(q_i, q_j)$ involves coupling constants $\phi_{ij}$ drawn from either a Gaussian or binary distribution, precisely analogous to exchange couplings in spin glass models. Frustration arises when random couplings create incompatible demands that cannot all be resolved on a manifold of finite extent. The geometry introduces additional frustration mechanisms: geodesic nonlinearity, topological constraints, and curvature effects.

Our model has a particularly close relationship to the \emph{spherical spin glass} \cite{kosterlitz1976}, where the Ising constraint $\sigma_i = \pm 1$ is replaced by a global spherical constraint $\sum_i \sigma_i^2 = N$. Our particles on the sphere $S^2$ satisfy individual spherical constraints $|\mathbf{x}_i|^2 = 1$ rather than a global constraint, with configuration space $(S^2)^N$. Key differences include: we use geodesic distance rather than spin products; each particle is independently constrained; we study overdamped Langevin dynamics; and the dimensional reduction we observe has no direct spin glass analogue.

A crucial distinction from both conventional and spherical spin glasses lies in the physical interpretation of thermal noise. In Ising spin glasses, thermal fluctuations drive discrete spin flips; in spherical spin glasses, noise moves the system on a high-dimensional hypersphere $S^{N-1}$, an abstract configuration space with no direct physical realization. In our model, by contrast, each particle undergoes ordinary Brownian motion from collisions with solvent molecules, the standard mechanism of heat exchange between a mesoscopic object and its thermal environment. The manifolds we consider (sphere, torus, cylinder) are familiar geometric objects that can be physically constructed or approximated in laboratory settings: colloidal particles on curved interfaces, molecules adsorbed on nanotubes or vesicles, or active matter confined to curved substrates. This concrete physical interpretation suggests that the phenomena we observe (frustration-induced dimensional reduction, slow collective modes, geometry-dependent anomalous diffusion) may be directly accessible to experiment in soft matter and biophysical systems.

\paragraph{Glassy Relaxation Phenomenology.}
Our simulations exhibit the hallmarks of glassy dynamics. The energy relaxation shows \emph{two-stage relaxation}: a fast initial drop ($\beta$-relaxation) followed by a slow approach to quasi-equilibrium ($\alpha$-relaxation), with dimensional reduction occurring during the $\alpha$ stage. \emph{History dependence} is evident: different initial conditions lead to different final configurations, indicating multiple degenerate NESS whose selection depends on the path through configuration space. \emph{Aging} is also present: cluster configurations continue to evolve slowly at long times, and two-time overlap functions $q(t, t_w)$ should exhibit the characteristic $t/t_w$ scaling of aging systems (detailed quantitative study remains for future work).

The band precession on the sphere, correlated cluster-pair oscillations on the cylinder, and dual ring rotations on the torus represent \emph{slow collective modes} that continue to evolve long after dimensional reduction is complete. These modes are analogous to the ``soft modes'' in structural glasses. Full time series analysis ($t = 0$ to $50$) reveals manifold-dependent behavior: on the \textbf{sphere}, the band forms by $t \approx 5$ with diffusive rotation ($\alpha = 0.9 \pm 0.3$); on the \textbf{cylinder}, four clusters form rapidly by $t \approx 1$ with ballistic rotation ($\alpha = 2.0 \pm 0.05$); on the \textbf{torus}, two rings form by $t \approx 13$ with superdiffusive rotation ($\alpha = 1.7 \pm 0.1$) and systematic positive drift ($\langle\bar{\omega}\rangle = +0.075$). The structure orientation analysis of Section~\ref{sec:orientation_summary} reveals that these slow modes have correlation times $\tau_c \approx 18$ on the sphere and torus, but only $\tau_c \approx 3$ on the cylinder, reflecting the different ``stiffness'' of continuous versus discrete structures against thermal reorientation.

\paragraph{Distribution Functions: $P(\Theta)$ vs.\ $P(q)$ in Spin Glasses.}
In mean-field spin glasses, the Parisi order parameter distribution $P(q)$ characterizes replica overlap and exhibits non-trivial structure below the glass transition \cite{mezard1987, parisi1980}. The cumulative rotation $\Theta(t)$ in our model is fundamentally different: it is a time-integrated quantity that grows diffusively, $\langle \Theta^2 \rangle \propto D_{\text{rot}} t$, rather than a bounded order parameter. At any fixed time $t$, the distribution $P(\Theta; t)$ is approximately Gaussian (from the central limit theorem applied to the sum of many particle contributions), with zero ensemble mean (from rotational symmetry) and variance that grows linearly with time.

A more direct analogue to the Parisi $P(q)$ is the distribution of instantaneous structure orientations $P(\hat{\mathbf{n}})$ analyzed in Section~\ref{sec:orientation_summary}. Unlike $\Theta(t)$, the structure normal $\hat{\mathbf{n}}(t)$ is bounded (it lives on $S^2$) and well-defined at each instant. Our simulations reveal geometry-dependent orientation statistics in the NESS: on the \textbf{sphere}, $\langle n_z \rangle \approx 0$ with $\sigma(n_z) \approx 0.11$, indicating isotropic exploration of the Goldstone manifold; on the \textbf{cylinder}, $\langle n_z \rangle \approx 0$ with $\sigma(n_z) \approx 0.07$, reflecting confinement to the $xy$-plane; on the \textbf{torus}, $\langle n_z \rangle \approx 0.73$ with $\sigma(n_z) \approx 0.44$, showing a strong $z$-bias with large fluctuations. The orientation autocorrelation $C(\tau) = \langle \hat{\mathbf{n}}(t) \cdot \hat{\mathbf{n}}(t+\tau) \rangle$ decays to $1/e$ at $\tau_c \approx 18$ on the sphere and torus, but at $\tau_c \approx 3$ on the cylinder. This order-of-magnitude difference in correlation times between continuous structures (band, rings) and discrete structures (clusters) is reminiscent of the distinction between continuous and discrete symmetry breaking in equilibrium systems, where the former supports only gradual Goldstone-mode evolution while the latter permits rapid domain-wall mediated transitions.

\subsection{Connection to Instanton Physics in Spin Glasses and QFT Models}
\label{sec:instantons}

The rapid structure formation that we observe, where the system transitions from a localized ``Big Bang'' initial state to an extended but organized configuration (band, rings, or clusters), bears resemblance to instanton-mediated transitions in quantum field theory and spin glass models.

\paragraph{Instantons in QFT and Spin Glasses.}
In quantum field theory (QFT), instantons are classical solutions of field equations in imaginary (Euclidean) time that interpolate between degenerate vacuum states separated by potential barriers. These configurations dominate the path integral in the semiclassical limit and describe tunneling processes that are exponentially suppressed by the barrier height but occur very rapidly (almost instantaneously) in the conventional (Minkowski) time, hence justifying their name ``instantons.''

The foundational work of Polyakov \cite{polyakov1975, polyakov1977} established instantons as essential for understanding non-perturbative phenomena in gauge theories such as quantum chromodynamics (QCD) and the O(3) nonlinear sigma model. In Polyakov's setting, instantons connect \emph{degenerate} classically stable vacuum states; local minima of the potential energy that would persist indefinitely in the absence of quantum fluctuations.

A different but related scenario was analyzed by Coleman in his celebrated paper ``Fate of the false vacuum'' \cite{coleman1977}. Coleman considered situations where the vacua are \emph{not} degenerate: a metastable ``false vacuum'' with higher energy density can decay through barrier penetration to a ``true vacuum'' with lower energy. Coleman originally proposed calling these solutions ``bounces'' rather than instantons, to distinguish them from the degenerate-vacuum case. However, the modern literature has adopted a broader usage, referring to all such solutions describing transitions between stable states via barrier penetration, whether induced by quantum or thermal fluctuations, and whether connecting degenerate or non-degenerate minima, as ``instantons.'' In Coleman's semiclassical theory, the decay rate per unit volume scales as $\Gamma/V \sim A e^{-B/\hbar}$, where $B$ is determined by the instanton action.

Zinn-Justin's comprehensive treatment \cite{zinnjustin2002} develops the instanton calculus for a wide range of quantum mechanical and QFT models, including the double-well potential, $\phi^4$ theory, and Yang-Mills theories, providing a unified framework for both the Polyakov and Coleman scenarios.

In disordered systems, the instanton picture becomes more complex due to the rugged energy landscape. Lopatin and Ioffe \cite{lopatin1999} developed an instanton approach for Langevin dynamics in spin glasses, showing that activated transitions between metastable states can be understood through saddle-point configurations of the stochastic action. In their formulation, the ``instanton'' is the most probable escape path from a metastable minimum, and the transition rate is controlled by the action evaluated on this path. The complexity of the spin glass landscape, with its hierarchical organization of barriers and ultrametric state structure, leads to a broad distribution of instanton actions and hence the characteristic stretched-exponential relaxation and aging phenomena.

\paragraph{Structure Formation as an Instanton-Like Process.}
Our ``Big Bang'' initial condition places the system in a state with nearly zero total energy: all particles are localized near a single point, so both the kinetic energy (negligible in overdamped dynamics) and potential energy (proportional to geodesic distances, which are nearly zero) vanish. The subsequent evolution carries the system through an intermediate high-energy ``expansion'' phase before settling into another near-zero energy configuration; the organized structures (bands, rings, clusters) we observe.

This process resembles an instanton transition in several key ways:
\begin{itemize}
    \item \textbf{Degenerate initial and final energies}: Both the initial localized state and the final organized state have approximately zero total energy, analogous to tunneling between degenerate vacua in QFT.
    \item \textbf{Barrier crossing}: The intermediate expansion phase involves configurations with higher potential energy, representing the ``barrier'' that must be traversed.
    \item \textbf{Rapid transition}: Once initiated (by thermal fluctuations), the structure formation proceeds relatively quickly compared to the subsequent slow evolution of the formed structures.
    \item \textbf{Exponential sensitivity}: The specific final configuration (band orientation, ring positions, cluster locations) depends sensitively on the noise realization, analogous to how instanton transitions select among degenerate final states.
\end{itemize}

\paragraph{Differences from Traditional Instantons.}
Despite these similarities, our scenario differs from the traditional instanton picture in important ways that make it more closely related to instantons in spin glasses than to instantons in QFT.

In QFT models with ``true'' instantons, such as QCD, the self-interaction potential remains a fixed function of the field for all field values; the potential landscape is static. In our model, by contrast, the potential acting on each particle is created by interactions with \emph{other} particles, and hence the potential landscape itself changes as particles move. This dynamical nature of the energy landscape is a feature our model shares with spin glasses, where the effective potential seen by each spin depends on the configuration of all other spins.

This fundamental difference has two important consequences.
First, our instanton-like transitions are \emph{not instantaneous}. In QFT, the instanton describes the tunneling of a single degree of freedom (or a coherent field configuration) through a fixed barrier, which occurs ``instantly'' in real time. In our model and in spin glasses, the transition involves the coordinated escape of many particles across a barrier, and this collective rearrangement takes time. The $\beta$-relaxation phase, during which the system settles into an intermediate metastable configuration, and the subsequent transition to the final structure both occur over finite timescales.

Second, the barrier is \emph{dynamic} and configuration-dependent. The most important consequence of this is for the stability of the final configuration. We observe that after the system settles into the final structure (bands, rings, or clusters), it does not return to the intermediate metastable state; the transition is effectively irreversible. The reason is \emph{not} primarily the energetic suppression of the reverse transition as in Coleman's false vacuum decay (where the reverse rate is suppressed by $e^{-\Delta E/T}$). Rather, the barrier that existed during the forward transition has \emph{self-destructed}: once the particles have rearranged into the final configuration, the intermediate metastable minimum from which they escaped no longer exists in the new energy landscape. The system cannot return because there is nowhere to return to. This phenomenon, where the instanton transition destroys the initial minimum, is characteristic of spin glasses and stands in contrast to QFT instantons, where both vacua persist as features of the fixed potential landscape.

Furthermore, the multiplicity of near-degenerate final states in our model (all band orientations on the sphere, all ring configurations on the torus, etc.) is not a discrete set as in simple double-well instantons, but a continuous manifold, the Goldstone manifold associated with the spontaneously broken rotational symmetry. The instanton-like transition thus lands the system somewhere on this manifold, and subsequent dynamics explores it through thermal fluctuations rather than through further instanton transitions. Section~\ref{sec:orientation_summary} quantifies this exploration: the structure normal $\hat{\mathbf{n}}(t)$ accumulates roughly $2300^\circ$ of precession on the sphere (at rate $\sim 61^\circ$/time), $2400^\circ$ on the torus ($\sim 65^\circ$/time), and $1200^\circ$ on the cylinder ($\sim 32^\circ$/time) during the NESS phase. The correlation time $\tau_c \approx 18$ on the sphere and torus sets the timescale for orientation decorrelation, far exceeding the microscopic thermal correlation time $\tau_v < 0.2$.

\paragraph{Implications.}
This connection to instanton physics suggests that analytical techniques developed for instantons in field theory and disordered systems \cite{polyakov1975, coleman1977, zinnjustin2002, lopatin1999} may be applicable to understanding the structure formation dynamics in our model. In particular, the formalism of Lopatin and Ioffe \cite{lopatin1999} for Langevin dynamics in spin glasses provides a natural framework for computing the most probable paths from the initial localized state to the various final configurations, and for understanding how the distribution of quenched couplings $\phi_{ij}$ influences the selection among degenerate final states.

\subsection{Comparison with Parisi-Sourlas Dimension Reduction}
\label{sec:parisi_sourlas}

The dimension reduction we observe invites comparison with the celebrated Parisi-Sourlas mechanism \cite{parisi1979_dimred, parisi1981}, one of the most remarkable results connecting disorder and dimensionality in statistical physics.

\paragraph{The Parisi-Sourlas Mechanism.}
Parisi and Sourlas discovered that the random-field Ising model (RFIM) in $d$ dimensions has the same critical exponents as the \emph{pure} Ising model in $d-2$ dimensions. This arises from hidden supersymmetry: the disorder average generates a field theory with equal bosonic and fermionic degrees of freedom, and the resulting supersymmetric cancellations effectively reduce dimensionality. The key ingredients are random fields coupling linearly to the order parameter, supersymmetric structure, equilibrium critical behavior, and exact reduction by 2.

\paragraph{Contrasts with Our Phenomenon.}
Our disorder-induced dimension reduction differs fundamentally: (1) We have random \emph{couplings} $\phi_{ij}$ (spin-glass structure) rather than random \emph{fields} (RFIM structure); random couplings create frustration while random fields destroy long-range order. (2) Our reduction is \emph{dynamical} (non-equilibrium relaxation to NESS), not equilibrium critical behavior. (3) Our reduction is \emph{approximate} (bands/rings have finite width), not an exact mapping. (4) Our reduction is \emph{geometry-dependent} ($S^2 \to S^1$, cylinder $\to \mathbb{Z}_2$, $T^2 \to \coprod_2 S^1$), while Parisi-Sourlas always gives exactly two-dimensional reduction. (5) The zero-torque property allows orientation to diffuse freely, enabling exploration of the Goldstone manifold; this has no analogue in Parisi-Sourlas.

\paragraph{Limitations and Common Threads.}
Parisi-Sourlas breaks down below the RFIM lower critical dimension and does not apply to dynamics. Our model, inherently about dynamics and non-equilibrium steady states, suggests disorder-induced dimension reduction may be a broader phenomenon. Common themes include: disorder reducing effective dimensionality; cancellation mechanisms (supersymmetric vs. balanced random forces); and robustness to distribution details. Understanding the deeper connections remains an intriguing open question.

\subsection{Large-$N$ Limit, Supersymmetry, and QFT}
\label{sec:susy}

The thermodynamic limit $N \to \infty$ of our model yields a statistical field theory on a Riemannian manifold. Large-$N$ methods have proven extraordinarily powerful in both quantum field theory \cite{zinnjustin2002} and disordered systems, where exact solutions of mean-field spin glass models \cite{mezard1987, bouchaud1998} have shaped our understanding of glassy dynamics. With appropriate scaling ($J^2 \to J^2/N$), our system is described by a single-particle distribution $P(q, t)$ evolving according to
\begin{equation}
    \frac{\partial P}{\partial t} = \nabla_q \cdot \left[ P \nabla_q \int d\mu(q') \, \bar{\phi}(q, q') P(q', t) \right] + D \Delta_q P,
\end{equation}
where $\bar{\phi}$ is an effective mean-field interaction and $\Delta_q$ is the Laplace-Beltrami operator. This infinite-$N$ limit is equivalent to a two-dimensional Euclidean-time quantum field theory on a Riemannian surface, where the manifold metric enters through covariant derivatives, Laplacian, and geodesic distance function. Appendix~\ref{app:analytical} provides further details on large-$N$ methods, dynamical mean-field theory, and related analytical approaches for non-equilibrium glassy dynamics.

\paragraph{Supersymmetry in Stochastic Dynamics.}
A deep structural connection exists between classical stochastic Langevin dynamics and supersymmetric quantum mechanics. Feigel'man and Tsvelik \cite{feigelman1982} discovered that the MSRJD path integral for Langevin equations possesses a hidden supersymmetry when the Jacobian is represented using Grassmann variables. This ``stochastic supersymmetry'' relates bosonic fields (the physical coordinates $q$) to fermionic ghost fields ($\bar{\psi}, \psi$), with the supersymmetric action
\begin{equation}
    S_{\text{SUSY}} = \int dt \left[ \frac{1}{4D}(\dot{q} + \nabla V)^2 + \bar{\psi}(\partial_t + \nabla^2 V)\psi \right].
\end{equation}

This connects directly to Witten's supersymmetric quantum mechanics \cite{witten1982susy}. For gradient dynamics, the Fokker-Planck operator can be written as the Hamiltonian of an $\mathcal{N}=2$ supersymmetric system:
\begin{equation}
    H = \{Q, Q^\dagger\} = -D\Delta + \frac{1}{4D}|\nabla V|^2 - \frac{1}{2}\Delta V,
\end{equation}
where $Q = \bar{\psi}^i(\partial_i + \frac{1}{2D}\partial_i V)$ is the supercharge. Supersymmetric ground states ($H|\Omega\rangle = 0$) correspond to equilibrium distributions, while spontaneous SUSY breaking signals the absence of normalizable steady states.

Parisi and Sourlas \cite{parisi1981} showed that SUSY in random field systems leads to dimensional reduction, connecting $(d+2)$-dimensional disordered systems to $d$-dimensional pure systems. Kurchan \cite{kurchan1992} extended these ideas to spin glass dynamics, showing that the fluctuation-dissipation theorem corresponds to unbroken SUSY, while its violation in aging systems reflects SUSY breaking.

\paragraph{SUSY on Curved Manifolds.}
On a Riemannian manifold $\mathcal{M}$, the SUSY structure must be compatible with the geometry. The Witten Hamiltonian generalizes to
\begin{equation}
    H = -D\Delta_{\mathcal{M}} + \frac{1}{4D}g^{ij}\partial_i V \partial_j V - \frac{1}{2}\Delta_{\mathcal{M}} V + \text{curvature terms}.
\end{equation}
On compact manifolds like $S^2$ and $T^2$, the supersymmetric ground states correspond to harmonic forms, with their counting determined by the Betti numbers. Witten's proof of the Morse inequalities \cite{witten1982morse} relied on this connection, interpreting critical points as instantons mediating tunneling between supersymmetric vacua.

Therefore, in the strict thermodynamic limit $N \to \infty$, our model can be formulated as a \emph{supersymmetric disordered field theory on a Riemannian surface}. The topological properties of the manifold (e.g., the Euler characteristic) should manifest in the structure of supersymmetric ground states and the pattern of SUSY breaking, potentially explaining the geometry-dependent phenomena observed in our simulations.

\subsection{Dimension Reduction as Confinement: An Analogy with QCD}
\label{sec:confinement}

The instanton interpretation developed in Section~\ref{sec:instantons} naturally leads to yet another perspective on the dimension reduction in our model: as a form of \emph{confinement} analogous to quark confinement in quantum chromodynamics (QCD). This analogy connects two seemingly unrelated phenomena: the localization of particles to lower-dimensional submanifolds in our classical stochastic system, and the permanent binding of quarks inside hadrons in a quantum gauge theory.

\paragraph{Instantons and Confinement in QCD.}
In the early development of QCD, instantons were proposed as a mechanism for quark confinement \cite{polyakov1977}. The idea was attractive: instantons represent tunneling between topologically distinct vacuum sectors, and their proliferation could disorder the vacuum sufficiently to prevent color flux from propagating freely, thereby confining quarks. While subsequent research established that instantons are crucial for the vacuum structure of QCD, explaining phenomena such as the $U(1)_A$ anomaly and chiral symmetry breaking, they are not directly responsible for confinement, which instead arises from the condensation of magnetic monopoles or center vortices in the dual superconductor picture \cite{thooft1978, mandelstam1976, polyakov1977}.

\paragraph{Confinement in Our Model.}
In our setting, ``confinement'' has a geometric meaning: particles that initially occupy the full 2D manifold become confined to lower-dimensional structures; great-circle bands on the sphere, rings on the torus, or localized clusters on the cylinder. This dimensional reduction is precisely what the instanton-like transitions described in Section~\ref{sec:instantons} accomplish: the system transitions from the initial near-zero energy state (localized ``Big Bang'' configuration) to a final near-zero energy state (organized low-dimensional structure) via a rapid restructuring process.

In this interpretation, our model realizes the original vision of instanton-driven confinement, albeit in a different physical context:
\begin{itemize}
    \item \textbf{Instantons mediate confinement}: The rapid structure formation process, analogous to the instantons in the Lopatin-Ioffe formalism for spin glass dynamics \cite{lopatin1999}, is directly responsible for confining particles to lower-dimensional submanifolds.
    \item \textbf{Degenerate vacuum structure}: Just as QCD instantons connect topologically distinct but energetically degenerate vacua, our instanton-like transitions connect the initial localized state to any one of the continuously degenerate final configurations (different band orientations, ring positions, etc.).
    \item \textbf{Non-perturbative phenomenon}: The confinement cannot be understood from small fluctuations around the initial state; it requires the full non-perturbative transition through the energy barrier.
\end{itemize}

\paragraph{Growing Forces and Strong Interactions.} A suggestive parallel with QCD concerns the distance dependence of forces. In QCD, the interquark potential grows linearly with separation at large distances (the ``string tension''), leading to confinement: pulling quarks apart requires infinite energy. In the opposite limit of very short inter-quark distances, QCD gives rise to \textit{asymptotic freedom}: quarks become non-interacting (free) particles at ultra-short distances. Our model features an analogous structure: the pairwise forces $\mathbf{F}_{ij} = \phi_{ij} \hat{\mathbf{t}}_{ij}$ have constant magnitude $|\phi_{ij}|$ regardless of geodesic distance. This means that the potential $U_{ij} = \phi_{ij} \cdot d_g(q_i, q_j)$ grows \textit{linearly} (if $\phi_{ij} > 0$) with geodesic separation, precisely the distance dependence associated with confinement in QCD, while vanishing at zero separation, mimicking asymptotic freedom.

However, the analogy has important subtleties. In our model, the coupling constants $\phi_{ij}$ have random signs:
\begin{itemize}
    \item For pairs with $\phi_{ij} > 0$ (attractive), the linear potential mimics both asymptotic freedom and confinement: the interaction vanishes at zero separation and grows linearly with distance, analogous to the QCD string.
    \item For pairs with $\phi_{ij} < 0$ (repulsive), asymptotic freedom is preserved (zero interaction at zero separation), but confinement is replaced by repulsion (``anti-confinement'' in QCD language): the potential \emph{decreases} (becomes more negative) with distance, driving particles apart rather than confining them.
\end{itemize}
The repulsive behavior of pairs with $\phi_{ij} < 0$ has no counterpart in QCD. However, it does not produce pathological dynamics in our setting due to the compactness of our Riemannian manifolds: particles on a sphere or torus cannot escape to infinity, so even maximally repulsive pairs remain at finite geodesic separation. The coexistence of confining (attractive) and repulsive interactions, randomly distributed among particle pairs, creates the frustration that drives the complex dynamics and ultimately the dimensional reduction. This mixture connects to the broader physics of frustrated systems and spin glasses.

\paragraph{The Role of Embedding Geometry.}
A final distinction concerns the relationship between the confining dynamics and the embedding space. In QCD, confinement is an intrinsic property of the gauge theory; it occurs in flat Minkowski spacetime and does not rely on any special geometry of the embedding space. In our model, the situation is more subtle: the particles live on 2D manifolds (sphere, cylinder, torus) embedded in $\mathbb{R}^3$, and the ``confinement'' to lower-dimensional structures occurs \emph{within} these manifolds, not in the embedding space.

The embedding geometry nevertheless plays a role through the definition of angular momentum and the projection of Brownian noise. The interplay between the intrinsic manifold geometry (which determines geodesics and hence forces) and the extrinsic embedding geometry (which defines angular momentum in $\mathbb{R}^3$) is essential for the specific structures that emerge. This geometric richness, absent in standard QCD, makes our model a laboratory for exploring how topology and curvature influence confinement-like phenomena.

\subsection{Connections to Astrophysical Structure Formation}
\label{sec:cosmology}

The dimensional reduction we observe (2D manifolds to 1D structures) has a superficial geometric similarity to phenomena in astrophysics, where 3D distributions collapse to 2D disks and 1D filaments \cite{zeldovich1970}. However, the underlying mechanisms are fundamentally different.

In astrophysical systems, the classical angular momentum $\mathbf{L}_{cl} = \sum_i m \mathbf{r}_i \times \mathbf{v}_i$ is approximately conserved in isolated systems. This conservation \emph{causes} dimensional reduction: a rotating gas cloud cannot collapse along the axis defined by $\mathbf{L}_{cl}$ without violating conservation, so it flattens into a disk perpendicular to this fixed axis \cite{springel2005, cuzzi2010}. Planetary rings form similarly: inelastic collisions dissipate energy while approximately conserving angular momentum, driving collapse from 3D to 2D.

Our model operates through an entirely different mechanism. Dimensional reduction arises from \emph{frustration} in the random interaction network and energy minimization on the curved manifold, not from angular momentum dynamics. The band on the sphere, rings on the torus, and clusters on the cylinder emerge because these configurations represent low-energy states in the disordered energy landscape, which persist indefinitely as non-equilibrium steady states (NESS). The cumulative rotation $\Theta(t)$ defined in Eq.~(\ref{eq:cumulative_rotation}) is a \emph{diagnostic} for how the orientation evolves after structure formation, not a causal agent. Indeed, our $\Theta(t)$ has dimensions [length]$^2$ (or dimensionless after normalization) and is zero in the noise-free deterministic limit.

The zero-torque result (\ref{eq:zero_torque}) establishes that pairwise forces contribute no deterministic drift to $\Theta(t)$, which is \emph{analogous} to approximate angular momentum conservation in the sense that orientation changes arise only from external perturbations (thermal noise in our case, tidal torques or mergers in astrophysics). This explains why the band orientation diffuses rather than drifts systematically. But the analogy should not be pushed further: astrophysical angular momentum is a nearly-conserved quantity with small fluctuations around a well-defined mean, while our cumulative rotation $\Theta(t)$ fluctuates without bound ($\sigma_\Theta \gg |\langle \Theta \rangle|$) and has no meaningful time-average.

\begin{table}[h!]
\centering
\footnotesize
\caption{Geometric versus mechanistic comparison with astrophysical systems. Despite similar structural outcomes, the underlying physics differs fundamentally. Here $\Theta$ denotes our cumulative rotation and $L$ denotes classical angular momentum.}
\label{tab:L_comparison_astro}
\begin{tabular}{@{}p{2.2cm}p{3.0cm}p{3.0cm}@{}}
\toprule
\textbf{Property} & \textbf{Our Model} & \textbf{Astrophysical Systems} \\ \midrule
Mechanism & Frustration, energy minimization & Angular momentum conservation \\[0.3em]
Role of $\Theta$ / $L$ & Diagnostic (orientation drift) & Causal (selects disk plane) \\[0.3em]
Fluctuations & Large: $\sigma_\Theta \gg |\langle \Theta \rangle|$ & Small: $\delta L/L \ll 1$ \\[0.3em]
Orientation & Random walk (diffusive) & Nearly fixed \\[0.3em]
Structure stability & NESS (persistent) & Stable (conserved $L$) \\ \bottomrule
\end{tabular}
\end{table}

In summary, the connection to astrophysics is primarily \emph{geometric} (both produce flattened, ring-like structures) rather than \emph{mechanistic}. Our model provides an alternative route to dimensional reduction that does not rely on angular momentum conservation, instead exploiting the interplay of disorder, dissipation, and manifold geometry.

\subsection{Comparison with Condensed Matter Systems}
\label{sec:condensed_matter}

Beyond spin glasses, our model connects to condensed matter systems exhibiting broken rotational symmetry and slow collective dynamics.

The isotropic-to-nematic transition in \emph{liquid crystals} breaks SO(3) $\to$ SO(2), with the director $\mathbf{n}$ selecting a preferred axis, directly analogous to our band normal on the sphere. Both systems exhibit relaxational dynamics toward free energy minima, and both have soft modes with quadratic dispersion characteristic of dissipative systems. Our structure orientation analysis (Section~\ref{sec:orientation_summary}) makes this analogy precise: the band normal $\hat{\mathbf{n}}(t)$ plays the role of the nematic director, exploring the Goldstone manifold $S^2 \cong \mathrm{SO}(3)/\mathrm{SO}(2)$ with correlation time $\tau_c \approx 18$ and accumulating roughly $2300^\circ$ of precession over the NESS phase. The key difference is that nematic ordering is an equilibrium phase transition with a frozen director below $T_c$, while our band normal never freezes but continues to diffuse indefinitely, driven by thermal noise in the NESS.

\emph{Ferromagnets} exhibit spontaneous magnetization that breaks SO(3) $\to$ SO(2), yielding one Type-II magnon with $\omega \propto k^2$. The reduction from two broken generators to one Goldstone mode occurs because the nonzero magnetization itself breaks symmetry between transverse directions: fluctuations of $\mathbf{M}$ perpendicular to $\langle \mathbf{M} \rangle$ are coupled through the commutation relations $[M_x, M_y] \propto M_z$. Our system exhibits a related structure. Once a band forms with normal $\hat{\mathbf{n}}$, rotations about different axes have inequivalent effects on the configuration. The measured isotropic orientation statistics ($\sigma(n_x) \approx \sigma(n_y) \approx \sigma(n_z) \approx 0.1$--$0.3$ on the sphere) suggest that our non-equilibrium steady state explores all directions equivalently over long times, in contrast to the frozen magnetization direction in a ferromagnet. The power spectrum $S_{n_z}(\omega) \propto \omega^{-1.6}$ on the sphere is shallower than the $1/\omega^2$ expected for type-A diffusion modes, possibly reflecting the bounded nature of the orientation variable $\hat{\mathbf{n}}$ on $S^2$.

Experimental systems of \emph{colloidal particles on curved interfaces} \cite{bausch2003} provide a direct physical realization of particles on 2D manifolds. These systems exhibit defect-mediated interactions from Gaussian curvature, frustration between local packing and global topology, and slow relaxation dynamics. Our model differs in having random rather than uniform interactions, but shares the geometric constraints and dissipative dynamics. The dimensional reduction we observe, quantified by the structure thickness $\mathcal{T}$ decreasing from $\sim 0.6$ to $\sim 0.3$ as the band forms (Table~\ref{tab:orientation_stats}), might have experimental signatures in colloidal systems with appropriately engineered interactions.

\emph{Active matter on curved surfaces} \cite{fily2012} exhibits collective phenomena including motility-induced phase separation and polar ordering. While active matter is intrinsically out of equilibrium like our model, the driving mechanism differs: active particles have internal energy sources, while our system is driven purely by thermal noise acting on passive particles. Both exhibit spontaneous symmetry breaking on curved spaces, and the interplay between activity/noise, geometry, and collective ordering represents an area for future comparative study.

\subsection{Relation to Self-Organizing Systems}

Our model exhibits spontaneous structure formation from initially disordered configurations, inviting comparison with self-organizing systems studied across physics, chemistry, and biology. The field encompasses several distinct paradigms with different mechanisms and requirements.

\paragraph{Dissipative structures.} The program initiated by Prigogine \cite{prigogine1977} and developed by Haken \cite{haken1983} and others \cite{cross1993} focuses on pattern formation in systems maintained far from equilibrium by continuous energy or matter flux. Classical examples (B\'enard convection rolls, Turing patterns, the Belousov-Zhabotinsky oscillator) require: (i) sustained thermodynamic gradients; (ii) nonlinear feedback that amplifies fluctuations; and (iii) control parameters tuned above critical thresholds. These ``dissipative structures'' are intrinsically non-equilibrium phenomena that vanish when the external driving ceases.

\paragraph{Self-organized criticality.} Bak and collaborators \cite{bak1987, bak1996} introduced a different paradigm: systems that spontaneously evolve toward a critical state without fine-tuning of parameters. The canonical example is a sandpile that, through slow addition of grains, organizes itself to a critical slope where avalanches of all sizes occur with power-law statistics. SOC systems share with our model the absence of parameter fine-tuning, but differ in requiring slow external input (grain addition) and producing scale-free, intermittent dynamics (avalanches) rather than persistent geometric structures.

\paragraph{Our model: disorder-induced organization.} The structures emerging in our model (equatorial bands, boundary clusters, poloidal rings) arise through a mechanism distinct from both paradigms above. Our model requires: \emph{no sustained thermodynamic gradients} (only thermal equilibrium with a heat bath); \emph{no slow external input} (the quenched couplings are fixed, not continuously supplied); and \emph{no parameter fine-tuning} (structure formation occurs for any non-zero coupling strength and temperature). The organizing agents are: (i) \textit{quenched disorder} in the random couplings $\phi_{ij}$, which creates frustration and a complex energy landscape; (ii) \textit{thermal noise}, which enables exploration of this landscape and drives instanton-like transitions; and (iii) \textit{manifold topology and geometry}, which constrain which low-dimensional configurations are kinetically accessible.

This combination suggests that \textit{disorder-induced self-organization} constitutes a distinct paradigm. The principle ``order from noise'' articulated by von Foerster \cite{vonfoerster1960} acquires a concrete realization: quenched disorder, thermal fluctuations, and geometric topology interact to produce ordered structures autonomously. Unlike dissipative structures, our patterns persist in thermal equilibrium (the NESS is maintained by the heat bath alone). Unlike SOC, our system produces specific geometric configurations determined by topology rather than scale-free avalanche statistics. The emergent order is geometrically simple (great-circle bands, minor-circle rings) but arises from minimal ingredients without external orchestration.

\subsection{Synthesis: Unifying Principles}

The preceding sections have drawn connections between our geometric model and diverse areas of physics: spin glass theory, instanton physics in QFT, quark confinement in QCD, astrophysical structure formation, and condensed matter systems with broken rotational symmetry. These connections point to universal mechanisms that transcend specific physical realizations:

\begin{enumerate}
    \item \textbf{Random interactions induce frustration}: Competing tendencies from quenched disorder prevent simple ground states and create complex energy landscapes with multiple metastable configurations.
    
    \item \textbf{Dissipation enables structure formation}: Energy dissipation through overdamped dynamics funnels evolution toward low-energy configurations, enabling persistent structures to form. This dissipation-driven route to ergodicity breaking follows the paradigm established in spin glass physics.
    
    \item \textbf{Zero-torque property enables orientation diffusion}: The vanishing of deterministic torque from pairwise interactions means that the orientation of emergent structures evolves as a pure random walk. The structure normal $\hat{\mathbf{n}}(t)$ diffuses on its Goldstone manifold with geometry-dependent correlation times: $\tau_c \approx 18$ on the sphere and torus (continuous structures), but $\tau_c \approx 3$ on the cylinder (discrete clusters). This order-of-magnitude difference reflects the greater susceptibility of discrete configurations to thermal reorientation.
    
    \item \textbf{Dimensional reduction is generic}: Across spin glasses, cosmology, and our model, final configurations have lower effective dimensionality than the starting point, suggesting that frustrated dissipative systems tend to concentrate on submanifolds. The structure thickness $\mathcal{T}$ quantifies this reduction: on the sphere, $\mathcal{T}$ decreases from $0.62$ (diffuse) to $0.29$ (well-defined band) during formation.
    
    \item \textbf{Spectral signatures are geometry-dependent}: Both MSD exponents and power spectrum exponents vary across geometries. The structure normal power spectrum $S_{n_z}(\omega) \propto \omega^{-\beta}$ shows $\beta \approx 1.6$ on the sphere, $\beta \approx 2.0$ on the cylinder, and $\beta \approx 1.4$ on the torus. Only the cylinder approaches the $1/\omega^2$ scaling predicted for type-A diffusion modes. The MSD exponents similarly vary (diffusive on sphere, ballistic on cylinder, superdiffusive on torus). This geometry dependence indicates that both local stochastic forcing and global constraints from boundaries, metric structure, and the bounded nature of $\hat{\mathbf{n}}$ on $S^2$ shape collective dynamics.
    
    \item \textbf{Geometry controls orientation statistics}: The structure normal exhibits geometry-dependent behavior in the NESS. On the sphere, $\hat{\mathbf{n}}$ explores all directions isotropically ($\langle n_z \rangle \approx 0$, $\sigma(n_z) \approx 0.11$). On the cylinder, $\hat{\mathbf{n}}$ is confined to the $xy$-plane ($\langle n_z \rangle \approx 0$, $\sigma(n_z) \approx 0.07$). On the torus, $\hat{\mathbf{n}}$ shows a strong $z$-bias with large fluctuations ($\langle n_z \rangle \approx 0.73$, $\sigma(n_z) \approx 0.44$). These statistics encode how manifold geometry constrains the accessible configurations of the symmetry-breaking order parameter.
    
    \item \textbf{Slow dynamics signal landscape complexity}: The persistence of aging, metastability, and slow collective modes indicates navigation through rugged energy landscapes, a universal feature connecting spin glasses, structural glasses, and our geometric model. The long correlation times $\tau_c \approx 18$ on the sphere and torus (compared to thermal velocity correlation times $\tau_v < 0.2$) confirm the adiabatic separation between fast microscopic dynamics and slow collective evolution.
\end{enumerate}

These principles point toward a unified framework for structure formation in dissipative systems with competing interactions. While different systems may share structural outcomes (dimensional reduction, symmetry breaking), the dynamics of collective variables can differ qualitatively, reflecting fundamentally different couplings to the environment.

\section{Conclusion}
\label{sec:conclusion}

\paragraph{Order Out of Noise and Disorder.}

The title of this paper, echoing Prigogine and Stengers' celebrated \emph{Order out of Chaos} \cite{prigogine1984}, encapsulates our central finding: \emph{order emerges from the combination of noise and disorder}, with the manifold's topology determining the fate of the frustrated system. As discussed in Section~\ref{sec:discussion}, our model represents a distinct paradigm of self-organization. Unlike dissipative structures, it requires no sustained thermodynamic gradients. Unlike self-organized criticality, it requires no slow external input and produces specific geometric configurations rather than scale-free avalanches. The only ``driving'' is thermal contact with a heat bath, and structure formation occurs without parameter fine-tuning. From three minimal ingredients (Brownian noise, quenched random couplings, and manifold geometry), ordered structures emerge spontaneously: bands on the sphere, rings on the torus, clusters on the cylinder. Geometry and topology convert randomness into order.

\paragraph{Theoretical Framework.}

Our model combines the essential physics of frustrated disordered systems with the geometry of compact Riemannian manifolds. Crucially, the model retains theoretical tractability: the large-$N$ limit (Section~\ref{sec:susy}, Appendix~\ref{app:analytical}) converts the particle system into a statistical field theory on a Riemannian surface, amenable to the MSRJD formalism, replica methods, and supersymmetric techniques. This opens connections to QFT on curved spacetimes, providing a classical dissipative analogue where manifold geometry controls emergent order. Systematic analytical exploration remains for future work.

\paragraph{Principal Findings.}

Our numerical simulations establish a universal phenomenon across three geometries: random interactions generically drive dimensional reduction from 2D manifolds to quasi-1D structures, with the specific character determined by topology. On \textbf{closed manifolds} (sphere $S^2$, torus $T^2$), particles form extended structures wrapping closed geodesics, specifically a great-circle band on the sphere, two minor-circle rings on the torus. On the \textbf{bounded cylinder}, particles localize into discrete clusters near boundaries. The symmetry breaking pattern reflects this topological distinction:
\begin{itemize}
    \item Sphere: $\text{SO}(3)\to\text{SO}(2)$ (continuous)
    \item Torus: $\text{SO}(2)\times\text{SO}(2)\to\text{SO}(2)\times\mathbb{Z}_2$ (mixed)
    \item Cylinder: $\text{SO}(2)\to\mathbb{Z}_2$ (discrete)
\end{itemize}

Unlike conventional spontaneous symmetry breaking, the symmetry-breaking direction slowly drifts via thermal noise while the reduced-dimensional structure persists. The slowly evolving collective variables (band orientation, ring positions, cluster locations) constitute classical realizations of dissipative Nambu-Goldstone modes, specifically the type-A diffusion modes predicted for symmetry breaking in open systems \cite{minami2018, hidaka2020}. This identification is supported by diffusive mean-square angular displacement ($\alpha \approx 0.9$) on the sphere. However, the power spectrum exponents are geometry-dependent: $S_{n_z}(\omega) \propto \omega^{-\beta}$ with $\beta \approx 1.6$ on the sphere, $\beta \approx 2.0$ on the cylinder, and $\beta \approx 1.4$ on the torus. Only the cylinder approaches the theoretical $1/\omega^2$ prediction for type-A diffusion modes. The bounded cylinder's ballistic MSD ($\alpha \approx 2.0$) and the torus's superdiffusive behavior ($\alpha \approx 1.7$) confirm that diffusive dynamics requires a continuous Goldstone manifold.

The relaxation dynamics exhibits hallmarks of glassy systems: two-stage relaxation with dramatic timescale separation and aging. Importantly, the final configurations are not metastable states but rather \emph{non-equilibrium steady states} (NESS): after the instanton-like transition completes, the energy barriers self-destruct, leaving a Mexican-hat-like potential with a continuum of degenerate minima that the system explores diffusively. The rapid structure formation from localized initial conditions can be understood as instanton-like transitions between near-zero energy configurations \cite{coleman1985, zinnjustin2002}, while the dimensional confinement to lower-dimensional submanifolds via a linear potential provides an analogy to quark confinement in QCD.

\paragraph{Analogy with Ferromagnetic Spin Alignment.}

The emergence of correlated rotational dynamics in our model bears a striking structural resemblance to spin alignment in three-dimensional ferromagnets. In a ferromagnet, individual atomic spins $\mathbf{s}_i$ interact through exchange coupling and spontaneously align below the Curie temperature, producing a macroscopic magnetization $\mathbf{M} = \sum_i \mathbf{s}_i$ that breaks SO(3) symmetry down to SO(2). The magnetization direction $\hat{\mathbf{M}}$ becomes the slow Goldstone mode, exploring the coset space $S^2 \cong \text{SO}(3)/\text{SO}(2)$. In our system, the structure normal $\hat{\mathbf{n}}(t)$ plays the precisely analogous role: before structure formation, particles are diffusely distributed (analogous to paramagnetic disorder above $T_c$); after structure formation, particles concentrate into a coherent band whose orientation $\hat{\mathbf{n}}$ slowly precesses on the same Goldstone manifold $S^2$. The symmetry breaking SO(3)$\to$SO(2) on the sphere directly parallels the ferromagnetic case.

The key difference lies in the nature of the ordered state. In equilibrium ferromagnets, the magnetization direction freezes and Goldstone modes are propagating spin waves with dispersion $\omega \propto k^2$ (Type-II). In our dissipative system, the structure normal never freezes but performs a random walk on the Goldstone manifold with correlation time $\tau_c \approx 18$---the hallmark of Type-A diffusive Goldstone modes in open systems \cite{minami2018, hidaka2020}. Both involve macroscopic order emerging from microscopic degrees of freedom through symmetry-breaking transitions, but the fluctuation character (propagating vs.\ diffusive) reflects whether the system is isolated or coupled to a thermal bath.

\paragraph{Outlook.}

Several directions merit future investigation. Systematic exploration of the phase diagram in $(T, J, N)$ parameter space could reveal non-equilibrium phase transitions and critical phenomena, including possible topological transitions where the number of emergent structures changes discontinuously. Quantitative characterization of aging through two-time correlation and response functions would establish the precise connection to spin glass phenomenology. Extension to higher-dimensional manifolds (e.g., $S^3$) could exhibit cascaded dimensional reduction, while manifolds with negative curvature may display qualitatively different clustering. Incorporating self-propulsion would connect to active matter on curved surfaces; adding external fields would enable study of driven glassy response.

The systematic positive angular velocity drift observed on the torus (Section~\ref{sec:torus}) exemplifies geometry-induced directed motion in stochastic systems. The stochastic ratchet mechanism we identified---where asymmetric ring positions and position-dependent noise amplitudes rectify thermal fluctuations into directed rotation---may have broader implications for molecular motors on curved membranes, active matter on non-uniformly curved surfaces, and other systems where stochastic dynamics unfold on Riemannian manifolds.

\paragraph{Universality and Broader Implications.}

The unifying principles identified in Section~\ref{sec:discussion} (frustration from disorder, dissipation enabling structure formation, zero-torque orientation diffusion, generic dimensional reduction, geometry controlling fluctuations) suggest that noise-plus-disorder-driven order may be far more common than previously recognized. Many physical, biological, and social systems involve heterogeneous conflicting interactions, stochastic dynamics, and geometric constraints on configuration space. Our model can serve as a prototype for such systems: molecules on curved membranes, order parameters on group manifolds, and collective coordinates in molecular dynamics. The explicit topology dependence (closed manifolds producing diffusive Goldstone modes, bounded manifolds producing discrete clusters) offers a clear prediction: the \emph{character} of emergent order should depend on the topology of the constraint space, not just its local geometry.

Beyond its theoretical interest, our model can serve as a \emph{benchmark system} for investigating collective behavior of particles on curved surfaces. Colloidal particles confined to curved liquid interfaces \cite{bausch2003}, molecules adsorbed on nanotubes or vesicles, and active matter on curved substrates all realize the essential ingredients of our model: Brownian motion from thermal contact with a solvent, pairwise interactions (which can be tuned experimentally), and geometric confinement to 2D manifolds. The phenomena we observe---frustration-induced dimensional reduction, slow collective modes, geometry-dependent anomalous diffusion, and the stochastic ratchet effect on the torus---should be directly accessible to experiment. The model can be extended in straightforward ways: incorporating hydrodynamic interactions, position-dependent coupling strengths, polydisperse particle sizes, or active self-propulsion. Such extensions would bridge from our idealized setting toward realistic soft matter and biophysical systems while retaining the essential geometric and topological constraints that generate the rich phenomenology we have documented.

\paragraph{A Toy Model for Non-Perturbative Phenomena in QFT and Condensed Matter Physics.}

The dynamics driving structure formation is fundamentally \emph{non-perturbative}: the dimensional reduction and symmetry breaking cannot be captured by small fluctuations around the initial configuration, but require the full nonlinear restructuring of the particle distribution. This non-perturbative character makes our model a potentially valuable toy model for studying topological and solitonic solutions in quantum field theory. While we have qualitatively identified and described instanton-like transitions in our model (Section~\ref{sec:instantons}), constructing explicit instanton solutions analytically using an approach similar to the Lopatin-Ioffe formalism \cite{lopatin1999} remains a task for future work.

The sphere $S^2$ provides the natural arena for magnetic monopole configurations: the 't~Hooft-Polyakov monopole \cite{thooft1974, polyakov1974} arises in non-Abelian gauge theories with adjoint Higgs fields, where the Higgs field at spatial infinity maps $S^2 \to S^2$ with nontrivial winding number. Our band formation on $S^2$, which selects a preferred great circle and breaks SO(3) $\to$ SO(2), creates precisely the symmetry-breaking pattern that characterizes monopole solutions. Extensions incorporating internal ``color'' or ``isospin'' degrees of freedom could provide a classical stochastic realization of monopole-like configurations.

Similarly, dyons (particles carrying both electric and magnetic charges \cite{julia1975, witten1979dyon}) arise in theories where the unbroken U(1) gauge group supports both types of charge. The torus geometry $T^2 = S^1 \times S^1$, with its product of two circle factors, provides a natural setting for exploring configurations with multiple conserved winding numbers. The weak anti-correlation we observe between the two rings on the torus ($\rho = -0.27$) hints at a constraint structure reminiscent of the Dirac-Schwinger-Zwanziger quantization condition relating electric and magnetic charges.

A particularly intriguing extension would introduce internal ``color'' degrees of freedom to particles, with couplings that depend on both position and color. This would create a geometric analogue of quarks in QCD, where confinement forces colored particles to form color-neutral bound states. Color-dependent couplings could produce novel clustering patterns where particles of different colors preferentially associate, potentially realizing geometric versions of mesons or baryons. The interplay between spatial confinement (to lower-dimensional submanifolds) and color confinement (to color-neutral clusters) could yield rich phase diagrams.

Finally, we are tempted to mention an apparent conceptual parallel with the AdS/CFT correspondence \cite{maldacena1999}, which relates gravity in anti-de Sitter spacetime to conformal field theory on its boundary. This duality exemplifies how geometry in one description can encode dynamics in another, and how dimensional reduction (from bulk to boundary) emerges from the mathematical structure. While our classical model is far removed from string theory, the general principle---that geometry and topology fundamentally constrain and organize physical phenomena---resonates with our findings. The boundary-dependence of our results (closed manifolds versus manifolds with boundary producing qualitatively different structures) hints that boundary conditions play a similarly fundamental role in our setting as they do in holographic dualities.

\paragraph{Relevance for Soft Matter and Biophysics.}

Beyond its connections to fundamental theory, our model of frustrated Brownian particles should be of direct interest to experimentalists working in soft matter, colloidal science, and biophysics. The model's ingredients---Brownian particles with tunable pairwise interactions confined to curved surfaces---are directly realizable in laboratory settings using existing techniques. Colloidal particles can be confined to curved liquid-liquid interfaces or the surfaces of emulsion droplets; functionalized nanoparticles can be adsorbed onto lipid vesicles or polymer capsules; and proteins naturally diffuse on the curved membranes of cells and organelles. In all these systems, pairwise interactions can be tuned through surface chemistry, depletion forces, or electrostatic screening, providing experimental control over the coupling strengths that drive structure formation in our model.

The phenomena we have documented---frustration-induced dimensional reduction, slow collective relaxation modes, geometry-dependent anomalous diffusion, and the stochastic ratchet effect on non-uniformly curved surfaces---should be directly accessible to experimental investigation. Modern particle tracking techniques can measure the MSD exponents, power spectra, and correlation functions that we have computed, enabling quantitative comparison between theory and experiment. The topology dependence we observe (closed manifolds producing extended structures with diffusive dynamics versus bounded manifolds producing localized clusters with ballistic dynamics) provides a clear experimental prediction that can be tested by comparing particle behavior on spherical droplets versus cylindrical capillaries. We hope that our work will stimulate experimental investigations of collective dynamics on curved surfaces, opening a dialogue between the theoretical framework developed here and the rich phenomenology of real soft matter and biological systems.

\paragraph{The Fate of the Frustrated Manifold.}

Returning to the title: what is the ``fate'' of a manifold populated by our \emph{frustrated Brownian particles}? The answer is \emph{spontaneous dimensional reduction and symmetry breaking, with the specific outcome determined by topology}. The sphere's fate is a great-circle band; the torus's fate is a pair of minor-circle rings; the cylinder's fate is localized clusters near the boundaries. This dimensional reduction appears to be a generic consequence of frustration on compact spaces: the manifold cannot accommodate all conflicting demands of the random couplings, so it ``projects'' the system onto a lower-dimensional subspace where conflicts are minimized.

Our work offers a new perspective on the ancient question of how order arises from chaos. The answer, in our setting, is that order does not oppose chaos: it emerges from it, channeled by the silent organizing principle of geometry. Noise provides the exploration; disorder provides the drive; topology provides the template. Together, they produce structures of regularity from ingredients of pure randomness.

The simplicity and tractability of the model, combined with its rich connections to fundamental physics (from spin glasses to QFT to astrophysics), make it a promising starting point for both analytical developments and applications to complex systems constrained to curved configuration spaces. We hope that this work will stimulate further investigation into the deep connections between disorder, noise, topology, and emergent order.

\appendix

\section{Geometry-Specific Derivations}
\label{app:geometry}

This appendix provides complete derivations of the Langevin dynamics for each of the three manifolds studied in this work. All derivations follow the Zinn-Justin projection method described in Section~\ref{sec:langevin_manifolds}: we start with the Langevin equation in the embedding Euclidean space $\mathbb{R}^3$ and project onto the manifold to obtain a covariant formulation consistent with the general result (\ref{eq:langevin_intrinsic}).

\subsection{Dynamics on the Sphere $S^2$}

The sphere is defined by $|\mathbf{x}|^2 = 1$ in the embedding space $\mathbb{R}^3$. Each particle position $\mathbf{x}_i \in S^2$ is a unit vector. The tangent space at $\mathbf{x}_i$ consists of vectors perpendicular to $\mathbf{x}_i$:
\begin{equation}
    T_{\mathbf{x}_i} S^2 = \{ \mathbf{v} \in \mathbb{R}^3 : \mathbf{v} \cdot \mathbf{x}_i = 0 \}
\end{equation}
with projection operator $P_{\alpha\beta}(\mathbf{x}_i) = \delta_{\alpha\beta} - (x_i)_\alpha (x_i)_\beta$. The geodesic distance is the arc length along the great circle: $d(\mathbf{x}_i, \mathbf{x}_j) = \arccos(\mathbf{x}_i \cdot \mathbf{x}_j)$.

In spherical coordinates $(\theta, \phi)$ with $\mathbf{x} = (\sin\theta\cos\phi, \sin\theta\sin\phi, \cos\theta)$, the induced metric is $g_{ij} = \mathrm{diag}(1, \sin^2\theta)$. For computational purposes, we use the embedding $\mathbb{R}^3$ representation which avoids coordinate singularities at the poles.

The unit tangent vector at $\mathbf{x}_i$ pointing toward $\mathbf{x}_j$ along the geodesic is $\hat{\mathbf{t}}_{ij} = \mathrm{Proj}_{T_{\mathbf{x}_i}}(\mathbf{x}_j)/|\mathrm{Proj}_{T_{\mathbf{x}_i}}(\mathbf{x}_j)|$, giving the force $\mathbf{F}_{ij} = \phi_{ij} \hat{\mathbf{t}}_{ij}$ with total force $\mathbf{F}_i = \sum_{j \neq i} \mathbf{F}_{ij} \in T_{\mathbf{x}_i} S^2$.

Applying the projection to the embedding-space Langevin equation yields:
\begin{equation}
    d\mathbf{x}_i = \mathbf{F}_i \, dt + \sqrt{2D} \, d\mathbf{W}_i^{\perp}
    \label{eq:langevin_sphere}
\end{equation}
where $d\mathbf{W}_i^{\perp} = (\mathbf{I} - \mathbf{x}_i \mathbf{x}_i^T) d\mathbf{W}_i$ is the projection of 3D Brownian motion onto the tangent space, with covariance $\langle (dW_i^\perp)_\alpha (dW_i^\perp)_\beta \rangle = P_{\alpha\beta}(\mathbf{x}_i) \, dt$.

\subsection{Dynamics on the Bounded Cylinder $S^1 \times [0,H]$}

The bounded cylinder of radius $R$ and height $H$ is embedded in $\mathbb{R}^3$ via $\mathbf{r}(\theta, z) = (R\cos\theta, R\sin\theta, z)$ with $\theta \in [0, 2\pi)$ and $z \in [0, H]$. The induced metric is $g_{ij} = \mathrm{diag}(R^2, 1)$ with inverse $g^{ij} = \mathrm{diag}(1/R^2, 1)$. The metric is constant (position-independent), indicating that the cylinder is intrinsically flat.

For points $(\theta_i, z_i)$ and $(\theta_j, z_j)$, the geodesic distance with reflective boundaries is:
\begin{equation}
    d_{\text{cyl}}(i,j) = \sqrt{(R \Delta\theta_{ij})^2 + (\Delta z_{ij})^2}
\end{equation}
where $\Delta\theta_{ij} = \theta_j - \theta_i$ (taken modulo $2\pi$ to lie in $(-\pi, \pi]$) and $\Delta z_{ij} = z_j - z_i$.

The force components are:
\begin{equation}
    F_i^{\theta} = \sum_{j \neq i} \phi_{ij} \frac{\Delta\theta_{ij}}{d_{\text{cyl}}(i,j)}, \quad
    F_i^{z} = \sum_{j \neq i} \phi_{ij} \frac{\Delta z_{ij}}{d_{\text{cyl}}(i,j)}
    \label{eq:force_cylinder}
\end{equation}

The covariant Langevin equation in coordinates $(\theta, z)$ becomes:
\begin{equation}
    \begin{aligned}
        d\theta_i &= F_i^{\theta} \, dt + \sqrt{\frac{2D}{R^2}} \, dW_i^{\theta} \\
        dz_i &= F_i^{z} \, dt + \sqrt{2D} \, dW_i^{z}
    \end{aligned}
    \label{eq:langevin_cylinder}
\end{equation}
where $dW_i^{\theta}$ and $dW_i^{z}$ are independent standard Wiener processes. The factor $1/R^2$ in the angular noise arises from $g^{\theta\theta} = 1/R^2$. Since the metric is constant, no additional drift terms from the It\^{o}-Stratonovich conversion are needed.

\subsection{Dynamics on the Torus $T^2$}
\label{sec:torus_appendix}

The torus with major radius $R$ and minor radius $r$ is embedded in $\mathbb{R}^3$ via:
\begin{equation}
    \mathbf{r}(\theta, \varphi) = \left( (R + r\cos\varphi)\cos\theta, (R + r\cos\varphi)\sin\theta, r\sin\varphi \right)
\end{equation}
where $\theta \in [0, 2\pi)$ is the toroidal angle and $\varphi \in [0, 2\pi)$ is the poloidal angle. The induced metric is:
\begin{equation}
    g_{ij} = \begin{pmatrix} (R + r\cos\varphi)^2 & 0 \\ 0 & r^2 \end{pmatrix}, \quad
    g^{ij} = \begin{pmatrix} 1/(R + r\cos\varphi)^2 & 0 \\ 0 & 1/r^2 \end{pmatrix}
    \label{eq:metric_torus}
\end{equation}
The Gaussian curvature $K = \cos\varphi/[r(R + r\cos\varphi)]$ is positive on the outer equator, negative on the inner equator, and zero at $\varphi = \pm\pi/2$.

The geodesic equations derived from the metric are:
\begin{equation}
    \begin{aligned}
        \frac{d^2\theta}{ds^2} + \frac{2r\sin\varphi}{R + r\cos\varphi} \frac{d\theta}{ds}\frac{d\varphi}{ds} &= 0 \\[0.5em]
        \frac{d^2\varphi}{ds^2} - \frac{(R + r\cos\varphi)\sin\varphi}{r^2} \left(\frac{d\theta}{ds}\right)^2 &= 0
    \end{aligned}
    \label{eq:geodesic_equations_torus}
\end{equation}
These coupled nonlinear ODEs do not admit closed-form solutions for general boundary conditions. Special geodesics include: meridians ($\theta = \text{const}$) with arc length $r|\Delta\varphi|$; the outer equator ($\varphi = 0$) with arc length $(R+r)|\Delta\theta|$; the inner equator ($\varphi = \pi$) with arc length $(R-r)|\Delta\theta|$. Other latitude circles ($\varphi = \text{const} \neq 0, \pi$) are not geodesics.

For simulation purposes, we consider several approximations. The constant-metric (flat) approximation:
\begin{equation}
    d_{\text{flat}}(i,j) = \sqrt{(R \Delta\theta_{ij})^2 + (r \Delta\varphi_{ij})^2}
    \label{eq:geodesic_torus_flat}
\end{equation}
The midpoint-metric approximation:
\begin{equation}
    d_{\text{mid}}(i,j) = \sqrt{\left[(R + r\cos\bar{\varphi}_{ij}) \Delta\theta_{ij}\right]^2 + (r \Delta\varphi_{ij})^2}
    \label{eq:geodesic_torus_midpoint}
\end{equation}
where $\bar{\varphi}_{ij} = (\varphi_i + \varphi_j)/2$. For force computation, we use the local metric at particle $i$'s position with $R_i = R + r\cos\varphi_i$:
\begin{equation}
    d_{\text{loc}}(i,j) = \sqrt{(R_i \Delta\theta_{ij})^2 + (r \Delta\varphi_{ij})^2}
    \label{eq:geodesic_torus_local}
\end{equation}

The force on particle $i$ from particle $j$ is:
\begin{equation}
    F_{ij}^{\theta} = \phi_{ij} \frac{\Delta\theta_{ij}}{d_{\text{loc}}(i,j)}, \qquad
    F_{ij}^{\varphi} = \phi_{ij} \frac{\Delta\varphi_{ij}}{d_{\text{loc}}(i,j)}
    \label{eq:force_torus_local}
\end{equation}

The covariant Langevin equation with position-dependent inverse metric takes the Stratonovich form:
\begin{equation}
    \begin{aligned}
        d\theta_i &= F_i^{\theta} \, dt + \sqrt{\frac{2D}{(R + r\cos\varphi_i)^2}} \circ dW_i^{\theta} \\
        d\varphi_i &= F_i^{\varphi} \, dt + \sqrt{\frac{2D}{r^2}} \, dW_i^{\varphi}
    \end{aligned}
    \label{eq:langevin_torus_strat}
\end{equation}
where $\circ$ denotes Stratonovich integration. Converting to It\^{o} form introduces a noise-induced drift:
\begin{equation}
    \begin{aligned}
        d\theta_i &= \left(F_i^{\theta} + \frac{D \sin\varphi_i}{(R + r\cos\varphi_i)^2}\right) dt + \sqrt{\frac{2D}{(R + r\cos\varphi_i)^2}} \, dW_i^{\theta} \\
        d\varphi_i &= F_i^{\varphi} \, dt + \sqrt{\frac{2D}{r^2}} \, dW_i^{\varphi}
    \end{aligned}
    \label{eq:langevin_torus_ito}
\end{equation}
The drift term $D\sin\varphi_i/(R + r\cos\varphi_i)^2$ ensures equilibrium is uniform with respect to the Riemannian area element $dA = (R + r\cos\varphi) \cdot r \, d\theta \, d\varphi$.

\subsection{Zero-Torque Property and Orientation Evolution}
\label{sec:angular_momentum_appendix}

As established in the main text (equation (\ref{eq:zero_torque})), pairwise interaction forces contribute zero net torque: the torque from particle $j$ on particle $i$ is $\boldsymbol{\tau}_{ij} = \phi_{ij} (\mathbf{x}_i \times \mathbf{x}_j)/\sin\theta_{ij}$, and since $\phi_{ij} = \phi_{ji}$, each pair's total torque vanishes. Consequently, the collective orientation evolves as a pure stochastic process driven only by thermal noise. See Appendix~\ref{app:rotational_diffusion} for the theoretical framework.

\section{Rotational Diffusion: Theoretical Framework}
\label{app:rotational_diffusion}

This appendix presents the theoretical basis for rotational diffusion in overdamped systems, following the treatment of van Kampen \cite{vankampen2007} and related work on rotational Brownian motion \cite{gardiner2009, hofling2025}. We first review the general theory for a rigid body, then adapt it to our multi-particle setting.

\subsection{Van Kampen's Derivation of Rotational Diffusion}

Consider a rigid body with moment of inertia tensor $\mathsf{I}$ immersed in a viscous fluid. In the body-fixed frame where $\mathsf{I}$ is diagonal, the angular momentum $\mathbf{L}$ satisfies the Langevin equation \cite{vankampen2007}:
\begin{equation}
    \frac{d\mathbf{L}}{dt} = \mathbf{L} \times (\mathsf{I}^{-1} \cdot \mathbf{L}) - \mathsf{A} \cdot \mathsf{I}^{-1} \cdot \mathbf{L} + \boldsymbol{\tau}_\xi(t)
    \label{eq:vk_langevin}
\end{equation}
where $\mathsf{A}$ is the friction tensor and $\boldsymbol{\tau}_\xi(t)$ is Gaussian white noise satisfying the fluctuation-dissipation relation $\langle \boldsymbol{\tau}_\xi(t) \otimes \boldsymbol{\tau}_\xi(t') \rangle = 2k_BT \mathsf{A} \, \delta(t-t')$.

The full state includes both angular momentum $\mathbf{L}$ and orientation angles $\Omega$ (e.g., Euler angles). The probability density $P(\mathbf{L}, \Omega, t)$ evolves according to a Kramers equation that couples the fast variable $\mathbf{L}$ to the slow variable $\Omega$.

\subsubsection{The Overdamped Limit}

In the limit of large friction ($\mathsf{A} \to \infty$), the angular momentum relaxation time $\tau_L \sim I/A$ becomes very short. Van Kampen shows that on timescales $t \gg \tau_L$, the angular momentum thermalizes instantaneously to its equilibrium distribution, and can be ``integrated out'' (adiabatic elimination). The resulting equation for the marginal distribution $w(\Omega, t)$ is the \emph{rotational diffusion equation}:
\begin{equation}
    \frac{\partial w}{\partial t} = \sum_{i,j} \hat{J}_i D_{ij}^{\text{rot}} \hat{J}_j \, w
    \label{eq:rot_diffusion}
\end{equation}
where $\hat{J}_i$ are the generators of infinitesimal rotations and $\mathsf{D}^{\text{rot}} = k_BT \mathsf{A}^{-1}$ is the rotational diffusion tensor (Einstein-Smoluchowski relation).

The corresponding overdamped Langevin equation for the orientation angles is:
\begin{equation}
    d\Omega = \sqrt{2\mathsf{D}^{\text{rot}}} \cdot d\mathbf{W}
    \label{eq:overdamped_orientation}
\end{equation}
in the Stratonovich interpretation (which arises naturally from the physical limit of the inertial dynamics \cite{vankampen2007, gardiner2009}).

\subsubsection{Summary: Orientation Diffuses, Not Angular Momentum}

In the overdamped regime:
\begin{enumerate}
    \item Angular momentum $\mathbf{L}$ becomes a fast variable with trivial dynamics (instantaneous equilibration)
    \item Orientation angles $\Omega$ are the slow variables that undergo diffusion
    \item The mean-squared angular displacement grows linearly: $\langle \Delta\Omega^2 \rangle = 2D_{\text{rot}} t$
\end{enumerate}

\subsection{Application to Multi-Particle Systems}

Our system of $N$ interacting particles is not a rigid body, but similar principles apply to the collective orientation of emergent structures (band, rings, clusters). After structure formation, the particles move approximately coherently, and the structure's orientation becomes a well-defined collective coordinate.

For particles on a manifold embedded in $\mathbb{R}^3$, the cumulative rotation about the $z$-axis is
\begin{equation}
    \Theta_z(t) = \int_0^t \sum_i \rho_i^2 \, d\theta_i
    \label{eq:Theta_z_def}
\end{equation}
where $\rho_i$ is the distance from the $z$-axis and $\theta_i$ is the azimuthal angle. This quantity has dimensions [length]$^2$; dividing by the total ``moment of inertia'' $\sum_i \rho_i^2$ gives a dimensionless rotation angle.

Using the overdamped Langevin equation for each particle and the zero-torque property (no deterministic drift), we obtain
\begin{equation}
    d\Theta_z = \sqrt{2D} \sum_i \rho_i \, dW_i^\theta
    \label{eq:dTheta_z}
\end{equation}
This is pure diffusion with effective coefficient $D_{\text{rot}} = D \sum_i \rho_i^2$.

\subsection{Geometry-Specific Results}

\subsubsection{Sphere $S^2$}

On the unit sphere with spherical coordinates $(\theta, \phi)$, the distance from the $z$-axis is $\rho = \sin\theta$. The cumulative azimuthal rotation satisfies:
\begin{equation}
    d\Theta_z = \sqrt{2D} \sum_{i=1}^{N} \sin\theta_i \, dW_i^{\phi}
    \label{eq:dTheta_sphere}
\end{equation}
For a uniform distribution, $\langle \sin^2\theta \rangle = 2/3$, giving $D_{\text{rot}} = \frac{2ND}{3}$.

\subsubsection{Cylinder $S^1 \times [0,H]$}

On a cylinder of radius $R$, the distance from the axis is constant: $\rho = R$. The cumulative rotation satisfies:
\begin{equation}
    d\Theta_z = \sqrt{2D} \cdot R \sum_{i=1}^{N} dW_i^{\theta}
    \label{eq:dTheta_cylinder}
\end{equation}
with $D_{\text{rot}} = DR^2 N$. Since the noise terms are independent, $\Theta_z$ is the sum of $N$ independent random walks, giving $\Theta_z \sim \sqrt{N}$ scaling.

\subsubsection{Torus $T^2$}

On the torus with major radius $R$ and minor radius $r$, the distance from the $z$-axis depends on the poloidal angle: $\rho(\varphi) = R + r\cos\varphi$. 

The Langevin equation in Stratonovich form is given by (\ref{eq:langevin_torus_strat}). Since the noise amplitude $b^\theta = \sqrt{2D}/\rho(\varphi)$ depends on $\varphi$, converting to It\^{o} form introduces the drift correction (\ref{eq:torus_drift_correction}). The correction arises from the Stratonovich-It\^{o} conversion:
\begin{equation}
    \Delta a^\theta = \frac{1}{2} b^\varphi \frac{\partial b^\theta}{\partial \varphi} = \frac{1}{2} \cdot \frac{\sqrt{2D}}{r} \cdot \sqrt{2D} \cdot \frac{r\sin\varphi}{\rho^2} = \frac{D\sin\varphi}{\rho^2}
\end{equation}
giving the It\^{o} form (\ref{eq:langevin_torus_ito}).

For the cumulative toroidal rotation, we define $\Theta_\theta := \int_0^t \sum_i \rho_i^2 \, d\theta_i$. Substituting (\ref{eq:langevin_torus_ito}):
\begin{equation}
    d\Theta_\theta = \sum_i \rho_i^2 F_i^\theta \, dt + D\sum_i \sin\varphi_i \, dt + \sqrt{2D} \sum_{i=1}^{N} \rho_i \, dW_i^{\theta}
    \label{eq:dTheta_torus_full}
\end{equation}
The first term represents the deterministic torque from pairwise forces. Unlike the sphere and cylinder, where the zero-torque property holds exactly, on the torus $\sum_i \rho_i^2 F_i^\theta = \sum_{i<j} \phi_{ij} (\Delta\theta_{ij}/d_{\text{loc}})(\rho_i^2 - \rho_j^2)$ vanishes only when interacting particles have equal $\rho$ values (same poloidal angle). For ring configurations where particles within each ring share the same $\varphi$, inter-particle forces within each ring produce zero torque. Inter-ring forces may contribute nonzero torque, but for symmetric two-ring configurations this contribution is small. We therefore approximate $\sum_i \rho_i^2 F_i^\theta \approx 0$ for the structured configurations observed in simulations.

The second term in (\ref{eq:dTheta_torus_full}) is a \emph{state-dependent drift} arising from the It\^{o} correction; it vanishes when particles are uniformly distributed in $\varphi$ (since $\langle \sin\varphi \rangle = 0$). For uniformly distributed particles with negligible force-induced torque, the evolution reduces to pure diffusion:
\begin{equation}
    d\Theta_\theta = \sqrt{2D} \sum_{i=1}^{N} (R + r\cos\varphi_i) \, dW_i^{\theta}
    \label{eq:dTheta_torus_diffusive}
\end{equation}
The variance is $\langle (d\Theta_\theta)^2 \rangle = 2D \sum_i \rho_i^2 \, dt$. For uniform distributions with respect to the Riemannian area element, $\langle \rho^2 \rangle = R^2 + 3r^2/2$, giving $D_{\text{rot}} = DN(R^2 + 3r^2/2)$.

The poloidal cumulative rotation $\Theta_\varphi := \int_0^t \sum_i r^2 \, d\varphi_i$ evolves as:
\begin{equation}
    d\Theta_\varphi = \sqrt{2D} \cdot r \sum_{i=1}^{N} dW_i^{\varphi}
\end{equation}
with $D_{\text{rot}} = Dr^2 N$, since the $\varphi$-direction has constant metric factor $g_{\varphi\varphi} = r^2$ and thus no It\^{o} correction.

\section{Covariant Integration via Cholesky Decomposition}
\label{app:covariant}

This appendix details the intrinsic covariant integration scheme. While the embed-and-project method is used in our simulations (and is preferable for the sphere due to coordinate singularities), the covariant formulation generalizes to more complex geometries.

\subsection{It\^{o}'s Formulation of Brownian Motion on Manifolds}

The mathematical foundation for Brownian motion on Riemannian manifolds was established by It\^{o} \cite{ito1962}, who showed that the infinitesimal generator of Brownian motion on a manifold $\mathcal{M}$ is one-half the Laplace-Beltrami operator:
\begin{equation}
    \mathcal{L} = \frac{1}{2}\Delta_{\mathcal{M}} = \frac{1}{2}g^{ij}\nabla_i\nabla_j = \frac{1}{2}g^{ij}\frac{\partial^2}{\partial \varphi^i \partial \varphi^j} - \frac{1}{2}g^{ij}\Gamma^k_{ij}\frac{\partial}{\partial \varphi^k}
    \label{eq:ito_generator}
\end{equation}
where $\Gamma^k_{ij}$ are the Christoffel symbols. The second term in (\ref{eq:ito_generator}) gives rise to a \emph{curvature-induced drift} in the stochastic differential equation. Defining the drift velocity $m^k = -\frac{1}{2}g^{ij}\Gamma^k_{ij}$, the SDE for free Brownian motion in intrinsic coordinates takes the form
\begin{equation}
    d\varphi^i = D \, m^i \, dt + \sqrt{2D} \, \sigma^i_k \, dW^k
    \label{eq:ito_sde}
\end{equation}
where $\sigma^i_k$ satisfies $\sum_k \sigma^i_k \sigma^j_k = g^{ij}$ (i.e., $\sigma$ is the Cholesky factor of the inverse metric).

\paragraph{Connection to the Embed-and-Project Method.}
The Zinn-Justin approach (Section~\ref{sec:langevin_manifolds}) and It\^{o}'s intrinsic formulation yield equivalent dynamics through the following correspondence. Starting from the embedding-space Langevin equation and projecting onto the manifold, we obtain the covariant form (\ref{eq:langevin_intrinsic}):
\begin{equation}
    d\varphi^i = -\frac{1}{\gamma} g^{ij} \partial_j A \, dt + \sqrt{2D} \, dW^i_{\mathcal{M}}
\end{equation}
where $\langle dW^i_{\mathcal{M}} dW^j_{\mathcal{M}} \rangle = g^{ij} dt$. This equation is written in the \emph{Stratonovich} sense, which is the physically correct interpretation for noise arising from projection of Euclidean thermal fluctuations. Converting to It\^{o} form using (\ref{eq:strat_to_ito}) introduces the drift correction:
\begin{equation}
    \Delta a^i = \frac{1}{2} \sum_{j,k} \sigma^{jk} \frac{\partial \sigma^{ik}}{\partial \varphi^j} = -\frac{D}{2} g^{jk} \Gamma^i_{jk} = D \, m^i
\end{equation}
which is precisely It\^{o}'s curvature-induced drift. Thus, the embed-and-project method with Stratonovich interpretation is mathematically equivalent to It\^{o}'s intrinsic formulation: both ensure that the equilibrium distribution is the Boltzmann measure $\rho_{\text{eq}} \propto \sqrt{g} \, e^{-U/k_BT}$, uniform with respect to the Riemannian volume element.

For our geometries, the curvature-induced drift vanishes for the cylinder (flat metric) and the $\varphi$-direction of the torus, but is non-zero for the sphere and the $\theta$-direction of the torus. For the sphere, the embed-and-project method used in our simulations has been shown to produce the correct equilibrium distribution without explicit drift correction \cite{ciccotti2008, lelievre2010}. For the torus, where we use intrinsic coordinates with forward Euler discretization, the drift correction must be included explicitly (see Section~\ref{sec:simulation}).

\subsection{General Scheme}

The manifold Brownian motion $dW^i_{\mathcal{M}}$ in equation (\ref{eq:langevin_intrinsic}) has covariance
\begin{equation}
    \langle dW^i_{\mathcal{M}} \, dW^j_{\mathcal{M}} \rangle = g^{ij} \, dt
    \label{eq:manifold_noise_cov}
\end{equation}
To generate such noise from independent standard Gaussians, we use Cholesky decomposition of the inverse metric:
\begin{equation}
    g^{ij} = \sum_{k=1}^{d} L^{ik} L^{jk} = (L L^T)^{ij}
    \label{eq:cholesky}
\end{equation}
where $L$ is a lower triangular matrix. Given independent $\eta^k \sim \mathcal{N}(0, \sqrt{dt})$, the correlated noise is $dW^i_{\mathcal{M}} = \sum_{k} L^{ik} \eta^k$.

Since Euler-Maruyama is an It\^{o}-type scheme (Section~\ref{sec:ito_stratonovich}), simulating Stratonovich dynamics requires including the curvature-induced drift $D m^i$. The complete covariant Euler-Maruyama scheme for Stratonovich dynamics reads:
\begin{equation}
    \varphi^{i,(n+1)} = \varphi^{i,(n)} + \left( -\frac{1}{\gamma} g^{ij} \partial_j A + D m^i \right) \Delta t + \sqrt{2D} \sum_{k} L^{ik}(\varphi^{(n)}) \, \eta^k
    \label{eq:covariant_euler}
\end{equation}
where $m^i = -\frac{1}{2}g^{jk}\Gamma^i_{jk}$ is the curvature-induced drift from (\ref{eq:ito_sde}). For manifolds with constant metric (like the cylinder), $m^i = 0$ and the drift term vanishes.

\subsection{Explicit Form for the Sphere}

In spherical coordinates $(\theta, \phi)$, the metric is $g_{ij} = \mathrm{diag}(1, \sin^2\theta)$ with inverse $g^{ij} = \mathrm{diag}(1, \csc^2\theta)$. The Cholesky factor is $L = \mathrm{diag}(1, \csc\theta)$. The curvature-induced drift is $m^\theta = \frac{1}{2}\cot\theta\csc\theta$ and $m^\phi = 0$.

The complete covariant Euler-Maruyama scheme for Stratonovich dynamics would be:
\begin{equation}
    \begin{aligned}
        \theta^{(n+1)} &= \theta^{(n)} + \left( F^\theta + \frac{D}{2}\cot\theta^{(n)}\csc\theta^{(n)} \right) \Delta t + \sqrt{2D \Delta t} \, \eta^\theta \\
        \phi^{(n+1)} &= \phi^{(n)} + \csc^2\theta^{(n)} \, F_\phi \Delta t + \sqrt{2D \Delta t} \, \csc\theta^{(n)} \, \eta^\phi
    \end{aligned}
    \label{eq:covariant_sphere}
\end{equation}
However, both the noise amplitude and the curvature drift diverge at the poles ($\theta = 0, \pi$), making this scheme numerically problematic. This is why the embed-and-project method (Section~\ref{sec:simulation}) is preferred for the sphere: it avoids coordinate singularities entirely and has been proven to produce correct equilibrium statistics without explicit drift computation \cite{ciccotti2008, lelievre2010}.

\subsection{Explicit Form for the Bounded Cylinder}

For a cylinder of radius $R$, the metric is $g_{ij} = \mathrm{diag}(R^2, 1)$ with inverse $g^{ij} = \mathrm{diag}(R^{-2}, 1)$. The Cholesky factor is $L = \mathrm{diag}(R^{-1}, 1)$. Since the metric is constant, the curvature-induced drift vanishes: $m^i = 0$. The covariant update equations are:
\begin{equation}
    \begin{aligned}
        \theta^{(n+1)} &= \left[ \theta^{(n)} + R^{-2} F_\theta \Delta t + \sqrt{2D \Delta t} \, R^{-1} \eta^\theta \right] \mod 2\pi \\
        z^{(n+1)} &= \mathcal{B}_z\left( z^{(n)} + F^z \Delta t + \sqrt{2D \Delta t} \, \eta^z \right)
    \end{aligned}
    \label{eq:covariant_cylinder}
\end{equation}
This coincides with the scheme in Section~\ref{sec:simulation}, confirming that no drift correction is needed for the flat cylinder geometry.

\subsection{Explicit Form for the Torus}

For a torus with major radius $R$ and minor radius $r$, the metric is $g_{ij} = \mathrm{diag}((R + r\cos\varphi)^2, r^2)$ with inverse $g^{ij} = \mathrm{diag}((R + r\cos\varphi)^{-2}, r^{-2})$. The Cholesky factor is $L = \mathrm{diag}((R + r\cos\varphi)^{-1}, r^{-1})$.

The $\theta$-component of the Cholesky factor depends on $\varphi$, so the curvature-induced drift is non-zero: $m^\theta = \sin\varphi/(R + r\cos\varphi)^2$ and $m^\varphi = 0$. Including this drift correction, the complete covariant Euler-Maruyama scheme is:
\begin{equation}
    \begin{aligned}
        \theta^{(n+1)} &= \left[ \theta^{(n)} + \left((R + r\cos\varphi^{(n)})^{-2} F_\theta + \frac{D\sin\varphi^{(n)}}{(R + r\cos\varphi^{(n)})^2}\right) \Delta t \right. \\
        &\qquad \left. + \sqrt{2D \Delta t} \, (R + r\cos\varphi^{(n)})^{-1} \eta^\theta \right] \mod 2\pi \\
        \varphi^{(n+1)} &= \left[ \varphi^{(n)} + r^{-2} F_\varphi \Delta t + \sqrt{2D \Delta t} \, r^{-1} \eta^\varphi \right] \mod 2\pi
    \end{aligned}
    \label{eq:covariant_torus}
\end{equation}
This matches equation (\ref{eq:euler_torus}) in Section~\ref{sec:simulation}.

\subsection{Extension to Non-Diagonal Metrics}

For non-diagonal metrics with $g^{12} \neq 0$, the Cholesky decomposition produces correlated noise:
\begin{equation}
    L = \begin{pmatrix} \sqrt{g^{11}} & 0 \\ g^{12}/\sqrt{g^{11}} & \sqrt{g^{22} - (g^{12})^2/g^{11}} \end{pmatrix}
\end{equation}
The covariant formulation handles such cases naturally, while embed-and-project would require re-derivation.

\section{Analytical Approaches for Non-Equilibrium Glassy Dynamics}
\label{app:analytical}

This appendix provides a concise overview of theoretical methods applicable to the non-equilibrium dynamics observed in our simulations. The goal is to identify analytical frameworks that can predict relaxation timescales, aging behavior, and the nature of the non-equilibrium steady states (NESS) that emerge from disorder-induced self-organization.

\subsection{Summary of Dynamical Regimes}

Our simulations reveal three distinct dynamical regimes, each requiring different theoretical tools:

\textbf{(i) $\beta$-relaxation} ($t \lesssim 1$): Rapid structure formation via an instanton-like transition from the localized initial state. Particles expand and reorganize, with the system navigating a complex energy landscape shaped by quenched disorder. The effective energy barriers present during this phase guide the system toward low-dimensional configurations.

\textbf{(ii) $\alpha$-relaxation} ($1 \lesssim t \lesssim 10$): Consolidation of emergent structures (bands, rings, clusters) accompanied by slow collective dynamics. The timescale separation between fast intra-structure fluctuations and slow inter-structure evolution ($\tau_\alpha/\tau_\beta \gtrsim 20$) is characteristic of glassy systems. Aging manifests as waiting-time-dependent relaxation.

\textbf{(iii) Non-equilibrium steady state} ($t \gtrsim 10$): The instanton transition destroys the barriers that existed during $\beta$-relaxation, leaving a Mexican-hat-like potential with a continuous manifold of degenerate minima. The system settles into a NESS characterized by persistent reduced-dimensional structure with slowly diffusing symmetry-breaking direction. This regime exhibits weak ergodicity breaking: the instantaneous configuration always breaks rotational symmetry, but the symmetry-breaking direction drifts continuously, exploring the entire Goldstone manifold over long times.

\subsection{Two-Time Correlation and Response Functions}

The hallmark of non-equilibrium glassy dynamics is the dependence of observables on two times rather than their difference \cite{bouchaud1998}. For our model, the natural correlation function is
\begin{equation}
    C(t, t_w) = \frac{1}{N} \sum_{i=1}^{N} \langle \cos d(q_i(t), q_i(t_w)) \rangle,
\end{equation}
where $d$ denotes geodesic distance and $t_w$ is the waiting time. In aging systems, $C(t, t_w)$ depends on both times separately, with characteristic scaling forms such as $C(t, t_w) \approx f(t/t_w)$.

The response function $R(t, t_w) = \delta \langle q_i(t) \rangle / \delta h_i(t_w)|_{h=0}$ measures sensitivity to perturbations. The fluctuation-dissipation theorem (FDT), $TR = \partial C/\partial t_w$, holds in equilibrium but is violated in aging systems. The fluctuation-dissipation ratio $X(t, t_w) \leq 1$, defined through $TR = X \, \partial C/\partial t_w$, quantifies departure from equilibrium. We expect $X \approx 1$ for fast intra-structure fluctuations and $X < 1$ for slow collective rearrangements, with the geometry entering through the specific form of $X(C)$.

\subsection{Theoretical Methods}

We now survey theoretical frameworks applicable to different aspects of the dynamics.

\paragraph{Martin-Siggia-Rose-Janssen-de Dominicis (MSRJD) Formalism}

The MSRJD path integral \cite{martin1973, janssen1976} provides a field-theoretic approach to stochastic dynamics. The generating functional introduces auxiliary response fields $\tilde{q}_i$ with action
\begin{equation}
    S[q, \tilde{q}] = \int dt \sum_i \tilde{q}_i \cdot (\dot{q}_i - \mathbf{F}_i + D\tilde{q}_i).
\end{equation}
Disorder averaging over random couplings $\phi_{ij}$ generates effective interactions non-local in time, directly responsible for aging and memory effects. This formalism enables systematic computation of two-time correlation and response functions, derivation of FDT violations, and renormalization group analysis of aging dynamics.

\paragraph{Dynamical Mean-Field Theory}

In the large-$N$ limit with $J^2 \to J^2/N$ scaling, dynamical mean-field theory becomes exact \cite{bouchaud1998}. The self-consistent equations involve a memory kernel $\Sigma(t, s)$ and correlation function $G(t, s)$ that depend on both times separately, encoding the mathematical signature of aging. The Cugliandolo-Kurchan equations, originally derived for the spherical $p$-spin model, have natural generalizations to our geometric setting that would predict aging solutions with geometry-dependent characteristics.

\paragraph{Trap Models}

Trap models \cite{bouchaud1998} provide an intuitive energy-landscape picture: the system occupies metastable ``traps'' with Arrhenius escape times $\tau(E) = \tau_0 \exp(E/T)$. For exponentially distributed trap depths, $\rho(E) \sim e^{-E/T_g}$, aging emerges below $T_g$ as the system falls into progressively deeper traps. In our model, traps correspond to different band orientations (sphere), cluster configurations (cylinder), or ring positions (torus). The continuous configuration space means transitions occur as diffusive motion on the Goldstone manifold rather than discrete jumps between isolated minima.

\paragraph{Mode-Coupling Theory}

Mode-coupling theory (MCT) \cite{gotze2009} predicts a dynamic glass transition where relaxation times diverge. For our model, MCT equations must be formulated using geometry-specific basis functions: spherical harmonics on $S^2$, double Fourier modes on $T^2$, and mixed representations on the cylinder. MCT describes the fast $\beta$-relaxation within structures; aging theories (DMFT, trap models) are needed for the slow $\alpha$-relaxation of collective modes and the subsequent NESS dynamics.

Linear stability analysis of the uniform distribution $P_{\text{unif}} = 1/\text{Vol}(\mathcal{M})$ determines when clustering instabilities occur. Perturbations expand in eigenfunctions of the Laplace-Beltrami operator, with unstable modes signaling the onset of structure formation.

\subsection{Geometry-Specific Approximations}

Each manifold admits natural approximation schemes exploiting its symmetries:

\textbf{Sphere $S^2$}: Spherical harmonics expansion with systematic $1/\ell$ treatment of slow modes. The $\ell = 1$ sector (center-of-mass motion) decouples via angular momentum conservation. An effective theory for band dynamics involves the orientation vector $\mathbf{n}$ and width $w$ as collective coordinates.

\textbf{Torus $T^2$}: Double Fourier expansion in $(\theta, \varphi)$. The position-dependent metric $(R + r\cos\varphi)^2$ introduces mode coupling, but for thin tori ($r/R \ll 1$) a perturbative expansion in aspect ratio is tractable.

\textbf{Bounded cylinder}: Fourier expansion in the azimuthal direction combined with real-space treatment of axial dynamics. Well-separated clusters admit description as interacting quasi-particles with effective potentials obtained by integrating out intra-cluster degrees of freedom.

%

\end{document}